\documentclass[%
  preprint,preprintnumbers,
  showpacs,showkeys,
  amsmath,amssymb,
  aps,
  prd,
]{revtex4-1}

\usepackage{subfigure}         
\usepackage{graphicx}          
\usepackage{dcolumn}           
\usepackage{multirow}
\usepackage{bm}                
\usepackage{hyperref}          
\usepackage[mathlines]{lineno} 
\usepackage[colorinlistoftodos,prependcaption,textsize=tiny]{todonotes}

\DeclareGraphicsExtensions{.eps,.pdf,.png,.jpg,.jpeg}

\newcommand{\eP}{\texttt{ePump }}

     \newcommand*{\LEinline}[1]%
      {\todo[inline, size=\footnotesize]{#1}}

\begin{document}


\preprint{MSUHEP-18-015}
\preprint{FERMILAB-PUB-18-389-PPD}

\vspace{1in}

\title{A New Method for Reducing PDF Uncertainties in the High-Mass Drell-Yan Spectrum \\ 14 January 2019} 

\author{C. G. Willis}
\author{R. Brock}
\author{D. Hayden}
\author{T.-J. Hou}
\author{J. Isaacson*}
\author{C. Schmidt}
\author{C.-P. Yuan}

\affiliation{Department of Physics and Astronomy, Michigan State University, East Lansing Michigan, 48823, USA \\
*Current Address: Theory Division, Fermi National Accelerator Laboratory, Batavia IL 60510-5011, USA }

\date{\today}

\begin{abstract}
 Uncertainties in the parametrization of Parton Distribution Functions (PDFs) are becoming a serious limiting systematic %
 uncertainty in Large Hadron Collider (LHC) searches for Beyond the Standard Model physics. This is especially true %
 for measurements at high scales induced by quark and anti-quark collisions, where Drell-Yan continuum backgrounds are %
 dominant. Tools are recently available which enable exploration of PDF fitting strategies and emulate the effects of new data in a future global fit.  \texttt{ePump} is such a tool and it is shown that judicious selection %
 of measurable kinematical quantities can reduce the assigned systematic PDF uncertainties by significant factors. This %
 will be made possible by the huge statistical precision of future LHC  Standard Model datasets.
\end{abstract}



\maketitle


\section{Introduction}
%
Beyond the Standard Model (BSM) physics at the Large Hadron Collider (LHC) would be found as deviations from Standard Model (SM) expectations, possibly in rates, but more typically in the kinematic distributions of final state objects or their combinations---of jets, leptons, and missing energy. Therefore the importance of accurately and precisely modeling SM physics cannot be overstated. While the electroweak properties of the SM are very precisely known, precision knowledge of Parton Distributions Functions (PDFs) is becoming a limiting factor for many BSM searches. This limitation comes from the theoretical uncertainties becoming so large at high-mass that a clear deviation from the SM becomes hard to distinguish, and even upon discovery of new physics the characterisation of this signal among various different theoretical models would be blurred.

As PDFs are not analytically calculable in the framework of perturbative Quantum Chromodynamics (QCD), their shapes must be modeled by globally fitting measured distributions from many combinations of varied experimental data. Most of these data come from legacy experiments, such as Deep Inelastic Scattering (DIS) experiments, various fixed target hadron experiments, and the Fermilab Tevatron. LHC experimental results are beginning to be used in global PDF fits, and in the coming decades new knowledge of PDFs will come from measurements at ATLAS~\cite{Aad:2008zzm}, CMS~\cite{Chatrchyan:2008aa}, and LHCb~\cite{Alves:2008zz}. We suggest that new strategies are worth exploring and we present one here. 

Constraining PDFs and their uncertainties is now an intense research program. The systematic uncertainty in the PDF models arises from the 1) experimental uncertainties of the input data used in a global fit, 2) any theoretical assumptions made by the fitting groups, and/or 3) the chosen parameterizations characterizing the functional forms of the PDFs themselves. All of the global PDF fitting groups (\texttt{CTEQ-TEA}~\cite{PhysRevD.93.033006}, \texttt{MMHT}~\cite{Martin2009}, and \texttt{NNPDF}~\cite{Ball2015}) characterize their fits with Hessian error matrices or Monte Carlo replicas so that experiments can legitimately include PDF uncertainties as a component to any theoretical error for any measurement or limit. 

In this paper we explore the PDF uncertainties as they apply to the BSM search for a resonant $Z'$ gauge boson in the dilepton invariant mass spectrum. The dominant and irreducible background process to this search is the Drell-Yan (DY) process. Both ATLAS~\cite{Aaboud2017} and CMS~\cite{Sirunyan:2018exx} have recently completed their searches for new high-mass phenomena from the first $\sqrt{s}=13$~TeV data-taking runs at the LHC. Both set comparable lower bounds on the mass of a hypothetical new vector boson and both publish extensive lists of their systematic uncertainties, including uncertainties attributed to our limited knowledge of PDF fitting. 

To date, only 5\% of the planned LHC data are in hand and yet these PDF uncertainties might already have limited future mass reaches for such searches. Not only are resonant $Z'$ boson searches ``at risk'' but also $W'$ boson searches and especially non-resonant (such as contact interactions) searches, which are very sensitive to sloped shape changes in the background. Furthermore, as we enter the new high integrated luminosity era of the LHC, experimental uncertainties will naturally be continually reduced, meaning that searches with even more complicated final states will eventually start to become limited predominantly by theoretical uncertainties.  Therefore, it is critical that we improve our understanding of PDFs and their associated uncertainties. 
\subsection{Our Strategy}
Experiments utilize PDF fits which are global and agnostic respecting a basic principle of the parton model: PDF sets and uncertainties originate from all data and are applicable to all scattering. But knowledge of the PDFs is not uniform nor are all reactions similarly dependent on them. For example, DY production is less sensitive to knowledge of the gluon PDF than many BSM searches. Instead, precision predictions of DY processes depend significantly on knowledge of both the valence and sea quark densities which largely come from deep inelastic scattering and DY experiments. And to that end, hadron collider DY experimental inputs have been a part of PDF global fitting for years. For example, the \texttt{CT14NNLO}~\cite{PhysRevD.93.033006} fits utilized inputs from the $W$ and $Z$ boson charge asymmetry  measurements from the Tevatron:  
  \cite{Abe:1994rj,Abe:1996us} and \cite{Acosta:2005ud} from CDF and  \cite{D0:2014kma,Abazov:2007pm} results from D\O.

And for the first time, in \texttt{CT14NNLO} the \texttt{CTEQ-TEA} group included LHC data from $W/Z$ cross sections
and the charged lepton asymmetry measurement from ATLAS \cite{Aad:2011dm},
the charged lepton asymmetry in the electron \cite{Chatrchyan:2012xt} and
muon decay channels \cite{Chatrchyan:2013mza} from CMS, and
the $W/Z$ lepton rapidity distributions and charged
lepton asymmetry from LHCb \cite{Aaij:2012vn}. But we will show that modern PDF global fits are not as potent for quark densities as are necessary for future precision measurements. 

The only remedy to this problem is the addition of qualitatively new experimental inputs to global fitting, but the LHC is currently the only PDF ``game in town.''  We propose  a way to  judiciously use LHC DY data itself as inputs to global fitting. The strategy would be to add $Z$ boson peak and DY continuum data  to global fitting from a well-measured, low-to-moderate invariant mass  control region ($M<1$~TeV). The resulting, ``boutique'' PDF sets could be used in an unbiased way to constrain the theoretical uncertainties in a kinematic search region relevant to modern BSM particle hunt, which is now in the $M > 5$~TeV region. 

We further show that the DY kinematics can be exploited to enhance the impact on LHC DY data, namely emphasizing well-understood up-quark densities and de-emphasize always limited sea-quark densities. This would require inputs which are differential in nature and not just asymmetry results near the $Z$ boson peak.

The machinery of PDF global fitting groups is very complex and for physicists outside of the PDF groups, testing new PDF analysis strategies can be cumbersome. This will change with the recent development of tools like \texttt{ePump}~\cite{Schmidt:2018hvu} (the Error PDF Updating Method Package, see Appendix~\ref{epumpapp} and \cite{Schmidt:2018hvu} for details), which makes it possible to explore the effects of new kinematic inputs to a global fit without requiring a full global analysis. \eP is not a substitute for full global fitting, but can be used as a tool to probe the effects of new data.

In essence one can consider \eP to be a simulation of global fitting in an approximation described in Appendix~\ref{epumpapp}. Pseudo-data can be added to an existing global fit in order to explore how that data might affect the central value and importantly, the uncertainties in the resulting candidate PDFs. All of the sum rules, QCD evolution, and uncertainties inherent in the ``parent'' global fit to which test data are added are preserved.  While other PDF profiling tools exist such as xFitter~\cite{Bertone:2017tig}, in this paper we choose to use \eP which has been thoroughly tested~\cite{Schmidt:2018hvu} against the \texttt{CT14NNLO}~\cite{PhysRevD.93.033006} global fits.

The work in this paper is the first published use of \texttt{ePump}. We demonstrate that new insight into  kinematics of the DY process has emerged, and that considerable reduction in the quark and anti-quark PDF uncertainties is possible with new data inputs to PDF global fitting.

\subsection{Our Goals}

Our goals in this paper are limited. We simply ask the optimistic questions: can qualitatively new data when combined with the current inputs of \texttt{CT14HERA2} reduce future PDF uncertainties and if so, by how much? And would any reduction improve the overall precision of high mass DY backgrounds relevant to future $Z'$ searches? We exploit the unprecedented statistical power of future LHC running and use DY kinematically motivated differential distributions to suggest that sensitivities to partons of special interest in DY production can be enhanced. 

Our \textit{ansatz} is to treat BSM DY searches as consisting of a \textbf{control region}---from which we envision mining DY data for global fitting---and a \textbf{signal region} to where those new global fits are extrapolated. Of course as in any control-signal region analysis, the assumption is that the control region contains only SM physics.  We specifically explore the possibility that LHC DY data in a safe control region might be useful to further constrain PDFs appropriate to high-mass BSM searches for which the continuum DY is the dominant background. Having determined that this is worth consideration, our ultimate proposal is that the LHC experiments and the PDF fitting teams work together to explore inclusion of LHC DY data into global fitting when prepared in a particularly useful way.

We chose to do our work using the most recent CTEQ PDF global fit, namely  \texttt{CT14HERA2}. This includes the most recent HERA1 and HERA2 data and utilizes an updated parametrization from the previous  \texttt{CT14NNLO} sets. Since the recent experimental ATLAS publication~\cite{Aaboud2017} limits were set using the the  \texttt{CT14NNLO} sets, we do make a brief comparison to show that the basic PDFs are very similar. 

Our goals are limited to asking and answering our two questions above. To that basic end, what we do not do here are the following:

\begin{itemize}
\item An important part of the theoretical uncertainties include exploration of the parameterization assumed and potentially additional parameterization choices. While exploring functional choices would be an interesting exercise when attempting to extrapolate into a new kinematical regime, we do not do that here. 
\item We do not attempt to optimize theoretical uncertainties associated with any other theoretical considerations like the strong coupling constant, electroweak couplings, or higher order electroweak and QCD effects. 
\item We also make no effort to optimize or explore the full set of possible experimental uncertainties. 
\end{itemize}

The paper is structured as follows. First, in Sec.~\ref{sec:Results} the current experimental results are briefly reviewed with an emphasis on the systematic uncertainties. Next, we review  the kinematics of the DY process in Sec.~\ref{sec:DYProcess}. Particular attention is paid to the role of the Collins-Soper (CS) angle ($\theta^{*}$)~\cite{PhysRevD.16.2219}, as this variable will be shown to possess hitherto unemphasized discrimination power between up- and down-type quark flavors which varies as a function of the invariant mass of the lepton pairs. Bearing this in mind, we then propose a new strategy for future PDF global fitting inspired by the use of \eP in Sec.~\ref{sec:approachToPDFErrorReduction}. Then, in Sec.~\ref{sec:PDFUpdateResults}, the results of such a strategy are assessed, first on the high-$x$ behavior of the post-fit \texttt{CT14HERA2} PDFs, and then on the expected event yields of the high-mass dilepton spectrum. Finally, concluding remarks are given in Sec.~\ref{sec:Outlook}. Appendix~\ref{epumpapp} describes \eP in more detail while Appendix~\ref{upapp} provides some kinematical explanatory details.
\section{Current $Z'$ Boson Search Results} \label{sec:Results}
\begin{figure}[!t]
  \centering
  \subfigure[]{
    \label{fig:ATLASmass}
    \includegraphics[width=0.43\textwidth]{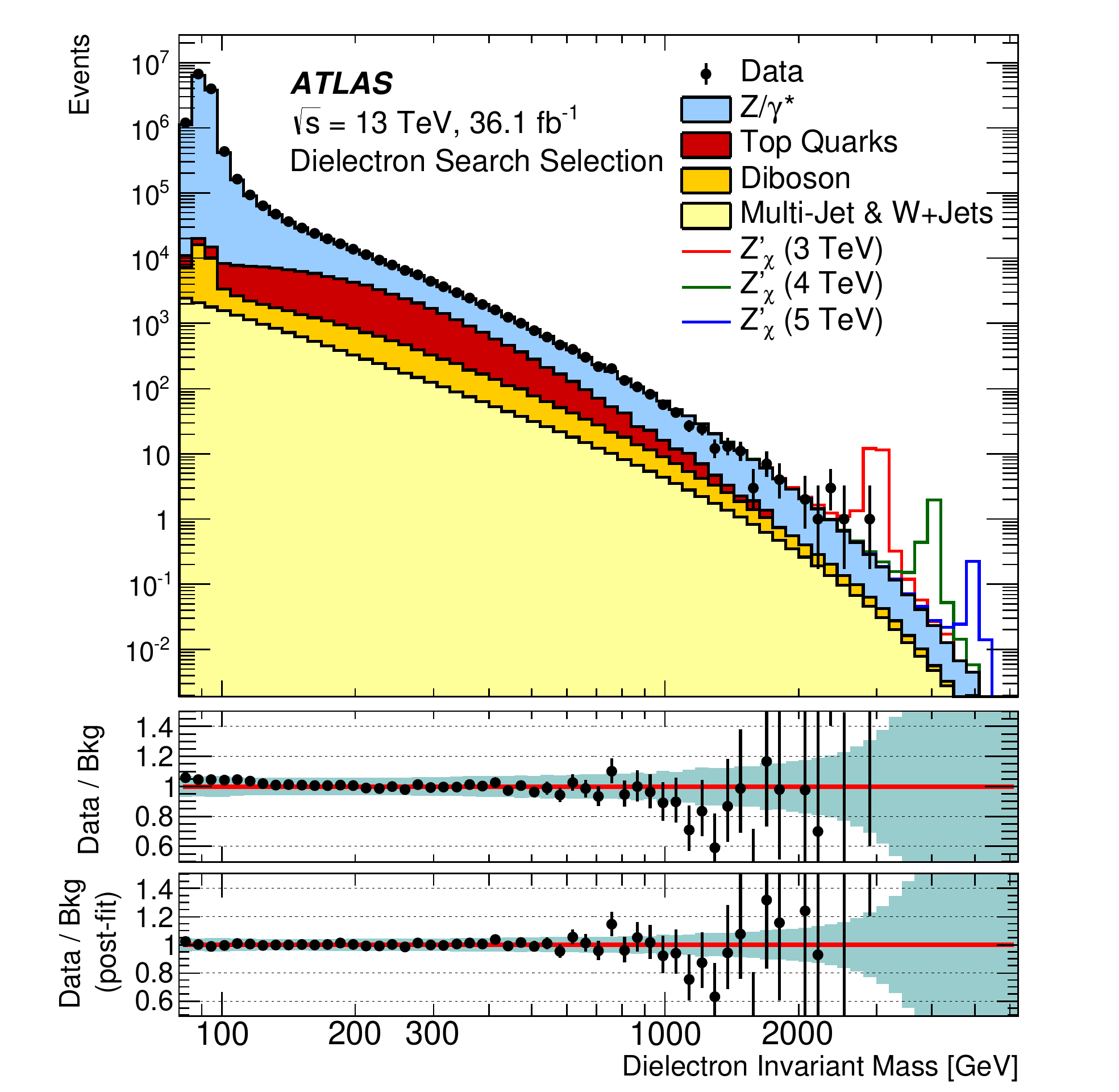}
  }
  \subfigure[]{
    \label{fig:CMSmass}
    \includegraphics[width=0.43\textwidth]{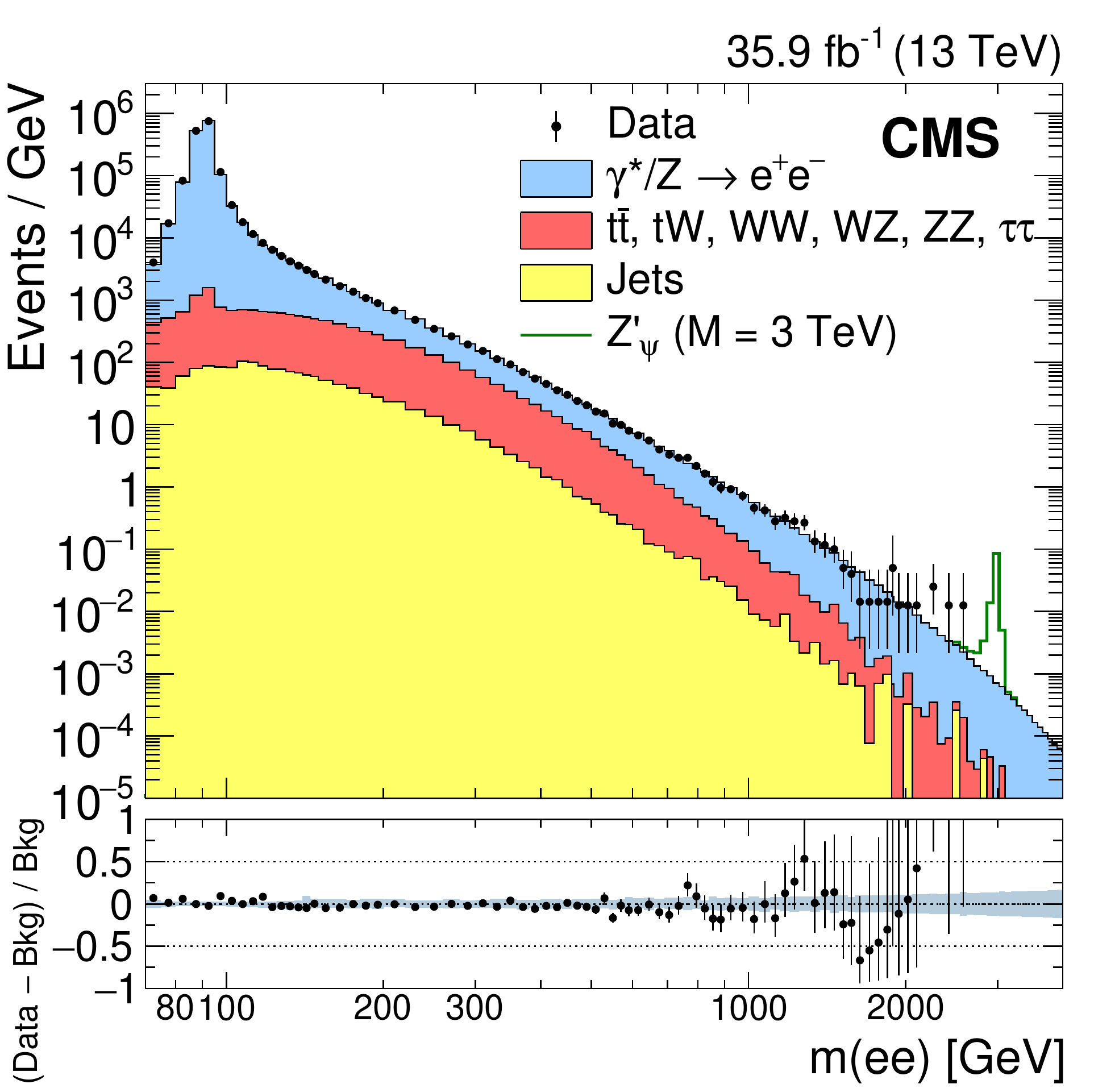}
  }
  \caption{Dielectron invariant mass search spectra in the (a) ATLAS~\cite{Aaboud2017} and CMS~\cite{Sirunyan:2018exx} %
    dilepton analyses at the LHC. }
  \label{fig:massPlots}
\end{figure}
\begin{table}[!t]
  \begin{ruledtabular}
    \begin{tabular}{c|c|c|c|c|c|c|c}
      Collaboration & $\sqrt{s}$ [TeV] & $\mathcal{L}$ $[\mathrm{fb}^{-1}]$ & Channel & \multicolumn{3}{c|}{Lower Limit on $M_{Z^{\prime}}$ [TeV]} & Reference \\
      \cline{5-7}
      &  &  &  & $Z^{\prime}_{\mathrm{SSM}}$ & $Z^{\prime}_{\psi}$ & $Z^{\prime}_{\chi}$ &  \\
      \hline
      CMS           & 13               & 36  & $\ell\ell$ & 4.5   & 3.9   & -     & \cite{Sirunyan:2018exx} \\
      ATLAS         & 13               & 36  & $\ell\ell$ & 4.5   & 3.8   & 4.1   & \cite{Aaboud2017} \\
    \end{tabular}
  \end{ruledtabular}
  \caption{Observed limits at 95\% C.L. on the mass of a $Z^{\prime}$ boson from the most recent LHC experimental searches. %
    The CMS analysis listed does not provide limits on the $Z^{\prime}_{\chi}$, which would otherwise be slightly higher than %
    what was obtained for the $Z^{\prime}_{\psi}$. The integrated luminosity for each analysis is rounded to the nearest whole %
    number. The $\ell\ell$ channel refers searches that combine individual electron and muon channels.}
  \label{tab:limitTablePrevious}
\end{table}
Both direct and indirect searches for $Z^{\prime}$ bosons have been conducted at several previous hadron collider experiments. Early results were obtained from the  D$\O$~\cite{ABAZOV201188} and CDF~\cite{PhysRevLett.106.121801} experiments at the Tevatron, and more recently, the ATLAS~\cite{PhysRevD.90.052005,2016372}  and CMS~\cite{Khachatryan2015,201757,Sirunyan:2018exx} experiments at the LHC. 

As the highest energy particle collider, the LHC experiments' ability to set $Z'$ limits is vastly improved compared to what was achievable at LEP and the Tevatron. The most stringent direct limits come from the ATLAS and CMS experiments where searches have been conducted at $\sqrt{s}=7, 8$ and $13$ TeV, with varying amounts of integrated luminosity.  The most recent results for the combined electron and muon pair invariant mass are shown in Fig.~\ref{fig:massPlots} for these two experiments. These searches usually consider two types of $Z^{\prime}$ models. The first model considered is a simple U(1) gauge extension called the Sequential Standard Model (SSM)~\cite{RevModPhys.81.1199}, where the coupling of the new gauge boson to SM particles is the same as the $Z$ boson. The second model considered is called the $E_6$ model~\cite{PhysRevD.34.1530}, and gives rise to the additional gauge boson through the decomposition of the $E_6$ grand unified theory gauge group. This can lead to a variety of different $Z^{\prime}$ scenarios and coupling to SM particles, depending on the mixings of two states. Of these possible scenarios the $Z^{\prime}_{\chi}$ has the widest width, and the $Z^{\prime}_{\psi}$ has the narrowest, leading to them often being used as two benchmarks to test both extremes of this model. 

Of particular interest are the systematic errors due to PDF fitting uncertainties. The two LHC experiments treat these quite differently. Table~\ref{tab:PDFuncertaintiesLHC} illustrates their assignments from CMS (\cite{Sirunyan:2018exx}) and ATLAS (\cite{Aaboud2017}). The PDF uncertainties for electron and muon pair backgrounds are shown for each experiment as are the total experimental uncertainties as quoted from each paper. The ATLAS experiment further assigns a ``PDF Choice'' uncertainty in accordance with the PDF4LHC forum~\cite{Butterworth:2015oua} to account for differences among the PDF fitting groups' predictions as excursions from the nominal choice and its full error matrix.

\begin{table}
  \begin{ruledtabular}
    \begin{tabular}{|c|ll|ll|}
      \multirow{2}{*}{Systematic Uncertainty} & \multicolumn{2}{l|}{CMS (\texttt{NNPDF2.3})} & \multicolumn{2}{l|}{ATLAS (\texttt{CT14NNLO})} \\
      & $ee$ [\%] & $\mu\mu$ [\%] & $ee$ [\%] & $\mu\mu$ [\%] \\
      \hline
      PDF Variation & - & - & 19 & 13 \\
      PDF Choice & - & - & 8.4 & 1.9 \\
      PDF Variation \& Choice & 7 & 7 & 20.8 & 13.1 \\
      Combined Experimental & 12 & 15 & 12.8 & 18.9 \\
    \end{tabular}
  \end{ruledtabular}
  \caption{Published uncertainties due to the lack of PDF knowledge on the DY backgrounds for CMS and ATLAS for the 13~TeV LHC running. Each experiment determines PDF uncertainties from a common nominal PDF choice which is parenthetically indicated. The uncertainties have a scale dependence and can differ according to the di-lepton channel. The results quoted here are evaluated at a mass of approximately 4~TeV, except the experimental results for CMS which are only quoted at 5~TeV in their paper. ``PDF choice'' for ATLAS results refers to the PDF4LHC forum recommendations~\cite{Butterworth:2015oua}. ``PDF variation'' is the result from the full error matrix for the nominal PDF set.}
  \label{tab:PDFuncertaintiesLHC}
\end{table}
The two experiments report different assignments for PDF uncertainties. For example ATLAS assigns large uncertainties derived from a detailed treatment. For the di-electron channel the reported overall uncertainty is 26.3\% which comes from: the combined  PDF (variation plus choice) uncertainties of 20.8\%,  other non-PDF theory uncertainties of 10\%, and total experimental uncertainties of 12.8\%. Di-muon uncertainties are not as large, but for both measurements the PDF uncertainties compete unfavorably with the experimental uncertainties. CMS reports smaller PDF uncertainties and comparable experimental uncertainties.

Experimental systematic uncertainties will likely be reduced with more data, but the PDF uncertainties at this point are largely irreducible in the absence of new data of a qualitatively different sort (new DIS experiments?) or new ideas. We propose new ideas to address this using LHC data itself.

\section{The Drell-Yan Process} \label{sec:DYProcess}
\begin{figure}[!t]
  \centering
  \includegraphics[width=1.0\textwidth]{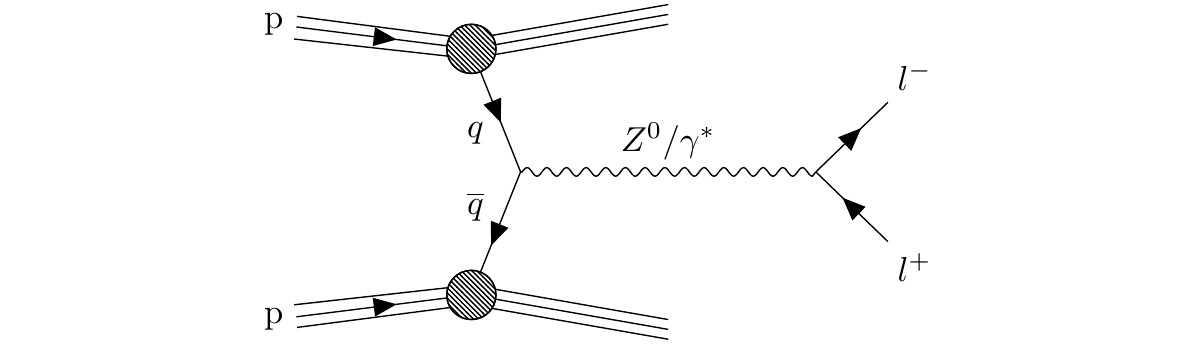}
  \caption{A Feynman diagram~\cite{Ellis:2016jkw} of the DY process initiated by a quark-antiquark pair as observed at the LHC.}
  \label{fig:DrellYans}
\end{figure}
The general Drell-Yan process~\cite{PhysRevLett.25.316} of $pp \rightarrow \ell^{+}\ell^{-} + X$ at leading order originates from an $s$-channel exchange of an electroweak boson
\begin{equation}\label{eq:s}
  q\overline{q} \rightarrow \gamma^{*}/Z \rightarrow \ell^{+}\ell^{-}.
\end{equation}
Here, $X$ denotes any additional final-state particles (radiated partons, the underlying event, multi-parton interactions, etc.). At next-to-leading order, the real corrections introduce three $t$-channel processes, listed in order of decreasing cross section at LHC energies,
\begin{align}
  qg &\rightarrow \gamma^{*}/Z \rightarrow \ell^{+}\ell^{-} + q \label{eq:t1}  \\ 
  \overline{q}g &\rightarrow \gamma^{*}/Z \rightarrow \ell^{+}\ell^{-} + q \label{eq:t2} \\
  q\overline{q} &\rightarrow \gamma^{*}/Z \rightarrow \ell^{+}\ell^{-} + g. \label{eq:t3} 
\end{align}
The leading order process is depicted in Fig.~\ref{fig:DrellYans}. \par
In each case, the vector boson decays into a pair of same-flavor, oppositely-charged leptons. For simplicity, our discussion will center on the leading order process, but all of our results are based on  Next to Leading Order (NLO) plus Next to Leading Log (NLL) calculations using the NLO-NLL  {\sc ResBos}~\cite{PhysRevD.50.R4239,PhysRevD.56.5558,PhysRevD.67.073016} package.

The DY triple-differential cross section can be represented as a function of the dilepton invariant mass $m_{\ell\ell}$, the dilepton rapidity $y_{\ell\ell}$, and the cosine of the lepton polar angle in the Collins-Soper rest frame $\cos\theta^{*}$. This was measured by ATLAS~\cite{Aaboud:2017exx} using data from  the  $\sqrt{s}=8$~TeV LHC running for $46< m_{\ell \ell} <150$~GeV.

For the LO $s$-channel process, the DY triple-differential cross section can be written as
\begin{equation} \label{eq:sigma3D}
  \frac{d^{3}\sigma}{dm_{\ell\ell}dy_{\ell\ell}d\cos\theta^{*}} = \frac{\pi\alpha^{2}}{3m_{\ell\ell}s} %
  \sum_{q} P_{q} \left[ f_{q/P_{1}}(x_{1},Q^{2}) f_{\bar{q}/P_{2}}(x_{2},Q^{2}) %
    + \left( q\leftrightarrow \overline{q} \right) \right].
\end{equation}
Here $\sqrt{s}$ is the centre of mass energy of the LHC, and $P_1$ and $P_2$ are the 4-momenta of protons 1 and 2. In the standard fashion, $x_{1}$ and $x_{2}$ are the incoming parton momentum fractions such that  $p_{1}=x_{1}P_{1}$ and  $p_{2}=x_{2}P_{2}$. We take our notation from~\cite{Aaboud:2017exx}.\par
The functions $f_{q/P_{1}}(x_{1},Q^{2})$ and $f_{\bar{q}/P_{2}}(x_{2},Q^{2})$ are the PDFs for quark flavors $q$ and $\bar{q}$, respectively. The term $(q\leftrightarrow \overline{q})$ accounts for the fact that either proton can carry a sea quark, as the LHC is a proton-proton collider. \par
Finally, the quantity $P_{q}$ accounts for the parton-level dynamics in terms of important electroweak parameters, and exhibits dependencies on both dilepton mass and $\cos\theta^\star$.  Each factor in this formula matters in a high-mass extrapolation, and are discussed in detail in Appendix~\ref{sec:relearning}. \par
\begin{figure}[t]
  \centering
  \subfigure[]{
    \label{fig:CT14nnlo_002GeV}
    \includegraphics[width=0.48\textwidth]{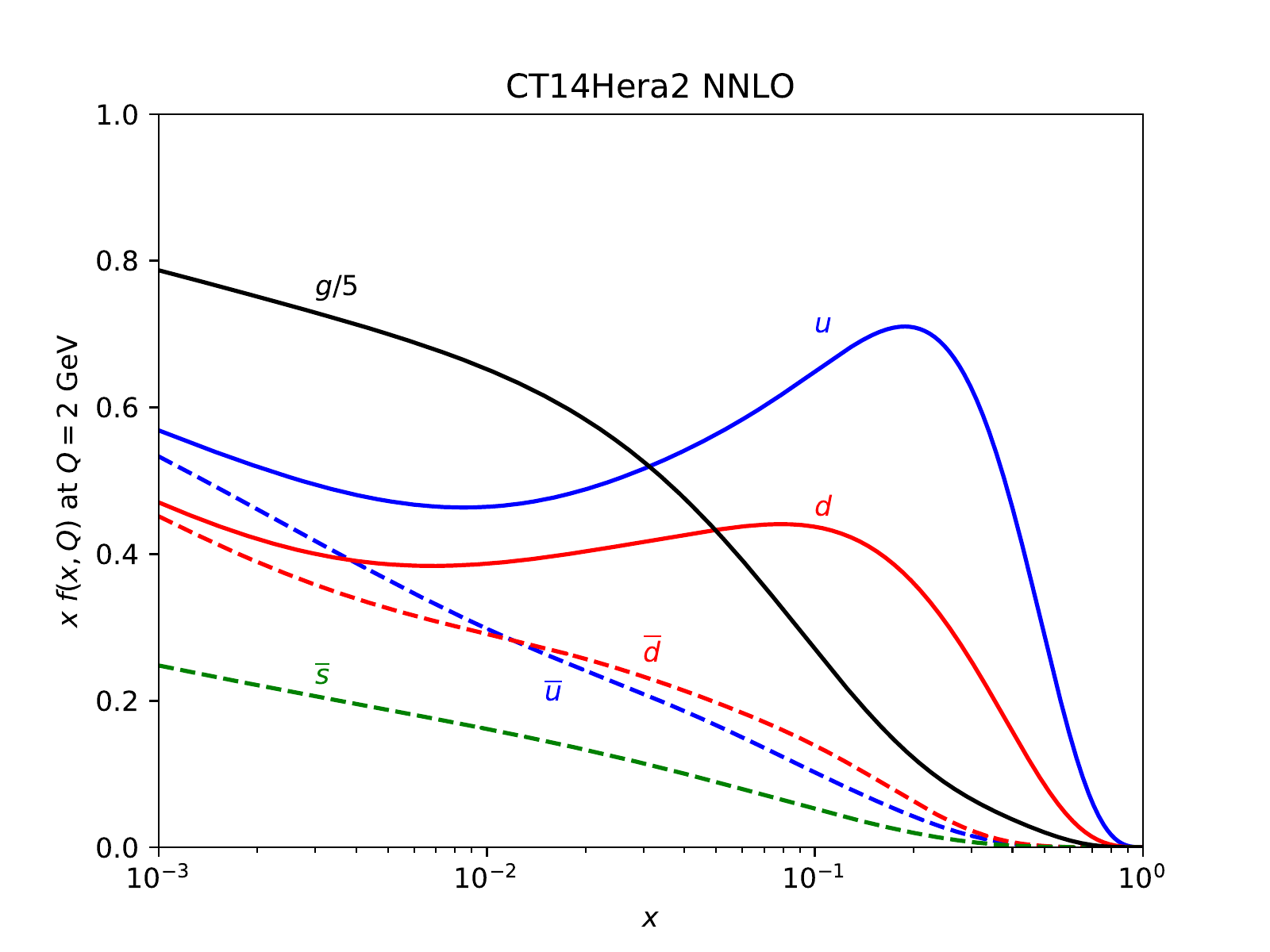}
  }
  \subfigure[]{
    \label{fig:CT14nnlo_100GeV}
    \includegraphics[width=0.48\textwidth]{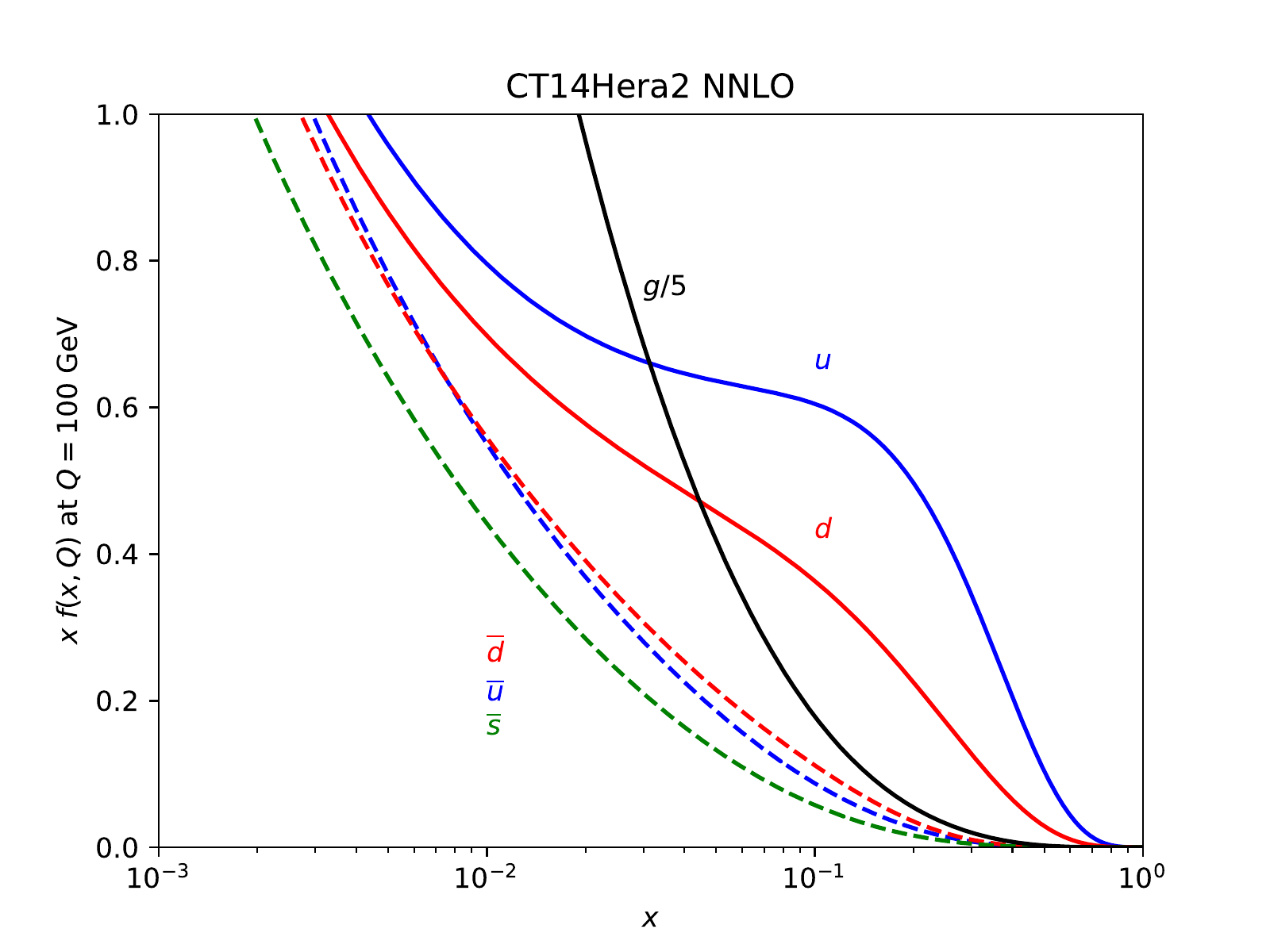}
  }
  \caption{The \texttt{CT14HERA2} PDFs of the CTEQ collaboration. Depicted are gluon, quark, and anti-quark PDFs as a %
    function of $x$, evaluated at a scale of $Q=2$ GeV~\subref{fig:CT14nnlo_002GeV} and $Q=100$ %
    GeV~\subref{fig:CT14nnlo_100GeV}~\cite{PhysRevD.93.033006}.}
  \label{fig:CT14nnlo}
\end{figure}

The energy scale of the collision is set by the transferred four-momentum squared $Q^{2}$, which can be identified with the square of the dilepton invariant mass $m^{2}_{\ell\ell}$. Well-known kinematic definitions include
\begin{align}
  Q^{2} = \left(p_{1}+p_{2}\right)^{2} = x_{1}x_{2}s,
\end{align}
and,
\begin{align}
  y_{\ell\ell} = \frac{1}{2} \ln \left( \frac{x_{1}}{x_{2}} \right),
\end{align}
which parametrizes the dilepton rapidity in terms of the $x$ fractions of the initial-state partons at LO. From these, the variables are related, also at LO, by
\begin{equation} \label{eq:xqxqb}
  x_{1}=\frac{m_{\ell\ell}}{\sqrt{s}}e^{+y_{\ell\ell}}, \quad x_{2}=\frac{m_{\ell\ell}}{\sqrt{s}}e^{-y_{\ell\ell}}.
\end{equation}
Eq.~(\ref{eq:xqxqb}) provides the first hint to the source of the large PDF uncertainty in high-mass DY production. The $\sqrt{s}=13$ TeV LHC is now probing extremely large values of $m_{\ell\ell}$, beyond a few TeV. As such, a central dilepton event with an invariant mass of $m_{\ell\ell}=3$~TeV and rapidity of $y_{\ell\ell}=0$ requires $x$ fractions beyond $x\simeq 0.2$.
This is beginning to probe regions of sea and even valence quark momentum fractions which are not well constrained by mostly DIS inputs. Figure~\ref{fig:CT14nnlo} shows quark, anti-quark and gluon momentum fractions from the \texttt{CT14HERA2} PDF set evaluated at two scales $Q^{2}$.
\subsection{Behavior of PDFs at high parton $x$}
\begin{figure}[th]
  \centering
  \includegraphics[width=0.6\textwidth]{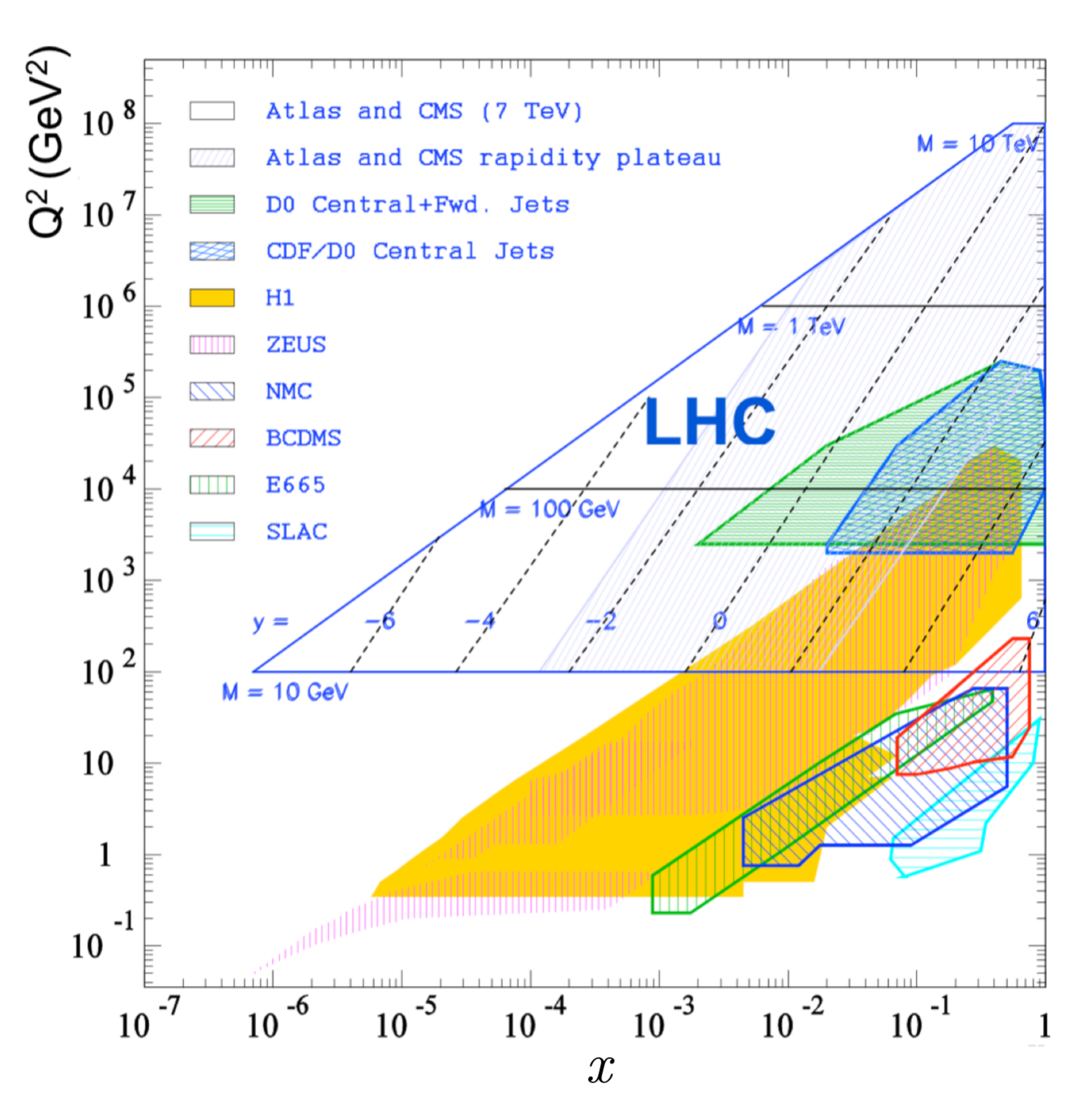}
  \caption{The transferred momentum squared $Q^{2}$ versus the parton momentum fraction $x$ at $\sqrt{s}=7$~TeV. The %
    regions probed by previous DIS, fixed-target, and collider-based experiments are labeled~\cite{Boonekamp:2009yd}.}
  \label{fig:xvsQ2New}
\end{figure}

The reason for this inherent high-$x$ uncertainty in the quark and anti-quark PDFs is due to the need to extrapolate experimental data---especially for quark and anti-quark fitting---as seen in Fig.~\ref{fig:xvsQ2New}. The only data which directly probe quark and anti-quark PDFs for $x \gtrsim 0.2$ come from legacy deep-inelastic scattering experiments and HERA measurements. PDFs relevant for current and future LHC DY production scales of interest require an extrapolation of almost three orders of magnitude in mass and this proves difficult to do precisely with the current world data. \par

\begin{figure}[!h]
  \centering
  \subfigure[]{
    \label{fig:uvl3000R}
    \includegraphics[width=0.43\textwidth]{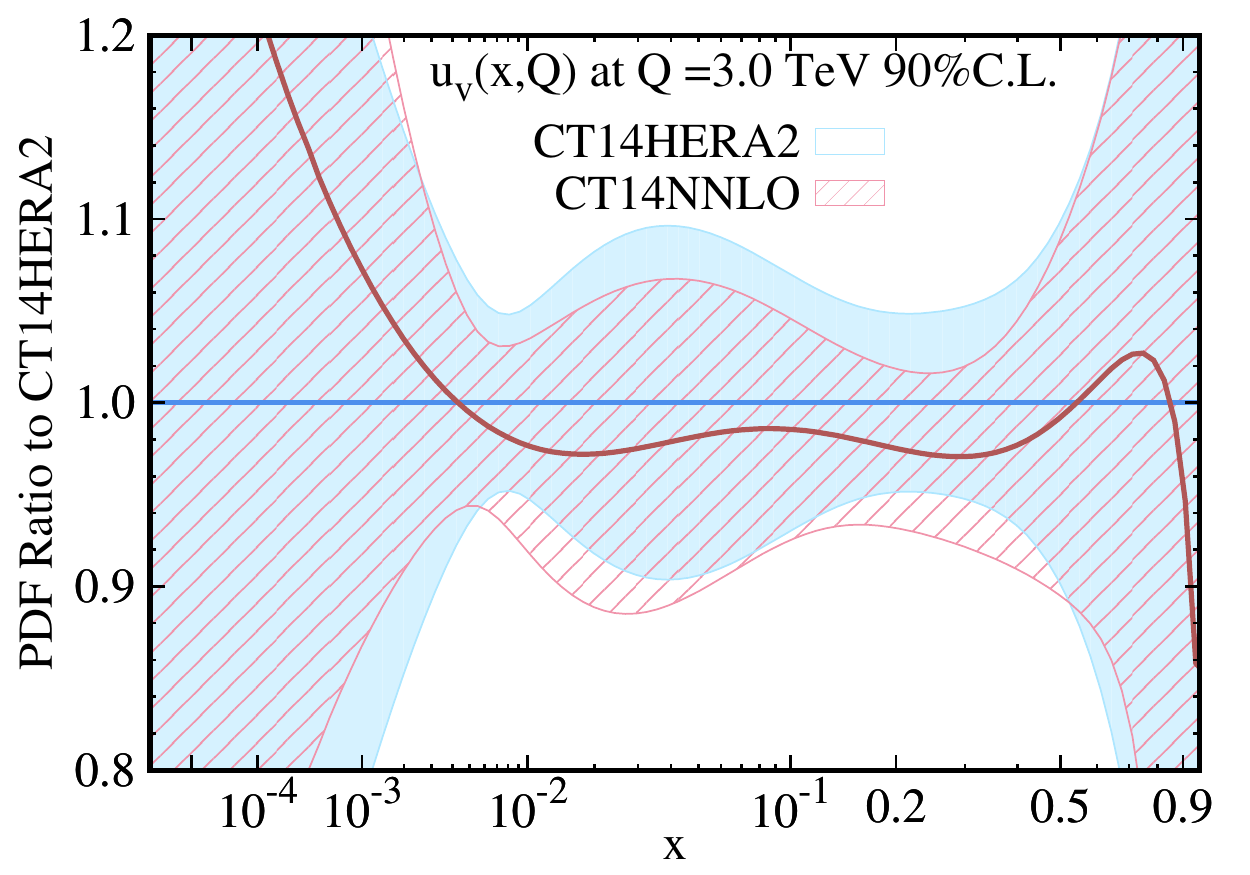}
  }
  \subfigure[]{
    \label{fig:dvl3000R}
    \includegraphics[width=0.43\textwidth]{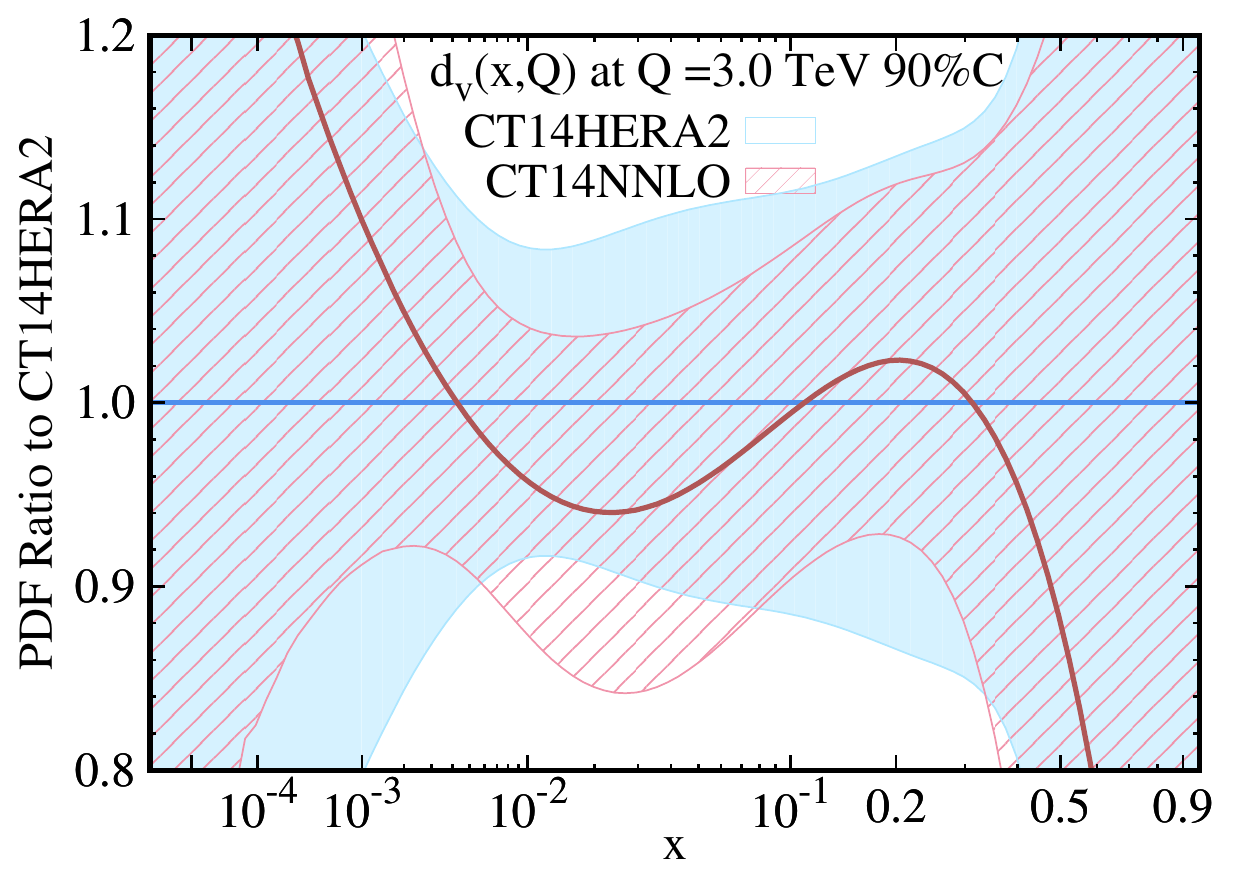}
  }\\
  \subfigure[]{
    \label{fig:ubar3000R}
    \includegraphics[width=0.43\textwidth]{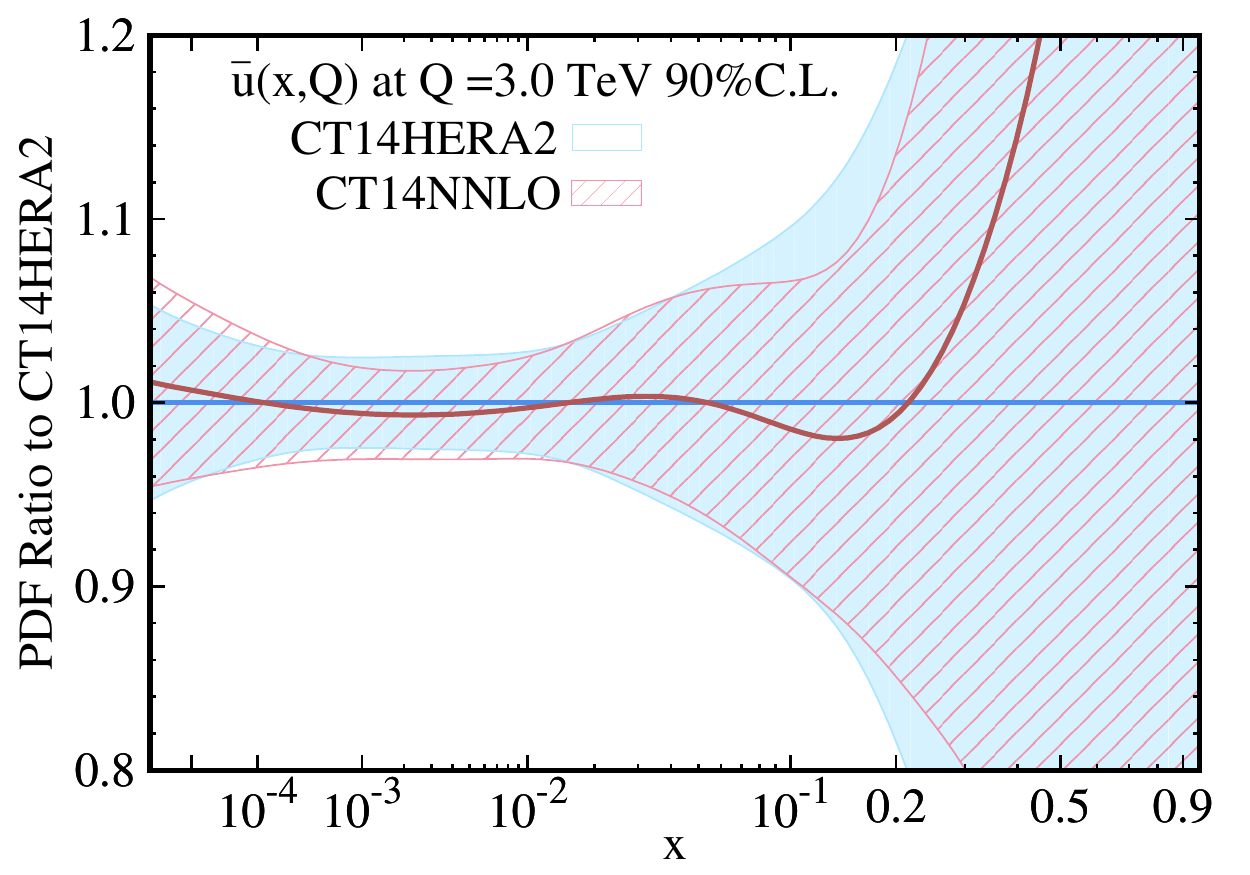}
  }
  \subfigure[]{
    \label{fig:dbar3000R}
    \includegraphics[width=0.43\textwidth]{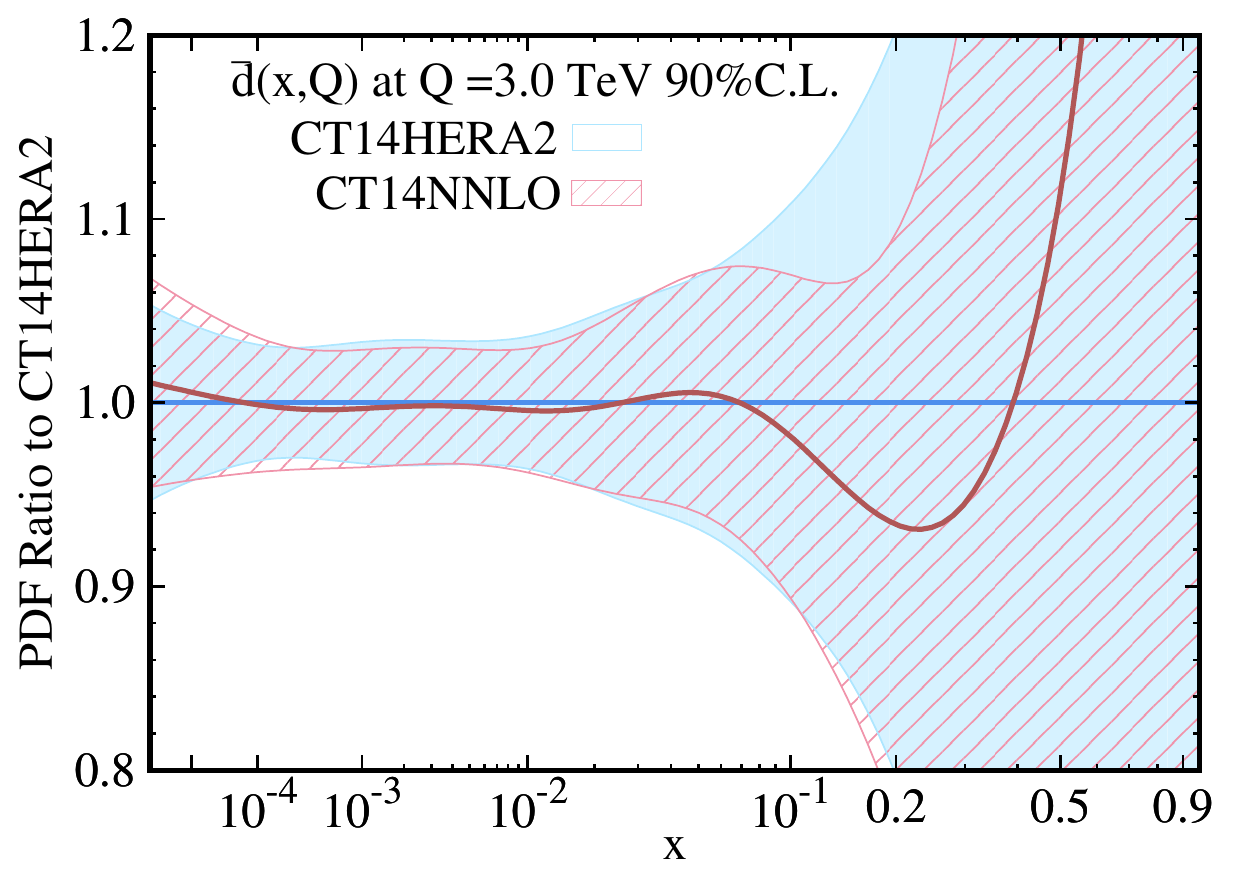}
  }
  \caption{PDF uncertainties associated with the ~\subref{fig:uvl3000R} $u_{v}(x)$, ~\subref{fig:dvl3000R} $d_{v}(x)$, %
    ~\subref{fig:ubar3000R} $\bar{u}(x)$, and ~\subref{fig:dbar3000R} $\bar{d}(x)$ distributions in the \texttt{CT14NNLO} %
    and \texttt{CT14HERA2} PDF sets evaluated at a scale of $Q=3$~TeV. The common denominator is the \texttt{CT14HERA2} central set.  
    At high values of $x$, such as $x\gtrsim 0.1$ %
    relevant for high-mass DY production, the PDF uncertainties begin to diverge. As noted in the introduction, the differences between these two recent fits is minimal, justifying our choice of  \texttt{CT14HERA2} in this analysis.}
  \label{fig:PDFUncertainties}
\end{figure}

\begin{figure}[th]
  \centering
  \includegraphics[width=0.6\textwidth]{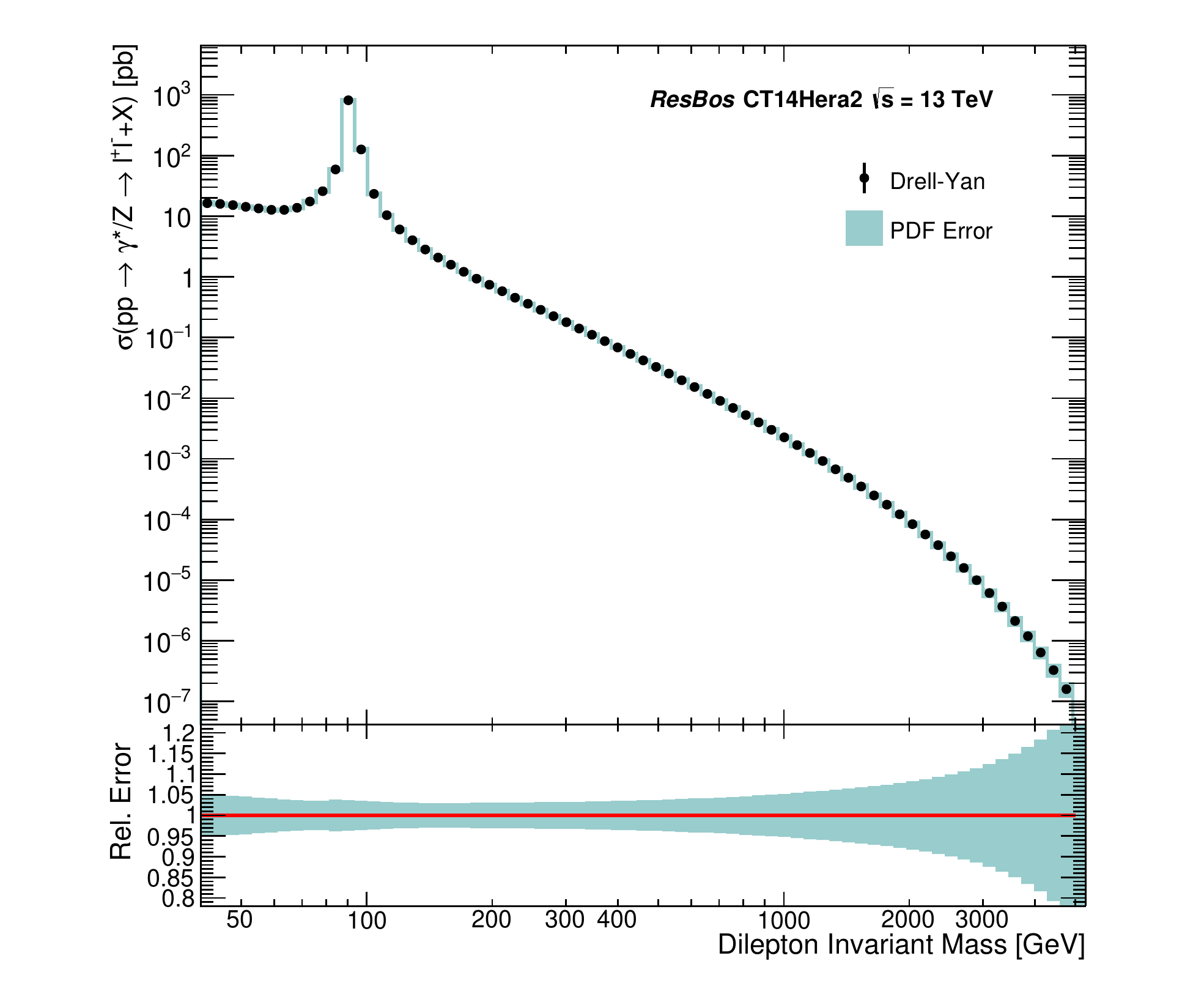}
  \caption{The dilepton invariant mass spectrum, generated with the {\sc ResBos} MC generator and the \texttt{CT14HERA2} %
    PDF set.}
  \label{fig:dileptonmass}
\end{figure}

Figure~\ref{fig:PDFUncertainties} shows the PDF uncertainties for several individual parton flavors in the \texttt{CT14NNLO} and \texttt{CT14HERA2} PDF sets. It's not surprising that the $\bar{u}(x)$ and $\bar{d}(x)$ distributions are least precisely known at moderate-to-high $x$ where input data are difficult to obtain and where their magnitudes have fallen to small fractions of their valence counterparts. But even the $u_{V}(x)$ and $d_{V}(x)$ distributions are poorly constrained in this region, although the up quark is much better determined than the down. Each of these distributions plays an important role in the initial-state quark-antiquark annihilation that results in the DY process. 
This significant lack of precision is the source of the large systematic uncertainties required in a high mass, dilepton $Z'$ search. Figure~\ref{fig:dileptonmass} shows the iconic invariant mass distribution of dilepton pairs calculated using the {\sc ResBos}~\cite{PhysRevD.50.R4239,PhysRevD.56.5558,PhysRevD.67.073016} Monte Carlo (MC) generator and the \texttt{CT14HERA2} PDF. The ratio band is the quoted \texttt{CT14HERA2}~\cite{PhysRevD.95.034003} PDF uncertainties of about 18\% at $m_{\ell\ell} = 4$~TeV consistent with that quoted in the ATLAS result of 19\% at $m_{ee} = 4$~TeV. \par
As DY data inputs are the only way to constrain high-$x$ PDFs, a strategy is explored here that turns this lack of sensitivity into an opportunity. The DY continuum is well-measured and reliably SM physics. If PDF global fits were to include LHC DY data well below any search region, but high enough in invariant mass to better constrain the fits, this uncertainty could be reduced. Moreover, the amount of LHC data that will become available in the coming years will be staggering, so we've decided to explore DY kinematics further in hopes of finding/discovering sensitivities that would help to enhance the potential of high-$x$ PDF fits.

We will show that there are DY observables, such as $\cos\theta^{*}$, that could in principle be incorporated in PDF global fitting, and the use of \eP tells us approximately how much reduction in PDF uncertainty is possible, as well as how much smaller the PDF systematic uncertainty might become in the DY differential mass spectrum. Due to the importance of $\cos\theta^{*}$, and the role it plays in our fitting strategy, a brief review is given in the next section.
\subsection{The Collins-Soper Polar Angle}
We have found particular power in $\cos\theta^{*}$ in Eq.~(\ref{eq:sigma3D}). This angle is defined in the Collins-Soper (CS)~\cite{PhysRevD.16.2219} rest frame of the lepton-pair with the polar and azimuthal angles defined relative to the two proton directions. The $z$ axis is defined in the $Z$ boson rest frame so that it bisects the angle formed by the momentum of one of the incoming protons and the negative of the momentum of the other incoming proton. The $y$ axis is constructed to be normal to the plane of the two proton momenta and the $x$ axis which is chosen in order to create a right-handed Cartesian coordinate system. 

The cosine of the polar angle $\theta^{*}$ defines the direction of the outgoing lepton $\ell^{-}$ relative to $\hat{z}$ in the CS frame and can be calculated directly from lab frame lepton quantities with
\begin{equation} \label{eq:cosine}
  \cos\theta^{*} = \frac{P_{z}}{|P_{z}|} %
  \frac{2\left(p^{+}_{1}p^{-}_{2}-p^{-}_{1}p^{+}_{2}\right)}{M_{\ell\ell}\sqrt{M_{\ell\ell}^{2} + P_{T}^{2}}}.
\end{equation}
The sign of the $z$ axis is defined on an event-by-event basis as the sign of the lepton pair momentum with respect to the $z$ axis in the laboratory frame. 
Here, $P_{T}$ and $P_{z}$ are the transverse and longitudinal momentum of the dilepton system, respectively, and,
\begin{equation}
  p^{\pm}_{i} = \frac{1}{\sqrt{2}}\left(E_{i} \pm p_{z,i}\right), \quad i=1,2,
\end{equation}
where the lepton (anti-lepton) energy and longitudinal momentum are $E_{1}$ and $p_{z,1}$ ($E_{2}$ and $p_{z,2}$), respectively. This definition requires the electric charge identification of each lepton. We define DY events as forward ($\cos\theta^{*}>0$) or backward ($\cos\theta^{*}<0$) according to the direction of the outgoing lepton in this frame of reference.
Our strategy was to explore the DY cross section with the goal of finding global PDF fitting inputs tailored specifically to DY physics. To that end we used the {\sc ResBos} MC (and the {\sc MadGraph} generator~\cite{Alwall:2014hca}
 as a check), configured with the \texttt{CT14HERA2} PDF set to study several kinematic distributions.

All simulation samples are produced in bins of true dilepton invariant mass in the range $m_{\ell\ell}=40$~GeV to $m_{\ell\ell}=1$~TeV at $\sqrt{s}=13$~TeV. In order to roughly correspond to ATLAS~\cite{Aaboud2017} and CMS~\cite{Sirunyan:2018exx} acceptances, the lepton pseudo-rapidities were restricted. Central-central (CC) events are required to have both leptons with $|y_\ell| < 2.47$, and the dilepton rapidity, $|y_{\ell\ell}| < 2.47$.  Central-forward (CF) events are required to have one lepton with $|y_\ell| < 2.47$, and the other with $2.47 < |y_\ell| < 3.6$, where the dilepton rapidity can extend out to $|y_{\ell\ell}| < 3.6$, which allows access to a wider range in $x$, c.f. Eq.~(\ref{eq:fiducial}).

\begin{figure}[!t]
  \centering
  \subfigure[]{
    \label{fig:cos_40_66}
    \includegraphics[width=0.43\textwidth]{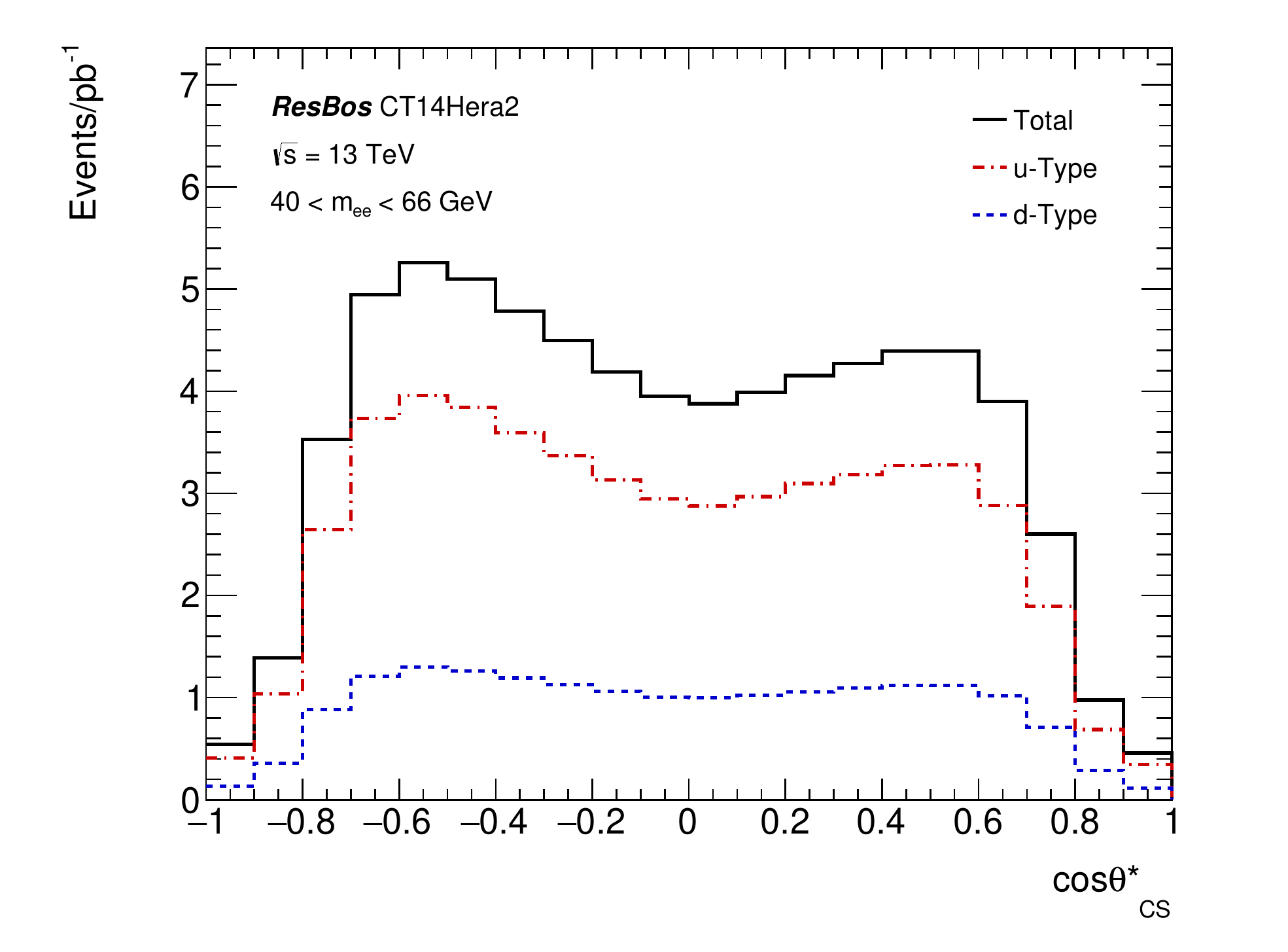}
  }
  \subfigure[]{
    \label{fig:cos_66_116}
    \includegraphics[width=0.43\textwidth]{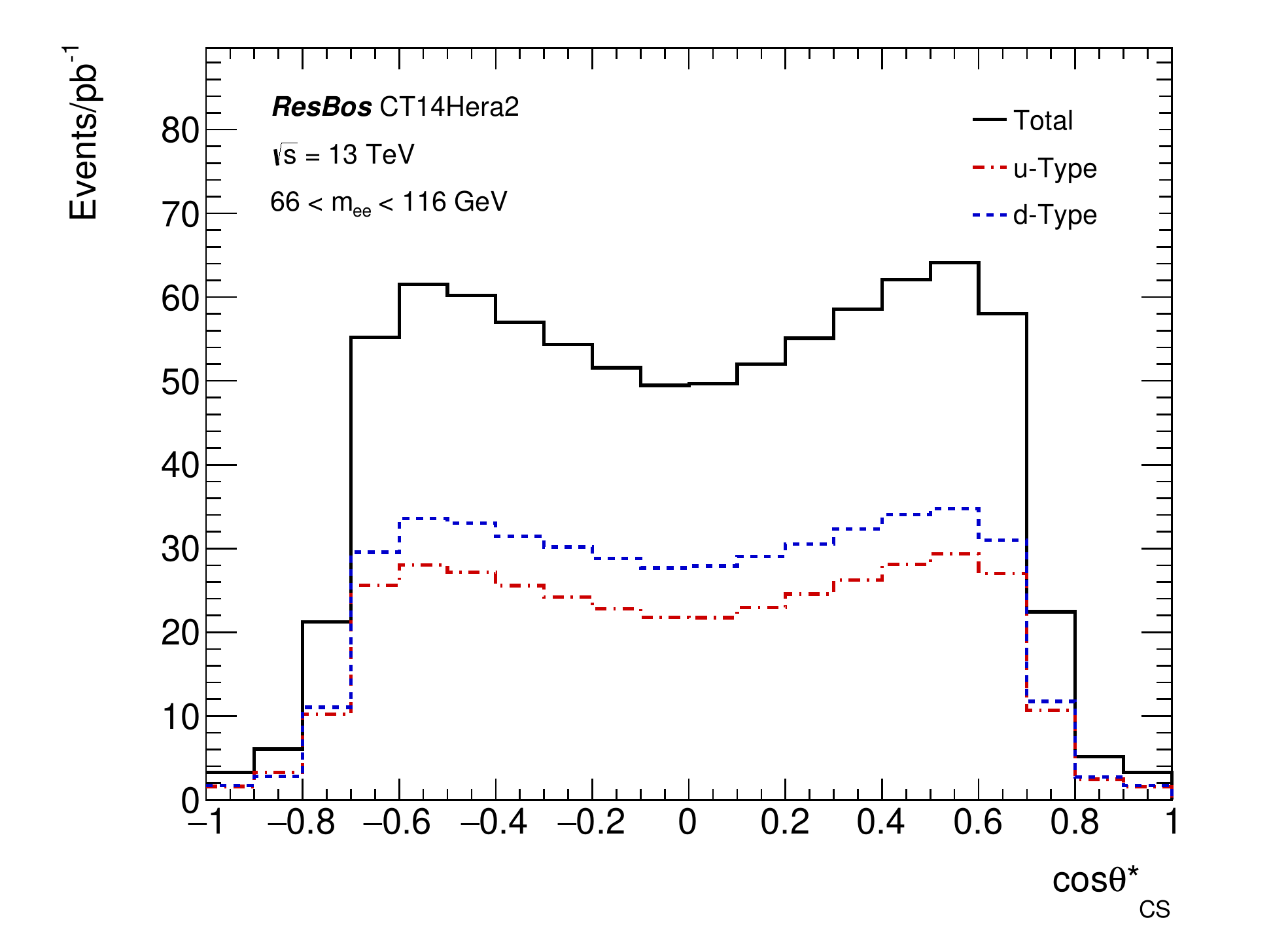}
  }\\
  \subfigure[]{
    \label{fig:cos_116_250}
    \includegraphics[width=0.43\textwidth]{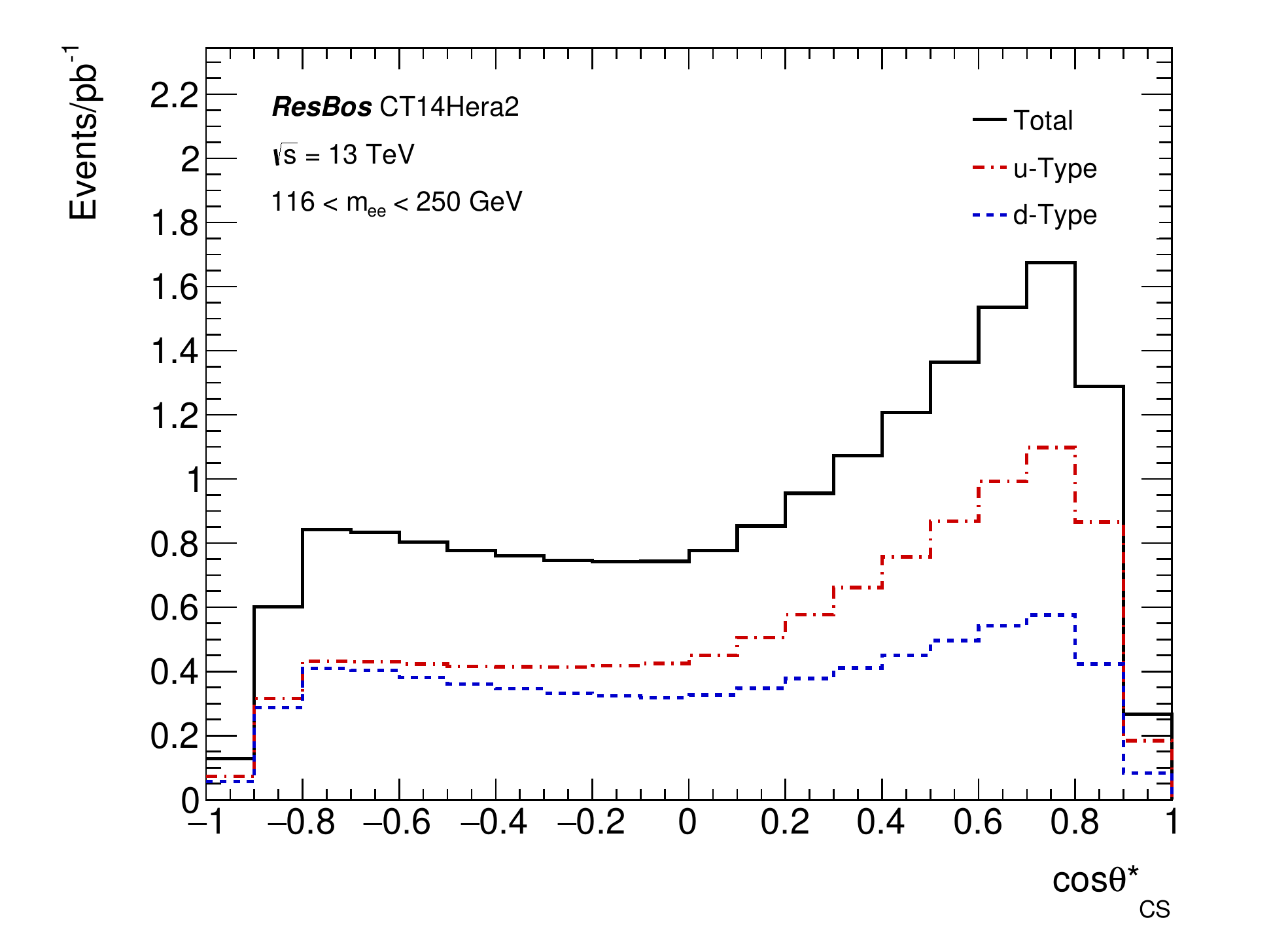}
  }
  \subfigure[]{
    \label{fig:cos_250_400}
    \includegraphics[width=0.43\textwidth]{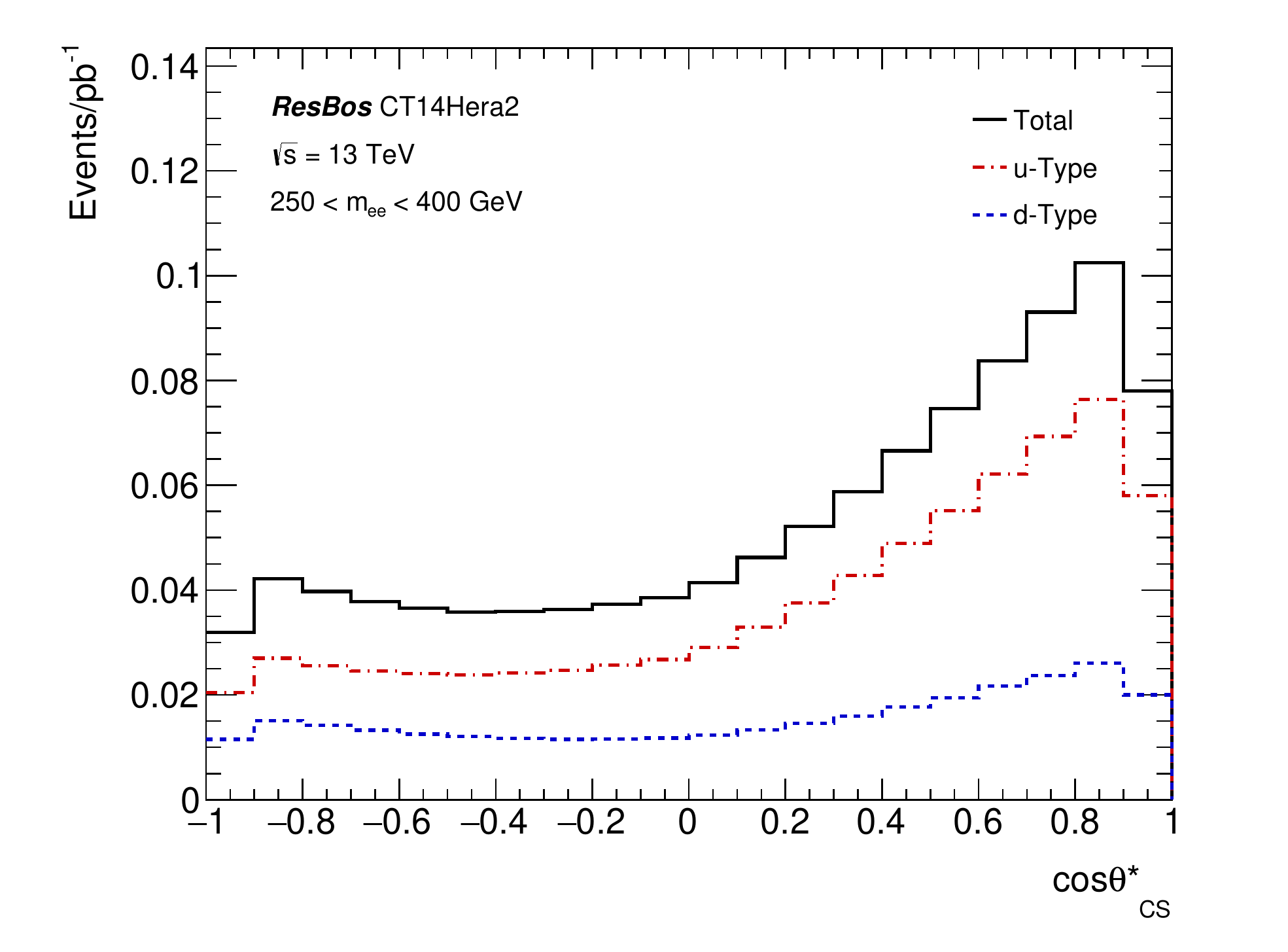}
  }\\
  \subfigure[]{
    \label{fig:cos_400_600}
    \includegraphics[width=0.43\textwidth]{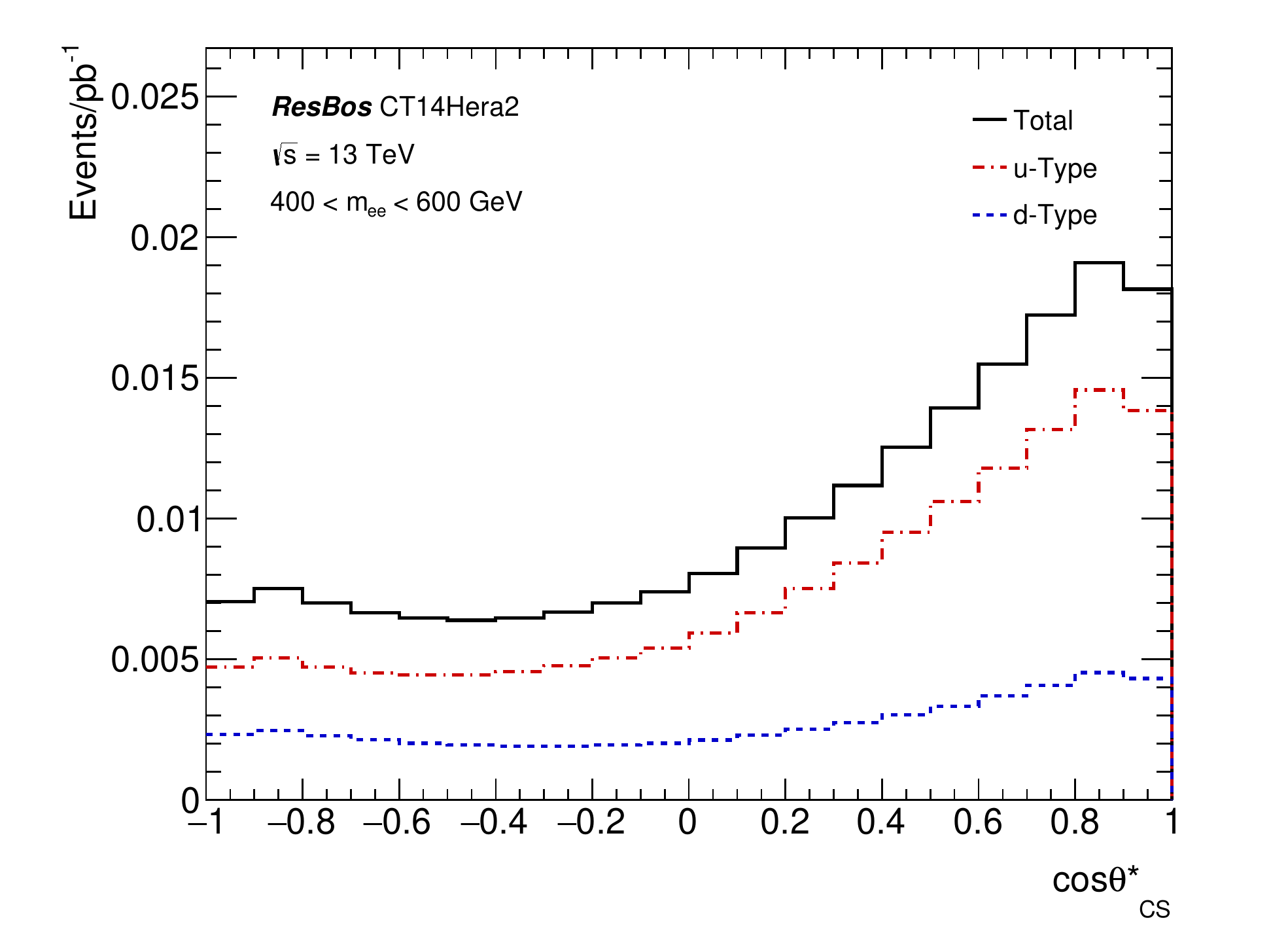}
  }
  \subfigure[]{
    \label{fig:cos_600_1000}
    \includegraphics[width=0.43\textwidth]{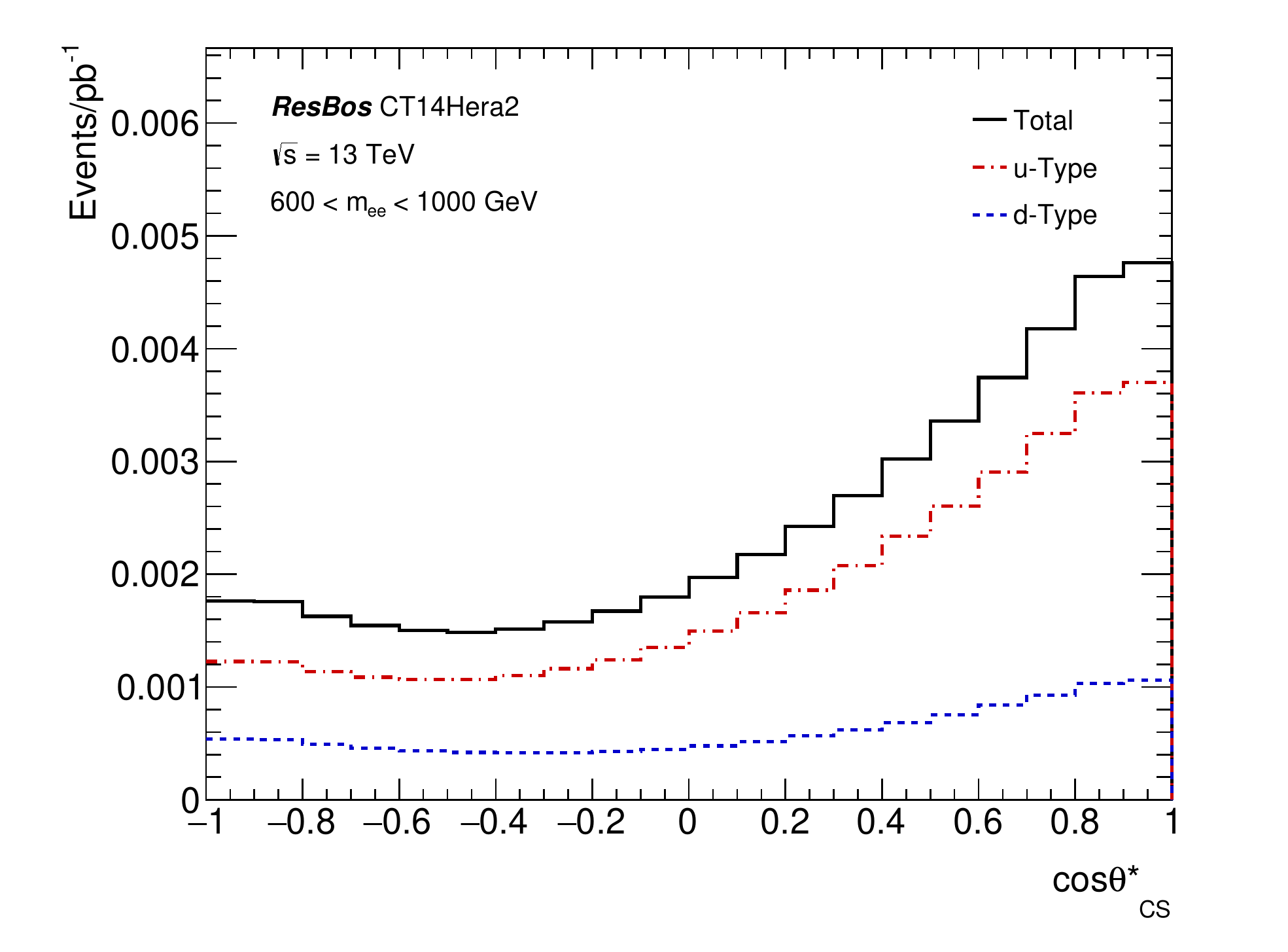}
  }
  \caption{The lepton angular distribution $\cos\theta^{*}$ in slices of dilepton invariant mass, $m_{\ell\ell}$, ranging %
    from 40~GeV to 1~TeV~\subref{fig:cos_40_66}-\subref{fig:cos_600_1000}. The up-type and down-type DY sub-processes %
    are shown as well, which exhibit a strong angular dependence, especially at high mass. The \texttt{CT14HERA2} PDF set is used.}
  \label{fig:massCosSlices}
\end{figure}
We found particular practical significance in focusing on the polar angle. Figure~\ref{fig:massCosSlices} shows several $\cos\theta^{*}$ distributions of Eq.~(\ref{eq:cosine}) in discrete slices of dilepton invariant mass. Each mass-slice is further decomposed into sub-processes that consist distinctly of up-type or down-type initial-state quarks. The up-type sub-processes include initial-states of $u\overline{u}$, $ug$, and $\overline{u}g$, where $u$ is the up quark or charm quark and $g$ is the
gluon. A similar definition applies to the $d$-type (down, strange, bottom) sub-processes, with $u$ replaced by $d$. This is in accordance with the four DY reactions in Eqs.~(\ref{eq:s}) and (\ref{eq:t3}). \par

The distributions in Figs.~\ref{fig:cos_40_66}, ~\ref{fig:cos_66_116}, and ~\ref{fig:cos_116_250} are essentially the regions covered by an ATLAS measurement of the triple differential cross section during the 8 TeV running.~\cite{Aaboud:2017exx} 
These are familiar as they show part of the source of the oft-measured Forward-Backward Asymmetry in both  $p-\bar{p}$ and $pp$ on-resonance $Z$ boson analyses~\cite{refId0}.

Intriguingly, the relative up-type and down-type sub-processes are highly dependent on both mass and polar angle $\theta^{*}$. This is especially true above the $Z$ boson mass peak, in which the forward region ($\cos\theta^{*}>0$) shows an increasing degree of separation between the rates associated with the up-type and down-type DY sub-processes. Indeed, in this region the contribution to the total cross section is due almost entirely to the up-type sub-process by itself: almost by a factor of four. \textit{At high mass and high polar angle, the LHC DY process proceeds almost entirely through the $u\bar{u}$ sub-process, effectively making the LHC a $u\bar{u}$ collider.} \par

Why is this the case? Appendix~\ref{upapp} explains this conclusion as a fortuitous conspiracy of electroweak couplings and parton luminosities which collectively favor up quarks and antiquarks over their down-like counterparts.  

Notice that we've not really learned anything new since DY kinematics is an old subject. But  high-mass behavior in regions only statistically available at the LHC  is revealing and the question is whether $\cos\theta^{*}$ behavior as a function of mass should be an important discrimination as an input to global PDF fitting. This is where \eP comes in. 
\section{A Proposed Strategy to PDF Error Reduction for DY} \label{sec:approachToPDFErrorReduction}
We attempt to shed light on two questions:
\begin{enumerate}
\item If $\cos\theta^{*}$ data were incorporated in global fitting, how significant might the reduction in PDF uncertainties be? 
\item Would those decreased errors be a significant reduction in the overall theoretical uncertainties in future BSM, high-mass DY searches?
\end{enumerate}

\begin{figure}[!t]
  \centering
  \includegraphics[width=\textwidth]{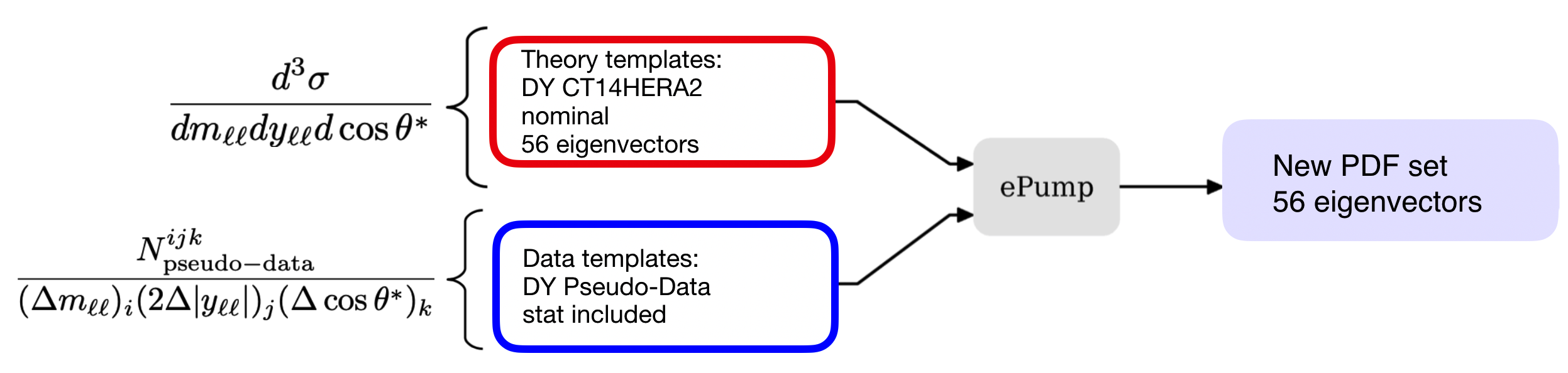}
  \caption{The \texttt{ePump} package requires two inputs to generate an updated PDF set: an existing Theory template of a PDF set %
    (parameters + uncertainties) and binned Data template of (pseudo-) data, including statistical uncertainties from integrated luminosity assumptions.}
  \label{fig:ePumpFlow}
\end{figure}

In order to answer Question 1, \eP  was used, which can update an existing PDF set with new experimental data (or pseudo-data) in order to produce an improved best-fit and Hessian Error PDFs. The \eP workflow can be seen in Fig.~\ref{fig:ePumpFlow}. \par

For this analysis, ``pseudo-data'' are used to mimic a possible future LHC dataset for PDF fitting. As any dataset has finite statistics, the resulting uncertainties in the new PDFs will reflect whatever statistical precision is modeled in the pseudo-data. The effects of new PDFs and uncertainties can then be used to re-evaluate the PDF systematic uncertainty on the high-mass dilepton event yield. \par

Furthermore, we imagine a Signal Region (SR) as $m_{\ell\ell}>1$~TeV and a Control Region (CR) to be  for $0.04<m_{\ell\ell}<1$~TeV. Since new physics should lie above the current limits of approximately $m_{\ell\ell} \sim 3$~TeV (as in Sec.~\ref{sec:Results}), it would be ``fair'' to use low-mass DY data to constrain the high-mass DY spectrum. 

\subsection{PDF Update Strategy} \label{sec:strategyPDFs}

\eP requires standard inputs to emulate the global fit---the templates in Fig.~\ref{fig:ePumpFlow}. We describe our strategy here. The analysis was performed at ``truth level,'' such that the acceptance and efficiency effects associated with the reconstruction and identification of prompt, high-$p_{T}$ leptons in an LHC detector are neglected. However, leptons are well measured at the LHC, so this is an acceptable first look at this technique. Additional dilepton backgrounds were neglected, but are well understood by the LHC experiments as can be seen in Fig.~\ref{fig:massPlots}. These backgrounds include $t\bar{t}$ production, $Wt$ Single Top production, $WW$, $WZ$, and $ZZ$ Diboson production, and $W$+jets \& Multi-jet production in the electron channel.
\subsection{\texttt{ePump} Template Construction} \label{sec:templateConstruction}
Naively, one might imagine only using $m_{\ell\ell}$ in the CR to predict the improvement in the SR, but our awareness of the significant differential quark sensitivities to $\cos\theta^{*}$ (and moderate sensitivity to $y_{\ell\ell}$) plus the knowledge that future LHC running will provide enormous continuum DY datasets led us to explore dividing pseudo-data into many bins of dilepton mass $m_{\ell\ell}$, as well as $y_{\ell\ell}$ and $\cos\theta^{*}$. \par
The fiducial region considered for our analysis is designed explicitly to probe the PDFs at high $x$, and is defined by
\begin{equation} \label{eq:fiducial}  
40\, \mathrm{GeV} < m_{\ell\ell} < 1000 \, \mathrm{GeV}, \quad |y_{\ell\ell}|<3.6, \quad -1<\cos\theta^{*}<1.
\end{equation}
DY samples were generated using the {\sc ResBos} MC generator with the \texttt{CT14HERA2} PDF set for the $\sqrt{s}=13$~TeV LHC. Events were further required to pass a loose event selection in order to construct the finalized Data templates. Dilepton events with an invariant mass of $m_{\ell\ell}>80$~GeV were required to satisfy $p_{T}>30$~GeV, while low-mass events in the interval of $40<m_{\ell\ell}<80$ must satisfy $p_{T}>15$~GeV. In addition, events must consist of leptons which are distributed as central-central or central-forward. \par
Events passing these selections were binned in \eP template histograms, which parametrize the triple-differential cross section of Eq.~(\ref{eq:sigma3D}), according to
\begin{equation} \label{eq:sigmaBins}
  \mathcal{L}_{\mathrm{int}} \left( \frac{d^{3}\sigma}{dm_{\ell\ell}d|y_{\ell\ell}|d\cos\theta^{*}} \right)_{ijk} %
  = \frac{N^{ijk}_{\mathrm{pseudo-data}}}{(\Delta m_{\ell\ell})_{i} (2\Delta |y_{\ell\ell}|)_{j} (\Delta\cos\theta^{*})_{k}},
\end{equation}
where $i$, $j$, and $k$ correspond to the bin indices of each distribution of interest.  Note that in a realistic measurement, the numerator of Eq.~(\ref{eq:sigmaBins}) would be replaced by $N^{ijk}_{\mathrm{data}}-N^{ijk}_{\mathrm{bkg}}$, where the background component arises from the standard dilepton background processes.

The total number of pseudo-data events are given by $N^{ijk}_{\mathrm{pseudo-data}}$, the integrated luminosity of the pseudo-dataset is $\mathcal{L}_{\mathrm{int}}$, and $(\Delta m_{\ell\ell})_{i}$, $(2\Delta |y_{\ell\ell}|)_{j}$, and $(\Delta \cos\theta^{*})_{k}$ are the corresponding bin widths. The factor of two in the denominator accounts for the modulus in the rapidity bin width. The bins used to parametrize Eq.~(\ref{eq:sigmaBins}) are
\begin{itemize}
\item $40<m_{\ell\ell}<1000: \{40,\ 66,\ 80,\ 91,\ 102,\ 116,\ 145,\ 200,\ 275,\ 381,\ 525,\ 725,\ 1000\} \,  \mathrm{GeV}$
\item $0<|y_{\ell\ell}|<2.4: \{0.0,\ 0.2,\ 0.4,\ 0.6,\ 0.8,\ 1.0,\ 1.2,\ 1.4,\ 1.6,\ 1.8,\ 2.0,\ 2.2,\ 2.4\}$
\item $2.4<|y_{\ell\ell}|<3.6: \{2.4,\ 2.6,\ 2.8,\ 3.0,\ 3.2,\ 3.4,\ 3.6\}$
\item $-1<\cos\theta^{*}<1: \{-1.0,\ -0.7,\ -0.4,\ 0.0,\ 0.4,\ 0.7,\ 1.0\}$.
\end{itemize}

where  two $|y_{\ell\ell}|$ regions  explicitly call out  the CC and CF selections. The total number of measurement bins is $N_{\mathrm{bins}} = 12 \times 18 \times 6 = 1296$ for the fiducial region considered and they define the $N_{\mathrm{new}}$ data points that supplement Eq.~(\ref{eq:chi2}). \par
Events were generated as if they came from a future integrated luminosity and so uncertainties in the \eP results are scattered according to the statistics of such a hypothetical LHC input dataset. For each bin the DY cross section estimate $\sigma^{ijk}_{\mathrm{Drell-Yan}}$ was scaled by a characteristic integrated luminosity $\mathcal{L}_{\mathrm{int}}$ to arrive at a definite DY event yield $N^{ijk}_{\mathrm{Drell-Yan}}$. The resulting yield was assumed to be the mean of a Poisson distribution, which was then used to throw a random number according to Poisson statistics, thereby populating the bin with $N^{ijk}_{\mathrm{pseudo-data}}$ pseudo-data events. 
Note that the pseudo-data were treated as those of one ``experiment,'' but in practice ATLAS, CMS, and LHCb would all be sources of fitting input data. For illustration we chose two future LHC scenarios for integrated luminosities: $\mathcal{L}_{\mathrm{int}}=300$~$\mathrm{fb}^{-1}$ approximating the data set for one experiment following Run-3 of the LHC, and $\mathcal{L}_{\mathrm{int}}=3000$~$\mathrm{fb}^{-1}$, approximating that of the final dataset for one experiment of the High Luminosity (HL) LHC. \par
\section{PDF Update Results} \label{sec:PDFUpdateResults}
We can answer Question 1 by re-evaluating the effect of the 3000~$\mathrm{fb}^{-1}$ DY pseudo-dataset on the \texttt{CT14HERA2} PDFs, as well as Question 2 by assessing the reduction of the PDF systematic uncertainty in the high-mass dilepton spectrum. 
\subsection{Impact on \texttt{CT14HERA2} PDFs}
\begin{figure}[!t]
  \centering
  \subfigure[]{
    \label{fig:ubar3000}
    \includegraphics[width=0.43\textwidth]{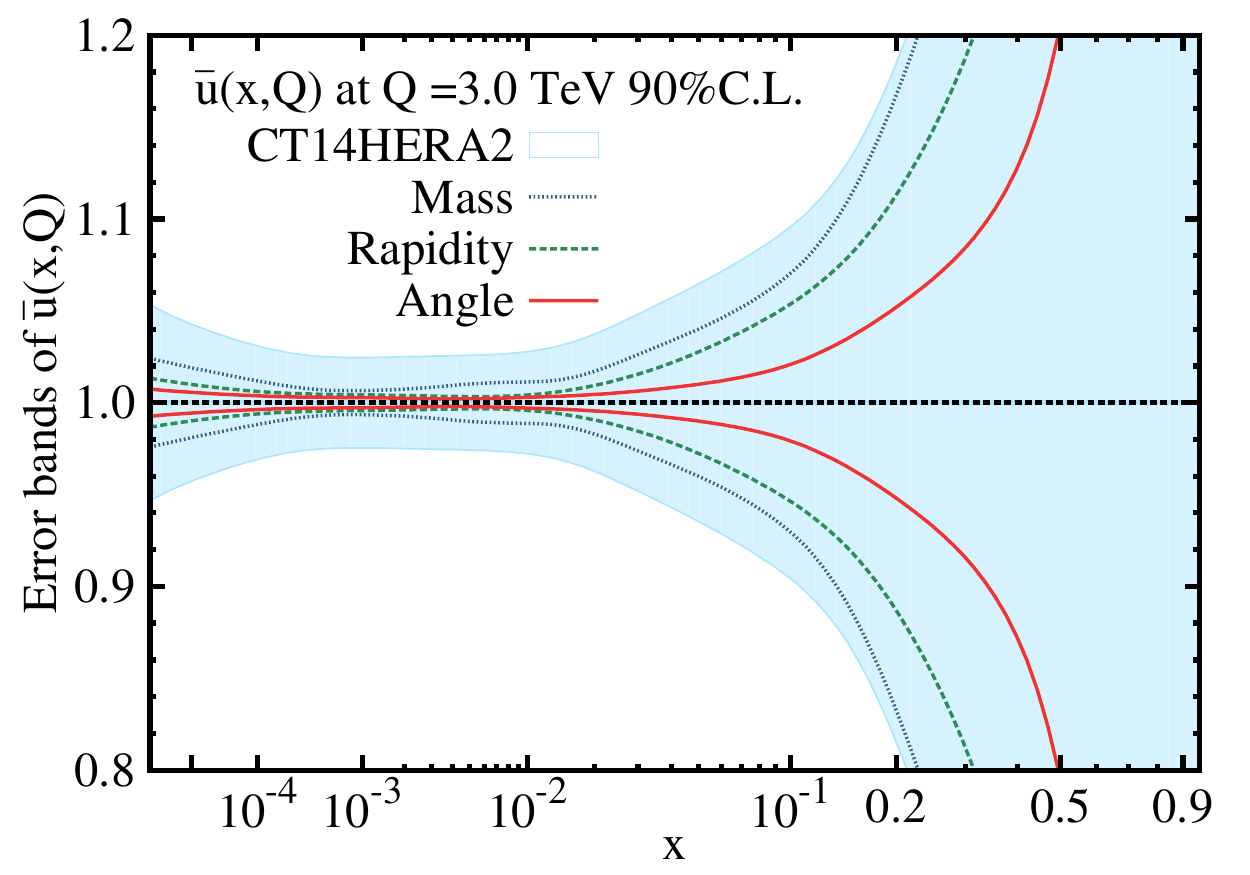}
  }
  \subfigure[]{
    \label{fig:dbar3000}
    \includegraphics[width=0.43\textwidth]{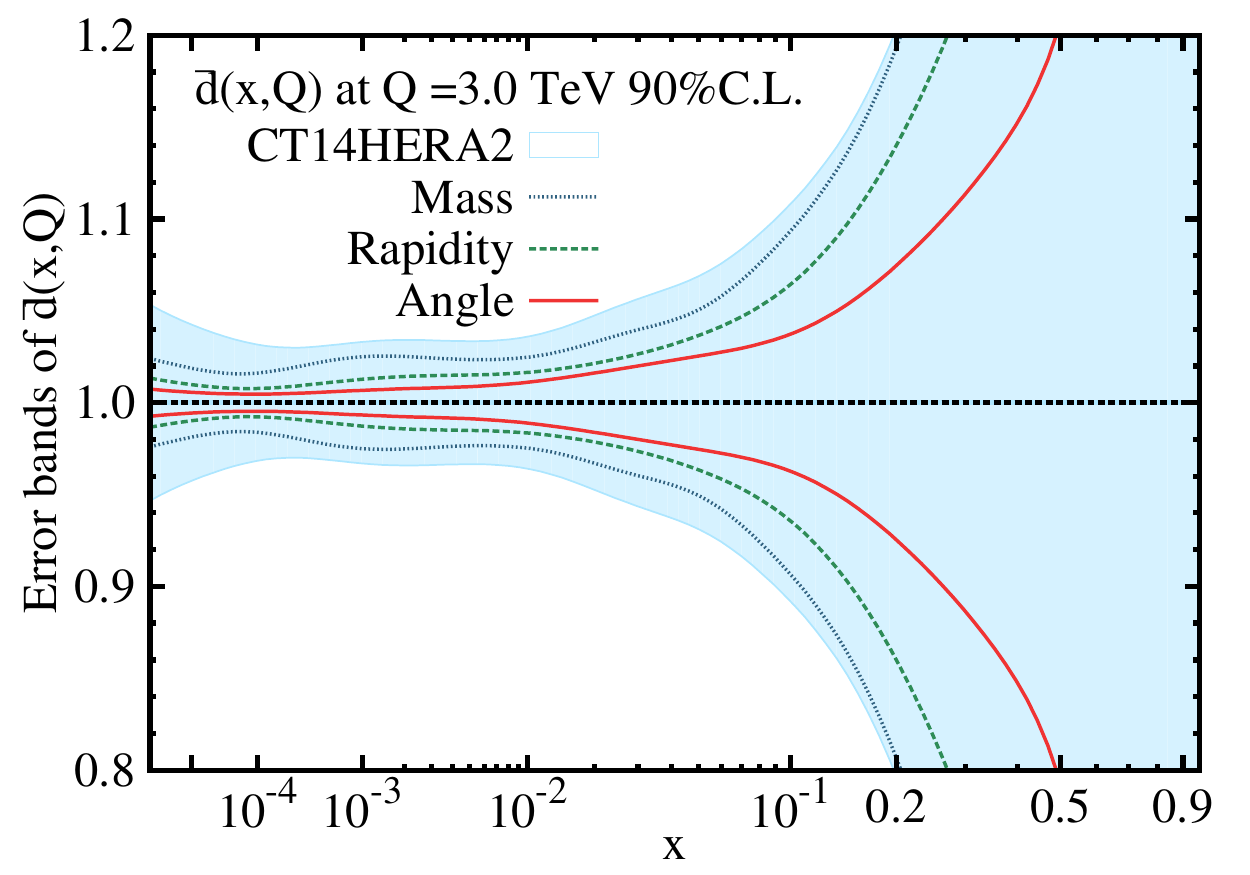}
  }
  \caption{Impact of the 3000~$\mathrm{fb}^{-1}$ update on the ~\subref{fig:ubar3000} \texttt{CT14HERA2} $\bar{u}(x)$ and ~\subref{fig:dbar3000} $\bar{d}(x)$. The shaded background shows the uncertainties resulting from the current \texttt{CT14HERA2} uncertainties. The dotted curve labeled ``mass'' corresponds to the error reduction by sending only binned $(\Delta m_{\ell\ell})$ to \texttt{ePump}; i.e., integrated over the $y_{\ell\ell}$ and $\cos\theta^{*}$ dimensions. The dashed curve labeled ``rapidity'' adds the cumulative effect of binned $(\Delta |y_{\ell\ell}|)$ and $(\Delta m_{\ell\ell})$ to \texttt{ePump}. Finally, the solid curve labeled ``angle'' adds the cumulative effect of binned $(\Delta \cos\theta^{*})$, $(\Delta |y_{\ell\ell}|)$, and $(\Delta m_{\ell\ell})$ to \texttt{ePump}.}
  \label{fig:fourCurvesSea3000}
\end{figure}

\begin{figure}[!t]
  \centering
  \subfigure[]{
    \label{fig:uvl3000}
    \includegraphics[width=0.43\textwidth]{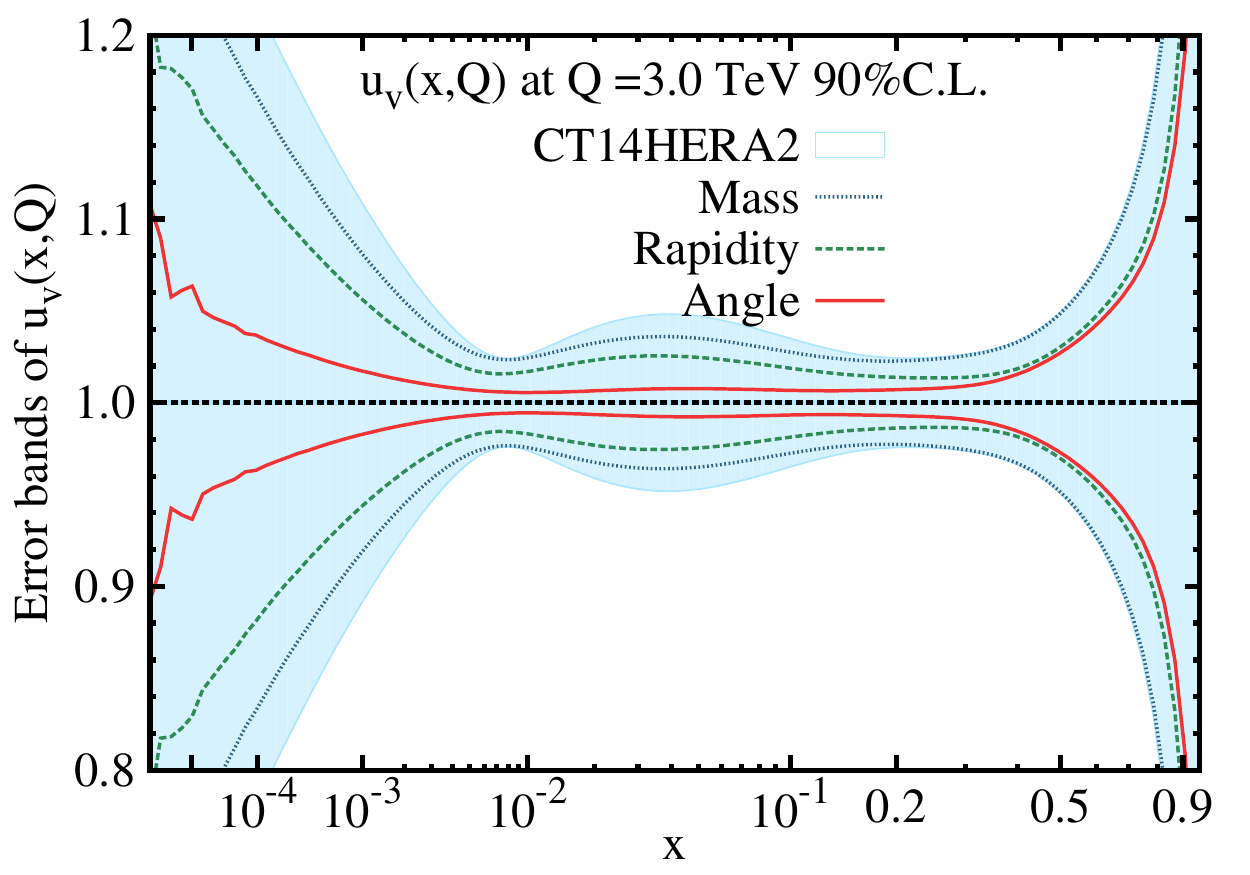}
  }
  \subfigure[]{
    \label{fig:dvl3000}
    \includegraphics[width=0.43\textwidth]{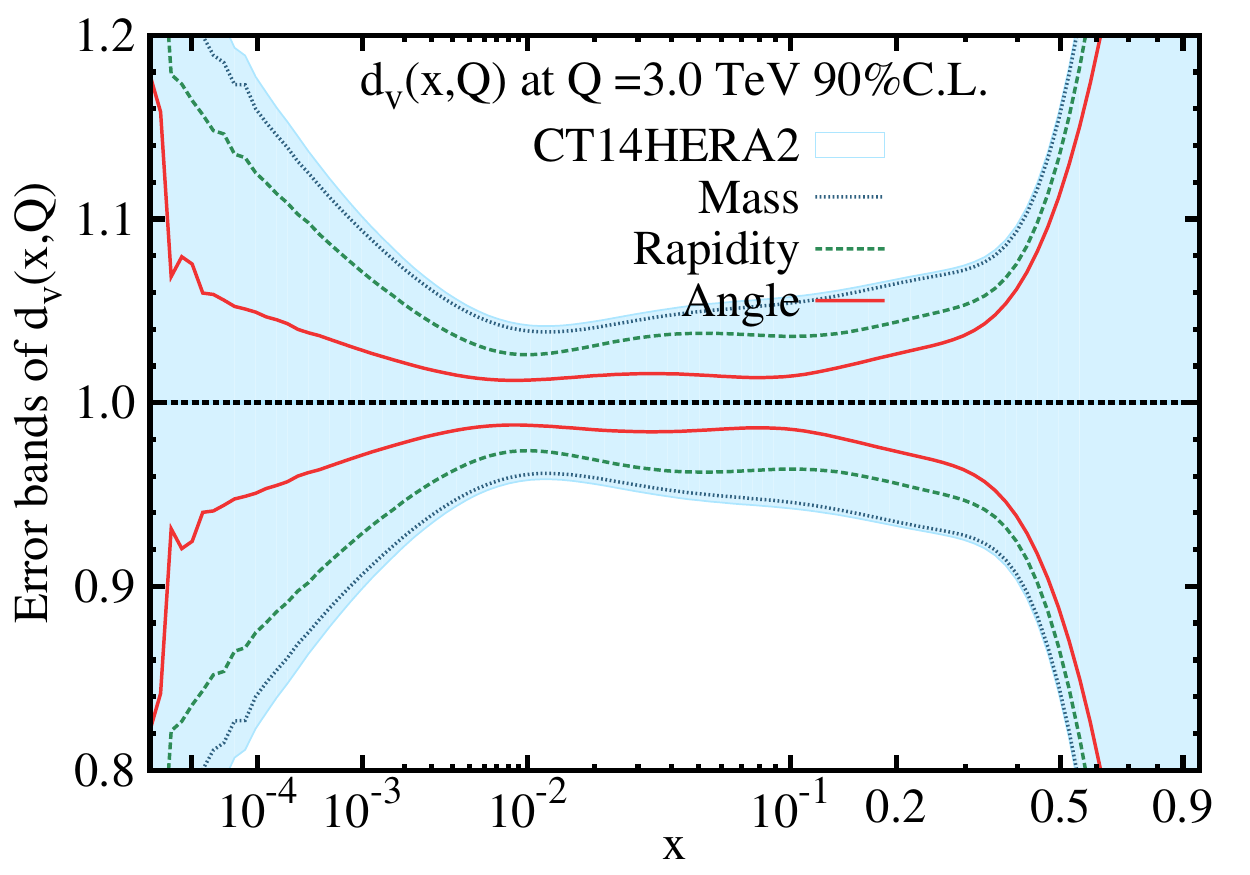}
  }
  \caption{Impact of the 3000~$\mathrm{fb}^{-1}$ update on the ~\subref{fig:uvl3000} \texttt{CT14HERA2} $u_{v}(x)$ and %
    ~\subref{fig:dvl3000} $d_{v}(x)$. The shaded background shows the uncertainties resulting from the current \texttt{CT14HERA2} uncertainties. The dotted curve labeled ``mass'' corresponds to the error reduction by sending only binned $(\Delta m_{\ell\ell})$ to \texttt{ePump}; i.e., integrated over the $y_{\ell\ell}$ and $\cos\theta^{*}$ dimensions. The dashed curve labeled ``rapidity'' adds the cumulative effect of binned $(\Delta |y_{\ell\ell}|)$ and $(\Delta m_{\ell\ell})$ to \texttt{ePump}. Finally, the solid curve labeled ``angle'' adds the cumulative effect of binned $(\Delta \cos\theta^{*})$, $(\Delta |y_{\ell\ell}|)$, and $(\Delta m_{\ell\ell})$ to \texttt{ePump}.}
  \label{fig:fourCurvesVal3000}
\end{figure}

Question 1 asked whether explicit inclusion of $\cos \theta^\star$ data might have a useful effect in reducing the uncertainties in the parton fits. The answer can be seen in the following four plots in Figs.~\ref{fig:fourCurvesSea3000} and~\ref{fig:fourCurvesVal3000}. In order to see the effect of each of the quantities in the \texttt{ePump}-simulated refitting, there are four sets of results in each plot.
Figure~\ref{fig:fourCurvesSea3000} shows the impact of the \texttt{ePump} update with the 3000~$\mathrm{fb}^{-1}$ scenario on the $\bar{u}(x)$ and $\bar{d}(x)$ sea distributions and Fig.~\ref{fig:fourCurvesVal3000}, the impact on the $u_{v}(x)$ and $d_{v}(x)$ valence distributions. \par
\begin{table}[!t]
  \centering
  \begin{ruledtabular}
    \begin{tabular}{c|cc|cc|cc|cc} 
      & \multicolumn{2}{c|}{$u_{v}(x)$} & \multicolumn{2}{c|}{$d_{v}(x)$} & \multicolumn{2}{c|}{$\bar{u}(x)$} & \multicolumn{2}{c}{$\bar{d}(x)$} \\
      x & $\delta_{pre}$ [\%] & $\delta_{post}$ [\%] & $\delta_{pre}$ [\%] & $\delta_{post}$ [\%] & $\delta_{pre}$ [\%] & $\delta_{post}$ [\%] & $\delta_{pre}$ [\%] & $\delta_{post}$ [\%] \\
      \hline
      0.1 & 3.4 & 0.7 & 5.8 & 1.5 & 9.8 & 2.2 & 11  & 3.8\\
      0.3 & 2.6 & 0.9 & 7.5 & 3.6 & 30  & 8.3 & 32  & 11\\
      0.5 & 4.8 & 2.6 & 16  & 11  & 71  & 20  & 69  & 20\\
      0.7 & 12  & 7.0 & 45  & 30  & 280 & 77  & 250 & 67\\
    \end{tabular}
  \end{ruledtabular}
  \caption{Impact of 3000~$\mathrm{fb}^{-1}$ update on the \texttt{CT14HERA2} $u_{v}(x)$ and $d_{v}(x)$ valence and $\bar{u}(x)$ %
    and $\bar{d}(x)$ sea distributions for several values of $x$ using the standard triple-differential templates at $Q=3$~TeV.
    To be compared with the ``Angle'' curves of Figs.~\ref{fig:fourCurvesSea3000} and~\ref{fig:fourCurvesVal3000}.}
  \label{tab:xTable3000}
\end{table}
The sea distributions show a considerable reduction in uncertainty at high $x$. For example, in both the $\bar{u}(x)$ and $\bar{d}(x)$ distributions, the PDF uncertainty is reduced from its pre-update value of approximately 70\% to 20\% at $x=0.5$. The improvement in the valence distributions at $x \gtrsim 0.5$ is less dramatic, but substantial improvement is observed in the ranges of $x \lesssim 0.5$.  The post-update $u_{v}(x)$ distribution remains better constrained than $d_{v}(x)$ at high $x$, where the uncertainty measures 2.6\% as compared to 11\% at $x=0.5$, respectively. Table~\ref{tab:xTable3000} lists the pre- and post-update uncertainties for several parton flavors and values of $x$ explicitly. \par
\begin{figure}[!t]
  \centering
  \subfigure[]{
    \label{fig:ubar0300}
    \includegraphics[width=0.43\textwidth]{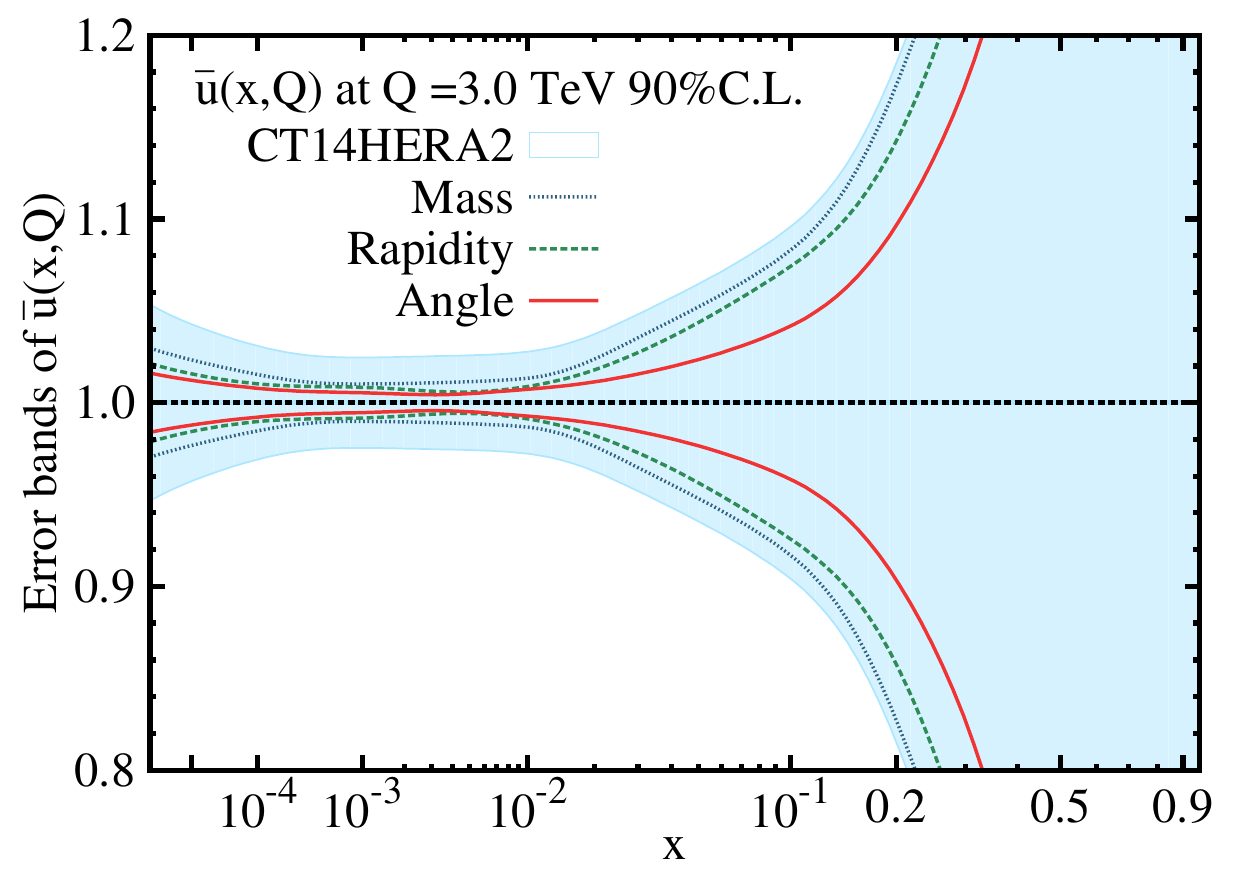}
  }
  \subfigure[]{
    \label{fig:dbar0300}
    \includegraphics[width=0.43\textwidth]{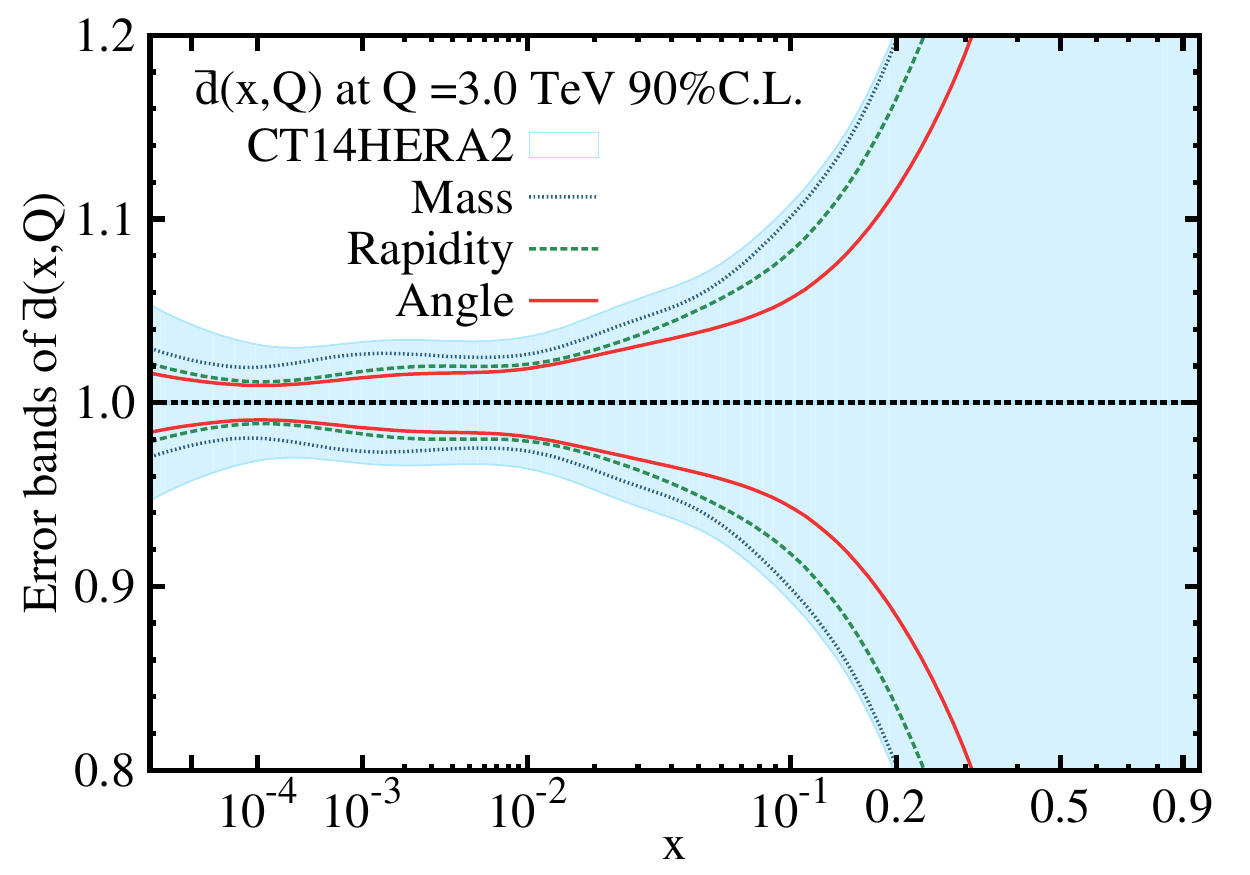}
  }
  \caption{Impact of the 300~$\mathrm{fb}^{-1}$ update on the ~\subref{fig:ubar0300} \texttt{CT14HERA2} $\bar{u}(x)$ and %
    ~\subref{fig:dbar0300} $\bar{d}(x)$. The shaded background shows the uncertainties resulting from the current \texttt{CT14HERA2} uncertainties. The dotted curve labeled ``mass'' corresponds to the error reduction by sending only binned $(\Delta m_{\ell\ell})$ to \texttt{ePump}; i.e., integrated over the $y_{\ell\ell}$ and $\cos\theta^{*}$ dimensions. The dashed curve labeled ``rapidity'' adds the cumulative effect of binned $(\Delta |y_{\ell\ell}|)$ and $(\Delta m_{\ell\ell})$ to \texttt{ePump}. Finally, the solid curve labeled ``angle'' adds the cumulative effect of binned $(\Delta \cos\theta^{*})$, $(\Delta |y_{\ell\ell}|)$, and $(\Delta m_{\ell\ell})$ to \texttt{ePump}.}
  \label{fig:fourCurvesSea0300}
\end{figure}
\begin{figure}[!t]
  \centering
  \subfigure[]{
    \label{fig:uvl0300}
    \includegraphics[width=0.43\textwidth]{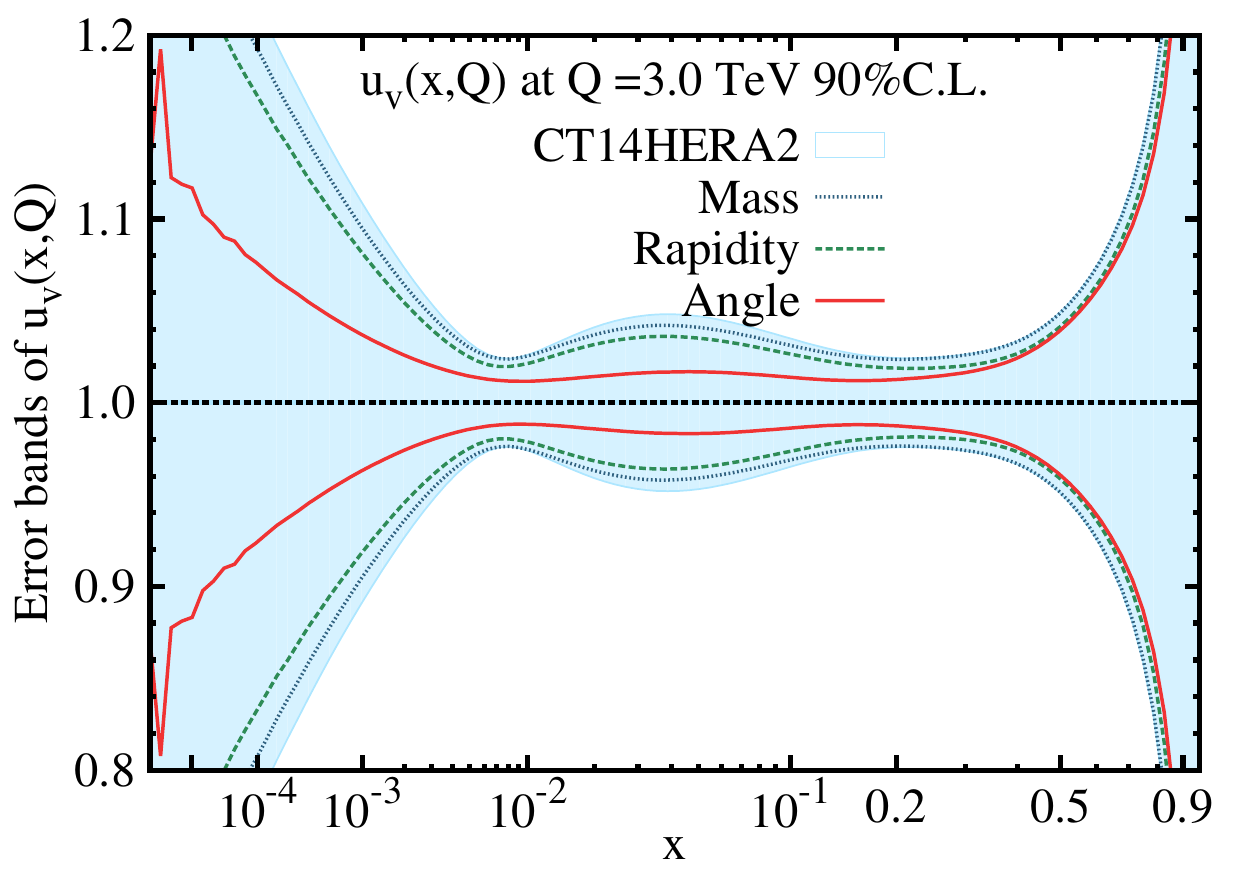}
  }
  \subfigure[]{
    \label{fig:dvl0300}
    \includegraphics[width=0.43\textwidth]{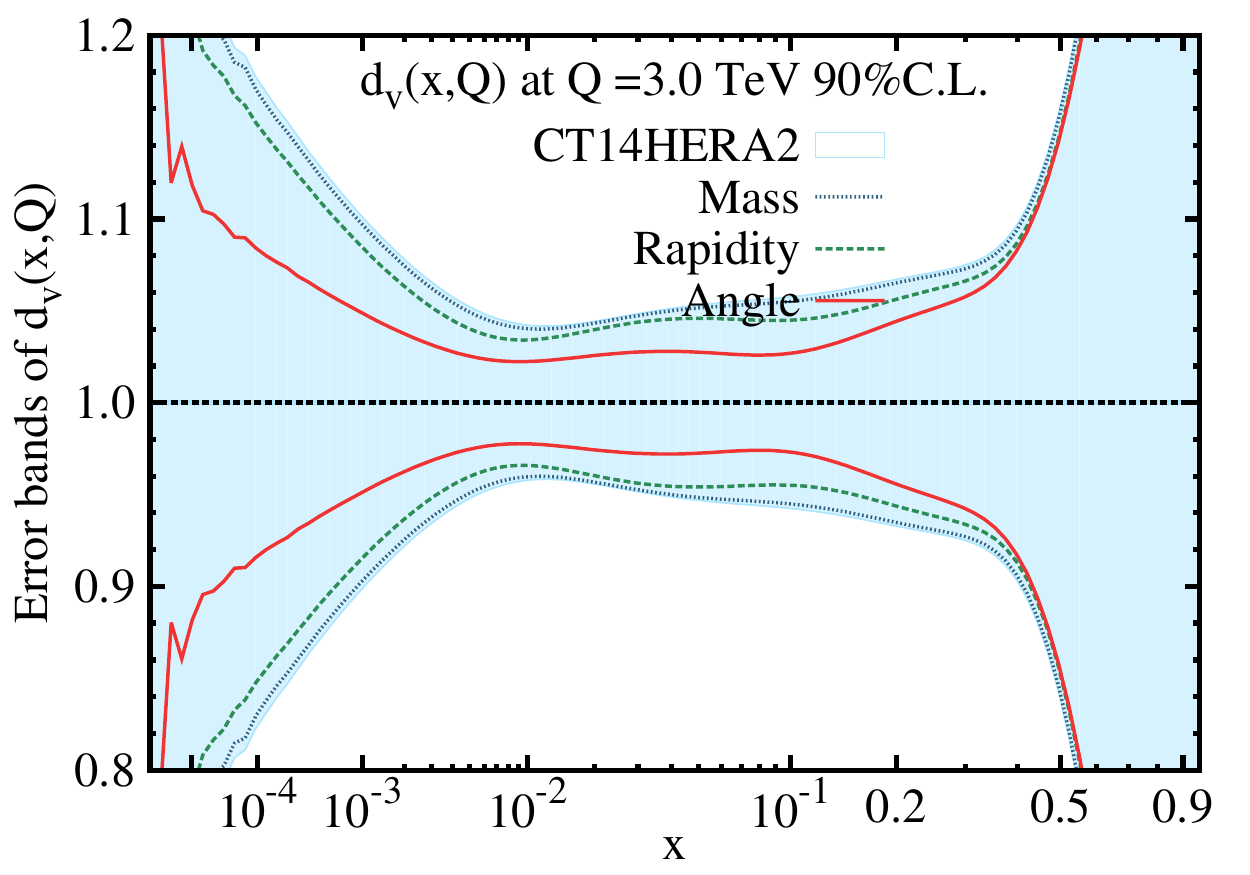}
  }
  \caption{Impact of the 300~$\mathrm{fb}^{-1}$ update on the ~\subref{fig:uvl0300} \texttt{CT14HERA2} $u_{v}(x)$ and %
    ~\subref{fig:dvl0300} $d_{v}(x)$. The shaded background shows the uncertainties resulting from the current \texttt{CT14HERA2} uncertainties. The dotted curve labeled ``mass'' corresponds to the error reduction by sending only binned $(\Delta m_{\ell\ell})$ to \texttt{ePump}; i.e., integrated over the $y_{\ell\ell}$ and $\cos\theta^{*}$ dimensions. The dashed curve labeled ``rapidity'' adds the cumulative effect of binned $(\Delta |y_{\ell\ell}|)$ and $(\Delta m_{\ell\ell})$ to \texttt{ePump}. Finally, the solid curve labeled ``angle'' adds the cumulative effect of binned $(\Delta \cos\theta^{*})$, $(\Delta |y_{\ell\ell}|)$, and $(\Delta m_{\ell\ell})$ to \texttt{ePump}.}
  \label{fig:fourCurvesVal0300}
\end{figure}

Figures~\ref{fig:fourCurvesSea0300} and ~\ref{fig:fourCurvesVal0300} show the reduction in uncertainties for the 300~$\mathrm{fb}^{-1}$ scenario and Table~\ref{tab:xTable0300} is the corresponding comparison for the 300~$\mathrm{fb}^{-1}$ scenario. \par
\begin{table}[!t]
  \centering
  \begin{ruledtabular}
    \begin{tabular}{c|cc|cc|cc|cc} 
      & \multicolumn{2}{c|}{$u_{v}(x)$} & \multicolumn{2}{c|}{$d_{v}(x)$} & \multicolumn{2}{c|}{$\bar{u}(x)$} & \multicolumn{2}{c}{$\bar{d}(x)$} \\
      x & $\delta_{pre}$ [\%] & $\delta_{post}$ [\%] & $\delta_{pre}$ [\%] & $\delta_{post}$ [\%] & $\delta_{pre}$ [\%] & $\delta_{post}$ [\%] & $\delta_{pre}$ [\%] & $\delta_{post}$ [\%] \\
      \hline
      0.1 & 3.4 & 1.4 & 5.8 & 2.7 & 9.8 & 4.3 & 11  & 6.0\\
      0.3 & 2.6 & 1.6 & 7.5 & 5.7 & 30  & 17  & 32  & 19\\
      0.5 & 4.8 & 3.9 & 16  & 14  & 71  & 43  & 69  & 41\\
      0.7 & 12  & 9.7 & 45  & 41  & 280 & 180  & 250 & 160\\
    \end{tabular}
  \end{ruledtabular}
  \caption{Impact of 300~$\mathrm{fb}^{-1}$ update on the \texttt{CT14HERA2} $u_{v}(x)$ and $d_{v}(x)$ valence and $\bar{u}(x)$ %
    and $\bar{d}(x)$ sea distributions for several values of $x$ using the standard triple-differential templates at $Q=3$~TeV.
    To be compared with the ``Angle'' curves of Figs.~\ref{fig:fourCurvesSea0300} and \ref{fig:fourCurvesVal0300}.}
  \label{tab:xTable0300}
\end{table}
The answer to Question 1 is that a global PDF fit which includes DY LHC data below 1 TeV in mass, and binned in rapidity and $\cos\theta^{*}$, would dramatically improve the precision in our knowledge of the up and down PDFs. During the LHC era DY measurements of this kind are likely the only way to reduce uncertainties on the PDFs at high x; no other input data are capable of achieving this improvement. 

\subsection{Impact on the High-Mass Drell-Yan Spectrum}
%
With an updated set of PDFs, we can answer Question 2: the effect of new PDFs on the systematic uncertainty on high-mass DY cross section.  Rather than the enormous extrapolation required of current-day PDFs, the extrapolation from our Control Region to our Signal Region is modest and impactful. In order to make contact with primarily the ATLAS dilepton analysis~\cite{Aaboud2017}, the invariant mass distribution assessed here utilizes leptons that originate in the central-central final state only. \par
\begin{figure}[!t]
  \centering
  \subfigure[]{
    \label{fig:Y21_3000}
    \includegraphics[width=0.43\textwidth]{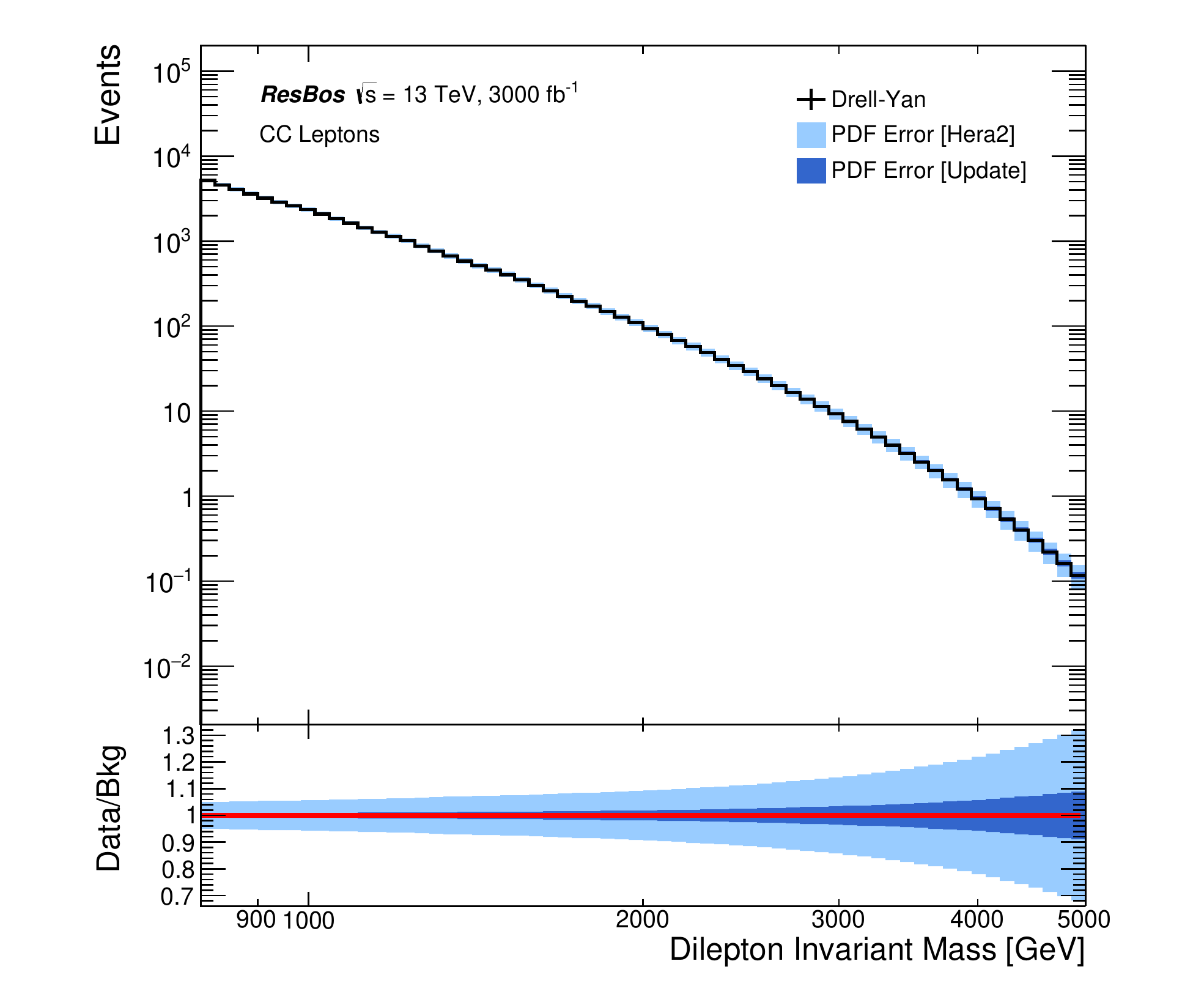}
  }
  \subfigure[]{
    \label{fig:Y23_3000}
    \includegraphics[width=0.43\textwidth]{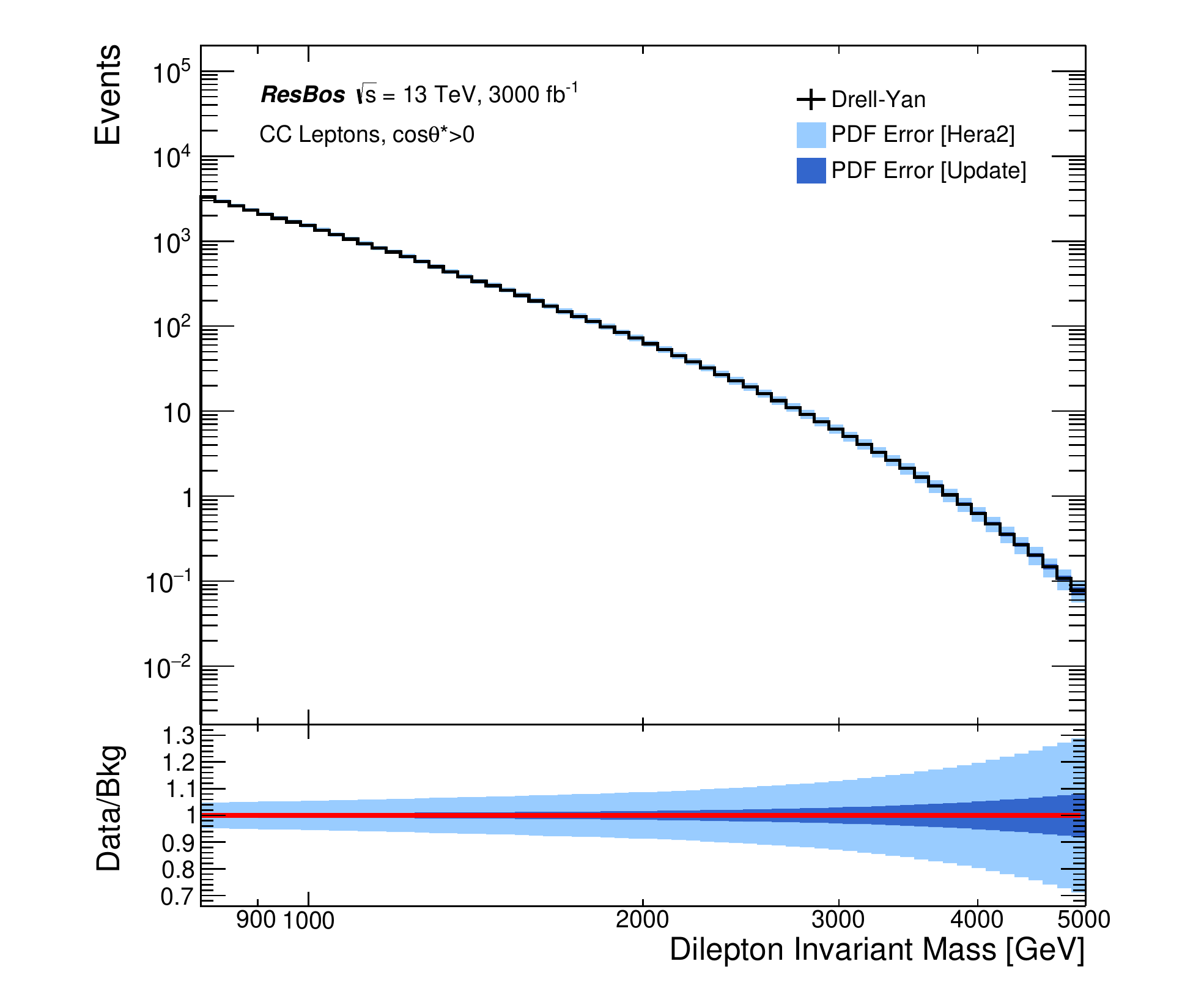}
  }
  \caption{The dilepton invariant mass distribution for ~\subref{fig:Y21_3000} central-central dilepton events, and ~\subref{fig:Y23_3000} with an %
    additional $\cos\theta^{*}>0$ requirement added to the selection. The ratio sub-plot depicts the %
    \texttt{CT14HERA2} PDF uncertainty before and after the 3000~$\mathrm{fb}^{-1}$ update.}
  \label{fig:updatedMass3000}
\end{figure}

The results are presented in Fig.~\ref{fig:updatedMass3000}, which shows the impact of the 3000~$\mathrm{fb}^{-1}$ pseudo-dataset on the high-mass PDF systematic uncertainty. The PDF uncertainty is evaluated at several characteristic values of dilepton mass, which are listed in Table~\ref{tab:updatedMassTable3000}. At $m_{\ell\ell}=5$~TeV, the PDF systematic uncertainty is reduced from 31\% to 8.9\%, a reduction of roughly a factor of 3.5. Similarly, at $m_{\ell\ell}=3$~TeV, the uncertainty is reduced from 15\% to 3.7\%, roughly a factor of 4. In each case, a substantial improvement is obtained compared to the current state-of-the-art predictions (as depicted in Fig.~\ref{fig:massCosSlices}). The PDF uncertainty assessed in the ATLAS dilepton analysis is, for example, 13\% and 29\% at $m_{\ell\ell}=3$ and $m_{\ell\ell}=5$~TeV, respectively.

It is worth remembering that many differential cross section analyses of DY data around the $Z$ peak have been performed over the years, including a triple differential cross section measurement by ATLAS~\cite{Aaboud:2017exx}. We found that because of the extremely high rate, including using the triple differential cross sections to global PDF fitting from the low mass region should indeed be important. However, because of the surprising sensitivities to the parton density flavors, and the enormous rates from the  3000~$\mathrm{fb}^{-1}$ running, about half of the above uncertainty improvement came from the high statistics, low mass region and about half came from the high mass continuum, but low cross section region. Therefore we advocate using the entire di-lepton invariant mass spectrum - from below the $Z$ peak to approximately 1~TeV - as the control region for inputs to future PDF global fitting. The only assumption this carries is that no new physics lurks in the continuum below that boundary.

\begin{table}[!t]
  \begin{ruledtabular}
    \begin{tabular}{c|cc|cc} 
      $m_{\ell\ell}$ [TeV] & \multicolumn{2}{c|}{CC Selection} & \multicolumn{2}{c}{CC+$\cos\theta^{*}$ Selection}\\
      & $\delta^{PDF}_{pre}$ [\%] & $\delta^{PDF}_{post}$ [\%] & $\delta^{PDF}_{pre}$ [\%] & $\delta^{PDF}_{post}$ [\%]\\
      \hline
      1 & 5.9 & 1.0 & 5.6 & 0.9\\
      2 & 9.6 & 2.0 & 8.9 & 1.7 \\
      3 & 15  & 3.7 & 13  & 3.2 \\
      4 & 22  & 6.0 & 20  & 5.3 \\
      5 & 31  & 8.9 & 28  & 8.0 \\
    \end{tabular}
  \end{ruledtabular}
  \caption{The estimated PDF uncertainty in several invariant mass bins for the distributions shown in %
    Fig.~\ref{fig:updatedMass3000}. 
    Two selections are tested: firstly the central-central selection, and secondly for the central-central selection with an additional $\cos\theta^{*} > 0$ requirement. For each selection, the current \texttt{CT14HERA2} uncertainty estimates are shown in the first column, and the result of the 3000 ${\rm fb}^{-1}$ update is shown in the second.
    The %
    pre-update values for the central-central selection are consistent with those assessed in the ATLAS dilepton analysis~\cite{Aaboud2017}.}
  \label{tab:updatedMassTable3000}
\end{table}

\section{Outlook} \label{sec:Outlook}

The impact of a future DY cross section measurement on the \texttt{CT14HERA2} PDF uncertainty was assessed using the 
\texttt{ePump} package at the $\sqrt{s}=13$~TeV LHC with 300~$\mathrm{fb}^{-1}$ and 3000~$\mathrm{fb}^{-1}$ of DY pseudo-data. The fiducial region 
considered for the PDF update was based on three variables: the dilepton mass ($m_{\ell\ell}$), the dilepton rapidity ($y_{\ell\ell}$), 
and the cosine of the polar angle in the CS-frame ($\cos\theta^{*}$). These regions were divided into 1296 histogram bins and used to construct 
\texttt{ePump} pseudo-data and signal templates, which were designed to probe the PDFs in the extreme kinematic regions of 
($x$,$Q^{2}$) only accessible at the LHC. \par
The \texttt{CT14HERA2} PDF set was used for the update, but similar effects would be observed in other PDF sets. The results showed a significant reduction in the uncertainties associated 
with all parton flavors, especially $\bar{u}(x)$ and $\bar{d}(x)$ sea distribution at high $x$. Likewise, these reduced PDF uncertainties, when propagated to the dilepton invariant mass spectrum, lead to a significantly improved description at high mass. \par
These proof-of-concept results indicate a great deal of improvement can still be obtained from precision PDF measurements at LHC.
The use of $\cos\theta^{*}$ as an additional dimension in future PDF global fits is absolutely crucial, as it supplements the more
standard double-differential measurements in invariant mass and rapidity; when used in conjunction, as was done here, the 
reduction in uncertainty can be dramatic. \par

\begin{table}[!t]
  \begin{ruledtabular}
  \begin{tabular}{c|c|ccccc} 
    $m_{\ell\ell}$ [TeV] & \multicolumn{1}{c|}{CC Selection} & \multicolumn{5}{c}{ATLAS Dilepton Analysis}\\
    & $\delta^{PDF}_{post}$ [\%] & $\delta^{PDF}$ [\%] & $\delta^{Choice}$ [\%] & $\delta^{Theory}$ [\%] & $\delta^{Exp}$ [\%] & $\delta^{Total}$ [\%]\\
    \hline
    2 & 2.0 & 8.7 & $<$ 1.0 & 9.8 & 11.0 & 14.7 \\
    4 & 6.0 & 19.0 & 8.4 & 23.0 & 12.8 & 26.3 \\
  \end{tabular}
  \end{ruledtabular}
  \caption{The post-update PDF uncertainty as compared to the current experimental and dominant theoretical uncertainties in the %
    electron channel of the dilepton analysis. As the PDF uncertainty will be reduced well below the current experimental %
    uncertainty, attention will be shifted to the reduction of others, such as the ``PDFChoice'' uncertainty, improving the %
    discovery potential of future iterations of the dilepton analysis.}
  \label{tab:futureUncertainties}
\end{table}

For these reasons, DY cross section measurements could be vital to the success of future searches and measurements at the 
LHC. Not only will the PDF uncertainty that affects the high-mass dilepton analysis be reduced, improving the discovery potential 
of many non-resonant new physics models, but also the inclusion of new and robust data into the modern PDF global fits might even  bring the 
uncertainty estimates of the various global fitting groups into better agreement.

Such an opportunity might result in a reduction of the ``PDF choice'' uncertainty when all PDF groups include triply differential DY data as discussed here. Obviously the goal would be to reach a stage in which the largest uncertainty would cease to be due to the PDFs. Table~\ref{tab:futureUncertainties} 
compares these uncertainties explicitly, where the uncertainty on the QCD background estimate is not included in 
calculating the post-update PDF uncertainty which will be reduced well below the current experimental uncertainty. 

Therefore, for the reasons outlined in this paper, experiments at the LHC and global fitting groups should seriously consider the inclusion of precision measurements of the DY triple-differential cross section over a large invariant mass region in order to further constrain the PDF uncertainties in future PDF global fits. 

\begin{acknowledgments}
	We thank our CTEQ-TEA colleagues for support and discussions.
	This work was supported by the U.S. National Science Foundation under Grants No. PHY-1719914 and PHY-1707812.
	C.-P. Yuan is also grateful for the support from the Wu-Ki Tung endowed chair in particle physics. This work was also supported by Fermi Research Alliance, LLC, under Contract No. DE-AC02-07CH11359 with the U.S. Department of Energy, Office of Science, Office of High Energy Physics.
\end{acknowledgments}

\appendix

\section{The \texttt{ePump} Package}\label{epumpapp}
In a standard PDF global fit, the PDFs are determined by minimizing the function,
\begin{equation} \label{eq:chi2}
  \chi^{2}_{\mathrm{global}} = \sum^{N_{exp}}_{n=1} \chi^{2}_{n}\, ,
\end{equation}
which consists of contributions from $N_{exp}$ fitted experiments, $\chi^{2}_{n}$.  In the simplest case with no correlations between
data points, the contribution from an experiment can be written
\begin{equation} \label{eq:chi2exp}
  \chi^{2}_{n} = \sum^{N_{n}}_{i=1}\frac{(T_{ni}({\bf z})-D_{ni})^2}{\sigma_{ni}^2},
\end{equation}
where $D_{ni}$ is the experimental data value, $\sigma_{ni}$ is the experimental error (combined systematic and statistical), and
$T_{ni}({\bf z})$ is the theory prediction, which depends on the PDFs, which in turn are described by a finite number
of parameters, ${\bf z}$.  In practice, the $\chi^2_n$ for modern experiments will include correlated errors among data points,
and there may be additional terms added to impose constraints on the theoretical parameters, but the general procedure is unchanged.
The central or best-fit PDFs are obtained  by minimizing $\chi^2_{\mathrm{global}}$ with respect to ${\bf z}$.  In addition, $\chi^2_{\mathrm{global}}$ is, to a good approximation, a quadratic function of the parameters around the minimum.  This is the basis for the Hessian approximation for PDF errors, which utilize PDF eigenvector sets, two for each PDF parameter, to evaluate the uncertainty due to the PDFs for any physical observable.  Each PDF eigenvector set corresponds to a movement in the parameter space along the eigenvector directions of the Hessian error matrix around the global minimum of $\chi^2_{\mathrm{global}}$ at a defined confidence level (C.L.). For \texttt{CT14HERA2} PDFs there are $56$ eigenvector sets defined at the 90\% C.L.\par
If the contribution from a new experiment, $\chi^2_{N_{exp}+1}$, is added to the global analysis, the exact solution of the problem
would require finding the new minimum of Eq.~(\ref{eq:chi2}), as well as diagonalizing the new Hessian matrix.  Since this requires
the full data sets from all experiments in the global analysis, as well as the theory calculations for every data point evaluated at many
parameter values, it is an onerous and time-consuming task even for
the global analysis teams that specialize in this endeavor. \par
This is where a tool such as \texttt{ePump} is advantageous. \texttt{ePump} works by using the fact that the original $\chi^2_{\mathrm{global}}$ is well-approximated by the known quadratic function and the fact that the theory predictions for the new observables, 
$T_{N_{exp}+1,i}({\bf z})$, can be approximated using the original Hessian error PDFs.
Under these approximations, the minimization and Hessian diagonalization can be performed algebraically~\cite{Paukkunen:2014zia, Schmidt:2018hvu}, with the numerical computations taking seconds, rather than hours or days.\par
Figure~\ref{fig:ePumpFlow} illustrates the use of \texttt{ePump}. In order to perform the PDF update, \texttt{ePump} requires 
two sets of inputs: data templates and theory templates. The data templates consist of the new experimental data values and
their statistical and systematic uncertainties, including correlations, exactly as would be included in a standard global analysis.
In the case of our present study these are the event counts of the new pseudo-data, along with their associated statistical uncertainties.
The theory templates consist of the corresponding theory predictions for the same observables, evaluated using the central PDF 
and each of the Hessian eigenvector PDFs.  Note that any number of new data sets can be included in the update by \texttt{ePump}, with any number of data points per new data set.\par
The output of \texttt{ePump} is an updated central and Hessian eigenvector PDFs, which
approximate the result that would be obtained from a full global re-analysis that includes the new data.  
As an additional benefit, \texttt{ePump} can also directly output the updated predictions and uncertainties for any other observables 
of interest (such as the cross section in the signal region), without the necessity to recalculate using the updated PDFs.
For more details about the use of \texttt{ePump}, see Ref.~\cite{Schmidt:2018hvu}.
The code for \texttt{ePump} and  more specific details of its usage can be obtained at the website \url{http://hep.pa.msu.edu/epump/}.

In the present study, we have used \texttt{ePump} to assess the reduction of PDF uncertainties from various kinematic selection choices on the Drell-Yan data.
It should be noted that if the included new data “deviate” more from the prediction (based on \texttt{CT14HERA2}),  the result of the ePump analysis will be less reliable. 
This is due to the nature of the ePump method which assumes a quadratic dependence of $\chi^2$ and a linear dependence of observables when the PDFs vary. Since the pseudodata (generated by MMHT14) and the theory predictions (from \texttt{CT14HERA2}) do not differ much, we expect the results of the ePump analysis in our study will hold to a very good approximation. However, should the future data deviate significantly from the theory predictions (from \texttt{CT14HERA2}), a full global analysis, probably with an extended non-perturbative parametrization form, must be carried out. 

\section{High Mass Favoring of $u\bar{u}$ in $pp$ collisions}\label{upapp}

That the $u\bar{u}$ contribution is nearly a factor of four more than the down quark contribution was not expected. In this appendix we show how this comes about.
\subsection{Relearning Drell-Yan Kinematics} \label{sec:relearning}
\begin{table}
  \begin{ruledtabular}
    \begin{tabular}{l|cc}
      & $v_{f}$ & $a_{f}$ \\
      \hline
      $u$ & $+\frac{1}{4}-\frac{2}{3}x_{W}$ & $-\frac{1}{4}$ \\ 
      $d$ & $-\frac{1}{4}+\frac{1}{3}x_{W}$ & $+\frac{1}{4}$ \\
      $\nu_{\ell}$ & $+\frac{1}{4}$ & $-\frac{1}{4}$ \\
      $\ell$ & $-\frac{1}{4}+x_{W}$ & $+\frac{1}{4}$ \\
    \end{tabular}
  \end{ruledtabular}
  \caption{Vector and axial-vector couplings of the SM fermions. Rows specify couplings within each respective fermion generation.}
  \label{tab:EWCharges}
\end{table}
We will exploit a novel feature of the DY subprocess cross section from Eq.~(\ref{eq:sigma3D}) and the definition of $\cos\theta^{*}$ of Eq.~(\ref{eq:cosine}), to form the basis of the PDF update with \eP in Sec.~\ref{sec:approachToPDFErrorReduction}. \par
The function $P_{q}$ of Eq.~(\ref{eq:sigma3D}) encodes the parton-level dynamics with
\begin{equation} \label{eq:PqFactored}
  P_{q} = C^{0}_{q}\left(1+\cos^{2}\theta^{*}\right) + C^{1}_{q}\cos\theta^{*},
\end{equation}
which is a weighted sum of an even function $\left(1+\cos^{2}\theta^{*}\right)$ and an odd function $\cos\theta^{*}$. In the calculation of the total inclusive cross section, the odd term integrates to zero, but is responsible for inducing the well-known $\gamma^{*}/Z$ forward-backward asymmetry $A_{FB}$. \par
The asymmetry coefficients $C^{0}_{q}$ and $C^{1}_{q}$ of Eq.~(\ref{eq:PqFactored}) include the electroweak couplings of the 
initial-state quarks and final-state leptons, and describe the $m_{\ell\ell}$ spectrum as
\begin{align} \label{eq:PqFactoredTerms}
  \begin{split}
    C^{0}_{q}(m_{\ell\ell}) &= Q^{2}_{\ell}Q^{2}_{q} + 2Q_{\ell}Q_{q}v_{\ell}v_{q}\chi_{1}(m_{\ell\ell})%
    + \left(a_{\ell}^{2}+v_{\ell}^{2}\right)\left(a_{q}^{2}+v_{q}^{2}\right)\chi_{2}(m_{\ell\ell})\\
    C^{1}_{q}(m_{\ell\ell}) &= 4Q_{\ell}Q_{q}a_{\ell}a_{q}\chi_{1}(m_{\ell\ell}) + 8a_{\ell}v_{\ell}a_{q}v_{q}\chi_{2}(m_{\ell\ell}).
  \end{split}
\end{align}
\noindent Where 
\begin{align} \label{eq:PqFactoredTerms2}
  \begin{split}
    \chi_{1}(m_{\ell\ell}) &= \frac{1}{\sin\theta_{W}\cos\theta_{W}} \frac{m_{\ell\ell}^{2}\left(m_{\ell\ell}^{2}-m_{Z}^{2}\right)}
        {\left(m_{\ell\ell}^{2}-m_{Z}^{2}\right)^{2}+\Gamma_{Z}^{2}m_{Z}^{2}},\\
        \chi_{2}(m_{\ell\ell}) &= \frac{1}{\sin^{2}\theta_{W}\cos^{2}\theta_{W}} \frac{m_{\ell\ell}^{4}}%
            {\left(m_{\ell\ell}^{2}-m_{Z}^{2}\right)^{2} + \Gamma_{Z}^{2}m_{Z}^{2}}.
  \end{split}
\end{align}
Here $m_{Z}$ and $\Gamma_{Z}$ are the mass and decay width of the SM $Z$ boson, and $Q_{f}$, $v_{f}$, and $a_{f}$ are the electric charge, and the electroweak vector and axial-vector couplings of each fermion, whose values are shown in Table~\ref{tab:EWCharges}. 
The function $\chi_{1}$ results from $\gamma^{*}/Z$ interference, while $\chi_{2}$ arises from pure $Z$ boson exchange. \par
\begin{figure}[!ht]
  \centering
  \includegraphics[width=0.6\textwidth]{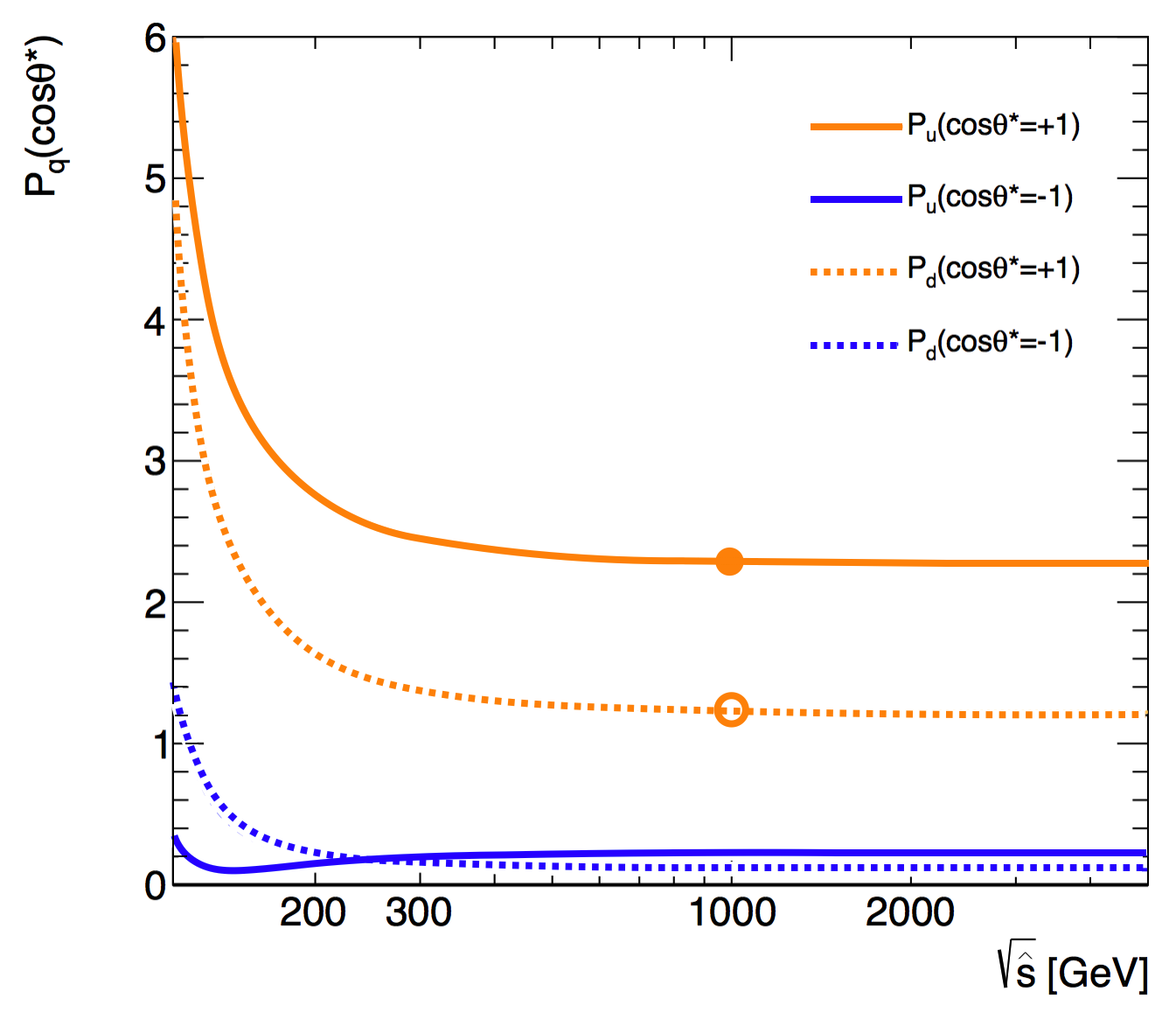}
  \caption{The DY parton-level kinematics as described Eq.~(\ref{eq:PqFactored}). The function $P_{q}$ is evaluated %
    at values of $\cos\theta^{*} = \pm 1$ for both up (solid)- and down (dashed)-type quarks. The dots indicate the values the $P_{u}$ and $P_{d}$ functions at $\cos\theta^{*} = 1.0$ and $\sqrt{\hat{s}} = 1$~TeV.}
  \label{fig:PqPlots}
\end{figure}
\begin{figure}[!ht]
  \centering
  \includegraphics[width=0.6\textwidth]{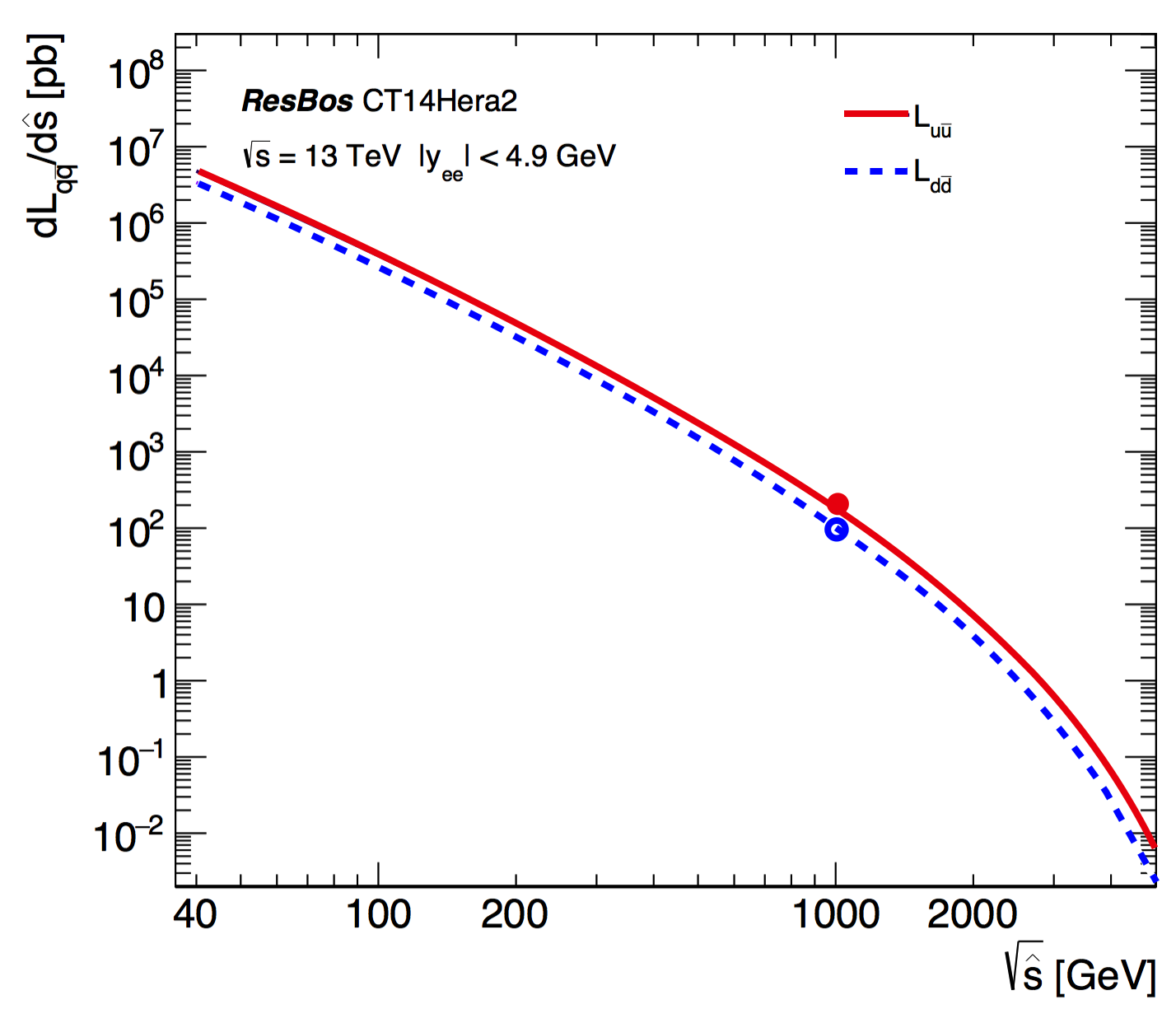}
  \caption{Parton luminosities for $u\bar{u}$ and $d\bar{d}$ DY sub-processes. The dots indicate the values of the luminosity functions at $\sqrt{\hat{s}} = 1$~TeV. }
  \label{fig:LumiPlots}
\end{figure}

Figure~\ref{fig:cos_600_1000} shows that as $\cos\theta^{*}$ nears $+1$, the up quark dominates DY production by almost a factor of four over that of the down quark. Taking apart Eqs.~(\ref{eq:sigma3D}), (\ref{eq:PqFactored}), (\ref{eq:PqFactoredTerms}), and~(\ref{eq:PqFactoredTerms2}) explains this observation. Figure~\ref{fig:PqPlots} shows the quantities $P_u$ and $P_d$ evaluated at $\cos\theta^{*}=\pm1$. The closed (open) circles tag $P_u$ ($P_d$) at a $\sqrt{\hat{s}}=1$~TeV for the  $\cos\theta^{*}=1.0$ curves. The ratio of $P_u/P_d$ is about 2. Figure~\ref{fig:LumiPlots} shows the separate parton luminosity functions $\mathcal{L}_{u\bar{u}}$ and $\mathcal{L}_{d\bar{d}}$ for the leading order $u\bar{u}$ and $d\bar{d}$ sub-processes in accordance with the \texttt{CT14HERA2} PDF set. Here too, the closed and open circles refer to the up-quark and down-quark parton luminosity functions and again, the ratio of $\mathcal{L}_{u\bar{u}}/\mathcal{L}_{d\bar{d}}$ is approximately 1.5. The product of these contributions (i.e., $P_{u}/P_{d} \times (\mathcal{L}_{u\bar{u}}/\mathcal{L}_{d\bar{d}}$) to the rates confirms the near factor of 4 ratio observed in Fig.~\ref{fig:massCosSlices} at $\cos\theta^{*}$ near 1. 

\bibliography{apssamp_20_final} 

\providecommand{\noopsort}[1]{}\providecommand{\singleletter}[1]{#1}%
\begin{thebibliography}{40}%
\makeatletter
\providecommand \@ifxundefined [1]{%
 \@ifx{#1\undefined}
}%
\providecommand \@ifnum [1]{%
 \ifnum #1\expandafter \@firstoftwo
 \else \expandafter \@secondoftwo
 \fi
}%
\providecommand \@ifx [1]{%
 \ifx #1\expandafter \@firstoftwo
 \else \expandafter \@secondoftwo
 \fi
}%
\providecommand \natexlab [1]{#1}%
\providecommand \enquote  [1]{``#1''}%
\providecommand \bibnamefont  [1]{#1}%
\providecommand \bibfnamefont [1]{#1}%
\providecommand \citenamefont [1]{#1}%
\providecommand \href@noop [0]{\@secondoftwo}%
\providecommand \href [0]{\begingroup \@sanitize@url \@href}%
\providecommand \@href[1]{\@@startlink{#1}\@@href}%
\providecommand \@@href[1]{\endgroup#1\@@endlink}%
\providecommand \@sanitize@url [0]{\catcode `\\12\catcode `\$12\catcode
  `\&12\catcode `\#12\catcode `\^12\catcode `\_12\catcode `\%12\relax}%
\providecommand \@@startlink[1]{}%
\providecommand \@@endlink[0]{}%
\providecommand \url  [0]{\begingroup\@sanitize@url \@url }%
\providecommand \@url [1]{\endgroup\@href {#1}{\urlprefix }}%
\providecommand \urlprefix  [0]{URL }%
\providecommand \Eprint [0]{\href }%
\providecommand \doibase [0]{http://dx.doi.org/}%
\providecommand \selectlanguage [0]{\@gobble}%
\providecommand \bibinfo  [0]{\@secondoftwo}%
\providecommand \bibfield  [0]{\@secondoftwo}%
\providecommand \translation [1]{[#1]}%
\providecommand \BibitemOpen [0]{}%
\providecommand \bibitemStop [0]{}%
\providecommand \bibitemNoStop [0]{.\EOS\space}%
\providecommand \EOS [0]{\spacefactor3000\relax}%
\providecommand \BibitemShut  [1]{\csname bibitem#1\endcsname}%
\let\auto@bib@innerbib\@empty
\bibitem [{\citenamefont {Aad}\ \emph {et~al.}(2008)\citenamefont {Aad} \emph
  {et~al.}}]{Aad:2008zzm}%
  \BibitemOpen
  \bibfield  {author} {\bibinfo {author} {\bibfnamefont {G.}~\bibnamefont
  {Aad}} \emph {et~al.} (\bibinfo {collaboration} {ATLAS}),\ }\href {\doibase
  10.1088/1748-0221/3/08/S08003} {\bibfield  {journal} {\bibinfo  {journal}
  {JINST}\ }\textbf {\bibinfo {volume} {3}},\ \bibinfo {pages} {S08003}
  (\bibinfo {year} {2008})}\BibitemShut {NoStop}%
\bibitem [{\citenamefont {Chatrchyan}\ \emph {et~al.}(2008)\citenamefont
  {Chatrchyan} \emph {et~al.}}]{Chatrchyan:2008aa}%
  \BibitemOpen
  \bibfield  {author} {\bibinfo {author} {\bibfnamefont {S.}~\bibnamefont
  {Chatrchyan}} \emph {et~al.} (\bibinfo {collaboration} {CMS}),\ }\href
  {\doibase 10.1088/1748-0221/3/08/S08004} {\bibfield  {journal} {\bibinfo
  {journal} {JINST}\ }\textbf {\bibinfo {volume} {3}},\ \bibinfo {pages}
  {S08004} (\bibinfo {year} {2008})}\BibitemShut {NoStop}%
\bibitem [{\citenamefont {Alves}\ \emph {et~al.}(2008)\citenamefont {Alves}
  \emph {et~al.}}]{Alves:2008zz}%
  \BibitemOpen
  \bibfield  {author} {\bibinfo {author} {\bibfnamefont {A.~A.}\ \bibnamefont
  {Alves}, \bibfnamefont {Jr.}} \emph {et~al.} (\bibinfo {collaboration}
  {LHCb}),\ }\href {\doibase 10.1088/1748-0221/3/08/S08005} {\bibfield
  {journal} {\bibinfo  {journal} {JINST}\ }\textbf {\bibinfo {volume} {3}},\
  \bibinfo {pages} {S08005} (\bibinfo {year} {2008})}\BibitemShut {NoStop}%
\bibitem [{\citenamefont {Dulat}\ \emph {et~al.}(2016)\citenamefont {Dulat},
  \citenamefont {Hou}, \citenamefont {Gao}, \citenamefont {Guzzi},
  \citenamefont {Huston}, \citenamefont {Nadolsky}, \citenamefont {Pumplin},
  \citenamefont {Schmidt}, \citenamefont {Stump},\ and\ \citenamefont
  {Yuan}}]{PhysRevD.93.033006}%
  \BibitemOpen
  \bibfield  {author} {\bibinfo {author} {\bibfnamefont {S.}~\bibnamefont
  {Dulat}}, \bibinfo {author} {\bibfnamefont {T.-J.}\ \bibnamefont {Hou}},
  \bibinfo {author} {\bibfnamefont {J.}~\bibnamefont {Gao}}, \bibinfo {author}
  {\bibfnamefont {M.}~\bibnamefont {Guzzi}}, \bibinfo {author} {\bibfnamefont
  {J.}~\bibnamefont {Huston}}, \bibinfo {author} {\bibfnamefont
  {P.}~\bibnamefont {Nadolsky}}, \bibinfo {author} {\bibfnamefont
  {J.}~\bibnamefont {Pumplin}}, \bibinfo {author} {\bibfnamefont
  {C.}~\bibnamefont {Schmidt}}, \bibinfo {author} {\bibfnamefont
  {D.}~\bibnamefont {Stump}}, \ and\ \bibinfo {author} {\bibfnamefont {C.-P.}\
  \bibnamefont {Yuan}},\ }\href {\doibase 10.1103/PhysRevD.93.033006}
  {\bibfield  {journal} {\bibinfo  {journal} {Phys. Rev. D}\ }\textbf {\bibinfo
  {volume} {93}},\ \bibinfo {pages} {033006} (\bibinfo {year}
  {2016})}\BibitemShut {NoStop}%
\bibitem [{\citenamefont {Martin}\ \emph {et~al.}(2009)\citenamefont {Martin},
  \citenamefont {Stirling}, \citenamefont {Thorne},\ and\ \citenamefont
  {Watt}}]{Martin2009}%
  \BibitemOpen
  \bibfield  {author} {\bibinfo {author} {\bibfnamefont {A.~D.}\ \bibnamefont
  {Martin}}, \bibinfo {author} {\bibfnamefont {W.~J.}\ \bibnamefont
  {Stirling}}, \bibinfo {author} {\bibfnamefont {R.~S.}\ \bibnamefont
  {Thorne}}, \ and\ \bibinfo {author} {\bibfnamefont {G.}~\bibnamefont
  {Watt}},\ }\href {\doibase 10.1140/epjc/s10052-009-1072-5} {\bibfield
  {journal} {\bibinfo  {journal} {The European Physical Journal C}\ }\textbf
  {\bibinfo {volume} {63}},\ \bibinfo {pages} {189} (\bibinfo {year}
  {2009})}\BibitemShut {NoStop}%
\bibitem [{\citenamefont {Ball}\ \emph {et~al.}(2015)\citenamefont {Ball},
  \citenamefont {Bertone}, \citenamefont {Carrazza}, \citenamefont {Deans},
  \citenamefont {Del~Debbio}, \citenamefont {Forte}, \citenamefont {Guffanti},
  \citenamefont {Hartland}, \citenamefont {Latorre}, \citenamefont {Rojo},\
  and\ \citenamefont {Ubiali}}]{Ball2015}%
  \BibitemOpen
  \bibfield  {author} {\bibinfo {author} {\bibfnamefont {R.~D.}\ \bibnamefont
  {Ball}}, \bibinfo {author} {\bibfnamefont {V.}~\bibnamefont {Bertone}},
  \bibinfo {author} {\bibfnamefont {S.}~\bibnamefont {Carrazza}}, \bibinfo
  {author} {\bibfnamefont {C.~S.}\ \bibnamefont {Deans}}, \bibinfo {author}
  {\bibfnamefont {L.}~\bibnamefont {Del~Debbio}}, \bibinfo {author}
  {\bibfnamefont {S.}~\bibnamefont {Forte}}, \bibinfo {author} {\bibfnamefont
  {A.}~\bibnamefont {Guffanti}}, \bibinfo {author} {\bibfnamefont {N.~P.}\
  \bibnamefont {Hartland}}, \bibinfo {author} {\bibfnamefont {J.~I.}\
  \bibnamefont {Latorre}}, \bibinfo {author} {\bibfnamefont {J.}~\bibnamefont
  {Rojo}}, \ and\ \bibinfo {author} {\bibfnamefont {M.}~\bibnamefont
  {Ubiali}},\ }\href {\doibase 10.1007/JHEP04(2015)040} {\bibfield  {journal}
  {\bibinfo  {journal} {Journal of High Energy Physics}\ }\textbf {\bibinfo
  {volume} {2015}},\ \bibinfo {pages} {40} (\bibinfo {year}
  {2015})}\BibitemShut {NoStop}%
\bibitem [{\citenamefont {Aaboud}\ \emph {et~al.}(2017)\citenamefont {Aaboud}
  \emph {et~al.}}]{Aaboud2017}%
  \BibitemOpen
  \bibfield  {author} {\bibinfo {author} {\bibfnamefont {M.}~\bibnamefont
  {Aaboud}} \emph {et~al.},\ }\href {\doibase 10.1007/JHEP10(2017)182}
  {\bibfield  {journal} {\bibinfo  {journal} {Journal of High Energy Physics}\
  }\textbf {\bibinfo {volume} {2017}},\ \bibinfo {pages} {182} (\bibinfo {year}
  {2017})}\BibitemShut {NoStop}%
\bibitem [{\citenamefont {Sirunyan}\ \emph {et~al.}(2018)\citenamefont
  {Sirunyan} \emph {et~al.}}]{Sirunyan:2018exx}%
  \BibitemOpen
  \bibfield  {author} {\bibinfo {author} {\bibfnamefont {A.~M.}\ \bibnamefont
  {Sirunyan}} \emph {et~al.} (\bibinfo {collaboration} {CMS}),\ }\href
  {\doibase 10.1007/JHEP06(2018)120} {\bibfield  {journal} {\bibinfo  {journal}
  {J. High Energ. Physics}\ }\textbf {\bibinfo {volume} {2018}},\ \bibinfo
  {pages} {120} (\bibinfo {year} {2018})},\ \Eprint
  {http://arxiv.org/abs/1803.06292} {arXiv:1803.06292 [hep-ex]} \BibitemShut
  {NoStop}%
\bibitem [{\citenamefont {Abe}\ \emph {et~al.}(1995)\citenamefont {Abe} \emph
  {et~al.}}]{Abe:1994rj}%
  \BibitemOpen
  \bibfield  {author} {\bibinfo {author} {\bibfnamefont {F.}~\bibnamefont
  {Abe}} \emph {et~al.} (\bibinfo {collaboration} {CDF}),\ }\href {\doibase
  10.1103/PhysRevLett.74.850} {\bibfield  {journal} {\bibinfo  {journal} {Phys.
  Rev. Lett.}\ }\textbf {\bibinfo {volume} {74}},\ \bibinfo {pages} {850}
  (\bibinfo {year} {1995})},\ \Eprint {http://arxiv.org/abs/hep-ex/9501008}
  {arXiv:hep-ex/9501008 [hep-ex]} \BibitemShut {NoStop}%
\bibitem [{\citenamefont {Abe}\ \emph {et~al.}(1996)\citenamefont {Abe} \emph
  {et~al.}}]{Abe:1996us}%
  \BibitemOpen
  \bibfield  {author} {\bibinfo {author} {\bibfnamefont {F.}~\bibnamefont
  {Abe}} \emph {et~al.} (\bibinfo {collaboration} {CDF}),\ }\href {\doibase
  10.1103/PhysRevLett.77.2616} {\bibfield  {journal} {\bibinfo  {journal}
  {Phys. Rev. Lett.}\ }\textbf {\bibinfo {volume} {77}},\ \bibinfo {pages}
  {2616} (\bibinfo {year} {1996})}\BibitemShut {NoStop}%
\bibitem [{\citenamefont {Acosta}\ \emph {et~al.}(2005)\citenamefont {Acosta}
  \emph {et~al.}}]{Acosta:2005ud}%
  \BibitemOpen
  \bibfield  {author} {\bibinfo {author} {\bibfnamefont {D.}~\bibnamefont
  {Acosta}} \emph {et~al.} (\bibinfo {collaboration} {CDF}),\ }\href {\doibase
  10.1103/PhysRevD.71.051104} {\bibfield  {journal} {\bibinfo  {journal} {Phys.
  Rev.}\ }\textbf {\bibinfo {volume} {D71}},\ \bibinfo {pages} {051104}
  (\bibinfo {year} {2005})},\ \Eprint {http://arxiv.org/abs/hep-ex/0501023}
  {arXiv:hep-ex/0501023 [hep-ex]} \BibitemShut {NoStop}%
\bibitem [{\citenamefont {Abazov}\ \emph {et~al.}(2015)\citenamefont {Abazov}
  \emph {et~al.}}]{D0:2014kma}%
  \BibitemOpen
  \bibfield  {author} {\bibinfo {author} {\bibfnamefont {V.~M.}\ \bibnamefont
  {Abazov}} \emph {et~al.} (\bibinfo {collaboration} {D0}),\ }\href {\doibase
  10.1103/PhysRevD.91.032007, 10.1103/PhysRevD.91.079901} {\bibfield  {journal}
  {\bibinfo  {journal} {Phys. Rev.}\ }\textbf {\bibinfo {volume} {D91}},\
  \bibinfo {pages} {032007} (\bibinfo {year} {2015})},\ \bibinfo {note}
  {[Erratum: Phys. Rev.D91,no.7,079901(2015)]},\ \Eprint
  {http://arxiv.org/abs/1412.2862} {arXiv:1412.2862 [hep-ex]} \BibitemShut
  {NoStop}%
\bibitem [{\citenamefont {Abazov}\ \emph {et~al.}(2008)\citenamefont {Abazov}
  \emph {et~al.}}]{Abazov:2007pm}%
  \BibitemOpen
  \bibfield  {author} {\bibinfo {author} {\bibfnamefont {V.~M.}\ \bibnamefont
  {Abazov}} \emph {et~al.} (\bibinfo {collaboration} {D0}),\ }\href {\doibase
  10.1103/PhysRevD.77.011106} {\bibfield  {journal} {\bibinfo  {journal} {Phys.
  Rev.}\ }\textbf {\bibinfo {volume} {D77}},\ \bibinfo {pages} {011106}
  (\bibinfo {year} {2008})},\ \Eprint {http://arxiv.org/abs/0709.4254}
  {arXiv:0709.4254 [hep-ex]} \BibitemShut {NoStop}%
\bibitem [{\citenamefont {Aad}\ \emph {et~al.}(2012)\citenamefont {Aad} \emph
  {et~al.}}]{Aad:2011dm}%
  \BibitemOpen
  \bibfield  {author} {\bibinfo {author} {\bibfnamefont {G.}~\bibnamefont
  {Aad}} \emph {et~al.} (\bibinfo {collaboration} {ATLAS}),\ }\href {\doibase
  10.1103/PhysRevD.85.072004} {\bibfield  {journal} {\bibinfo  {journal} {Phys.
  Rev.}\ }\textbf {\bibinfo {volume} {D85}},\ \bibinfo {pages} {072004}
  (\bibinfo {year} {2012})},\ \Eprint {http://arxiv.org/abs/1109.5141}
  {arXiv:1109.5141 [hep-ex]} \BibitemShut {NoStop}%
\bibitem [{\citenamefont {Chatrchyan}\ \emph {et~al.}(2012)\citenamefont
  {Chatrchyan} \emph {et~al.}}]{Chatrchyan:2012xt}%
  \BibitemOpen
  \bibfield  {author} {\bibinfo {author} {\bibfnamefont {S.}~\bibnamefont
  {Chatrchyan}} \emph {et~al.} (\bibinfo {collaboration} {CMS}),\ }\href
  {\doibase 10.1103/PhysRevLett.109.111806} {\bibfield  {journal} {\bibinfo
  {journal} {Phys. Rev. Lett.}\ }\textbf {\bibinfo {volume} {109}},\ \bibinfo
  {pages} {111806} (\bibinfo {year} {2012})},\ \Eprint
  {http://arxiv.org/abs/1206.2598} {arXiv:1206.2598 [hep-ex]} \BibitemShut
  {NoStop}%
\bibitem [{\citenamefont {Chatrchyan}\ \emph {et~al.}(2014)\citenamefont
  {Chatrchyan} \emph {et~al.}}]{Chatrchyan:2013mza}%
  \BibitemOpen
  \bibfield  {author} {\bibinfo {author} {\bibfnamefont {S.}~\bibnamefont
  {Chatrchyan}} \emph {et~al.} (\bibinfo {collaboration} {CMS}),\ }\href
  {\doibase 10.1103/PhysRevD.90.032004} {\bibfield  {journal} {\bibinfo
  {journal} {Phys. Rev.}\ }\textbf {\bibinfo {volume} {D90}},\ \bibinfo {pages}
  {032004} (\bibinfo {year} {2014})},\ \Eprint {http://arxiv.org/abs/1312.6283}
  {arXiv:1312.6283 [hep-ex]} \BibitemShut {NoStop}%
\bibitem [{\citenamefont {Aaij}\ \emph {et~al.}(2012)\citenamefont {Aaij} \emph
  {et~al.}}]{Aaij:2012vn}%
  \BibitemOpen
  \bibfield  {author} {\bibinfo {author} {\bibfnamefont {R.}~\bibnamefont
  {Aaij}} \emph {et~al.} (\bibinfo {collaboration} {LHCb}),\ }\href {\doibase
  10.1007/JHEP06(2012)058} {\bibfield  {journal} {\bibinfo  {journal} {JHEP}\
  }\textbf {\bibinfo {volume} {06}},\ \bibinfo {pages} {058} (\bibinfo {year}
  {2012})},\ \Eprint {http://arxiv.org/abs/1204.1620} {arXiv:1204.1620
  [hep-ex]} \BibitemShut {NoStop}%
\bibitem [{\citenamefont {C.~Schmidt}\ and\ \citenamefont
  {Yuan}(2018)}]{Schmidt:2018hvu}%
  \BibitemOpen
  \bibfield  {author} {\bibinfo {author} {\bibfnamefont {C.~P.~Y.}\
  \bibnamefont {C.~Schmidt}, \bibfnamefont {J.~Pumplin}}\ and\ \bibinfo
  {author} {\bibfnamefont {P.}~\bibnamefont {Yuan}},\ }\href {\doibase
  10.1007/JHEP12(2014)100} {\bibfield  {journal} {\bibinfo  {journal} {J. High
  Energ. Phys.}\ } (\bibinfo {year} {2018}),\ 10.1007/JHEP12(2014)100},\
  \Eprint {http://arxiv.org/abs/1806.07950} {arXiv:1806.07950 [hep-ph]}
  \BibitemShut {NoStop}%
\bibitem [{\citenamefont {Bertone}\ \emph {et~al.}(2018)\citenamefont {Bertone}
  \emph {et~al.}}]{Bertone:2017tig}%
  \BibitemOpen
  \bibfield  {author} {\bibinfo {author} {\bibfnamefont {V.}~\bibnamefont
  {Bertone}} \emph {et~al.} (\bibinfo {collaboration} {xFitter Developers'
  Team}),\ }\bibfield  {booktitle} {\emph {\bibinfo {booktitle} {{Proceedings,
  25th International Workshop on Deep-Inelastic Scattering and Related Topics
  (DIS 2017): Birmingham, UK, April 3-7, 2017}}},\ }\href@noop {} {\bibfield
  {journal} {\bibinfo  {journal} {PoS}\ }\textbf {\bibinfo {volume}
  {DIS2017}},\ \bibinfo {pages} {203} (\bibinfo {year} {2018})},\ \Eprint
  {http://arxiv.org/abs/1709.01151} {arXiv:1709.01151 [hep-ph]} \BibitemShut
  {NoStop}%
\bibitem [{\citenamefont {Collins}\ and\ \citenamefont
  {Soper}(1977)}]{PhysRevD.16.2219}%
  \BibitemOpen
  \bibfield  {author} {\bibinfo {author} {\bibfnamefont {J.~C.}\ \bibnamefont
  {Collins}}\ and\ \bibinfo {author} {\bibfnamefont {D.~E.}\ \bibnamefont
  {Soper}},\ }\href {\doibase 10.1103/PhysRevD.16.2219} {\bibfield  {journal}
  {\bibinfo  {journal} {Phys. Rev. D}\ }\textbf {\bibinfo {volume} {16}},\
  \bibinfo {pages} {2219} (\bibinfo {year} {1977})}\BibitemShut {NoStop}%
\bibitem [{\citenamefont {Abazovai}\ \emph {et~al.}(2011)\citenamefont
  {Abazovai} \emph {et~al.}}]{ABAZOV201188}%
  \BibitemOpen
  \bibfield  {author} {\bibinfo {author} {\bibfnamefont {V.~M.}\ \bibnamefont
  {Abazovai}} \emph {et~al.} (\bibinfo {collaboration} {D\O}),\ }\href
  {\doibase https://doi.org/10.1016/j.physletb.2010.10.059} {\bibfield
  {journal} {\bibinfo  {journal} {Physics Letters B}\ }\textbf {\bibinfo
  {volume} {695}},\ \bibinfo {pages} {88 } (\bibinfo {year}
  {2011})}\BibitemShut {NoStop}%
\bibitem [{\citenamefont {Aaltonen}\ \emph {et~al.}(2011)\citenamefont
  {Aaltonen} \emph {et~al.}}]{PhysRevLett.106.121801}%
  \BibitemOpen
  \bibfield  {author} {\bibinfo {author} {\bibfnamefont {T.}~\bibnamefont
  {Aaltonen}} \emph {et~al.} (\bibinfo {collaboration} {CDF}),\ }\href
  {\doibase 10.1103/PhysRevLett.106.121801} {\bibfield  {journal} {\bibinfo
  {journal} {Phys. Rev. Lett.}\ }\textbf {\bibinfo {volume} {106}},\ \bibinfo
  {pages} {121801} (\bibinfo {year} {2011})}\BibitemShut {NoStop}%
\bibitem [{\citenamefont {Aad}\ \emph {et~al.}(2014)\citenamefont {Aad} \emph
  {et~al.}}]{PhysRevD.90.052005}%
  \BibitemOpen
  \bibfield  {author} {\bibinfo {author} {\bibfnamefont {G.}~\bibnamefont
  {Aad}} \emph {et~al.} (\bibinfo {collaboration} {ATLAS}),\ }\href {\doibase
  10.1103/PhysRevD.90.052005} {\bibfield  {journal} {\bibinfo  {journal} {Phys.
  Rev. D}\ }\textbf {\bibinfo {volume} {90}},\ \bibinfo {pages} {052005}
  (\bibinfo {year} {2014})}\BibitemShut {NoStop}%
\bibitem [{\citenamefont {Aaboud}\ \emph {et~al.}(2016)\citenamefont {Aaboud}
  \emph {et~al.}}]{2016372}%
  \BibitemOpen
  \bibfield  {author} {\bibinfo {author} {\bibfnamefont {M.}~\bibnamefont
  {Aaboud}} \emph {et~al.} (\bibinfo {collaboration} {ATLAS}),\ }\href
  {\doibase https://doi.org/10.1016/j.physletb.2016.08.055} {\bibfield
  {journal} {\bibinfo  {journal} {Physics Letters B}\ }\textbf {\bibinfo
  {volume} {761}},\ \bibinfo {pages} {372 } (\bibinfo {year}
  {2016})}\BibitemShut {NoStop}%
\bibitem [{\citenamefont {Khachatryan}\ \emph {et~al.}(2015)\citenamefont
  {Khachatryan} \emph {et~al.}}]{Khachatryan2015}%
  \BibitemOpen
  \bibfield  {author} {\bibinfo {author} {\bibfnamefont {V.}~\bibnamefont
  {Khachatryan}} \emph {et~al.} (\bibinfo {collaboration} {CMS}),\ }\href
  {\doibase 10.1007/JHEP04(2015)025} {\bibfield  {journal} {\bibinfo  {journal}
  {Journal of High Energy Physics}\ }\textbf {\bibinfo {volume} {2015}},\
  \bibinfo {pages} {25} (\bibinfo {year} {2015})}\BibitemShut {NoStop}%
\bibitem [{\citenamefont {Khachatryan}\ \emph {et~al.}(2017)\citenamefont
  {Khachatryan} \emph {et~al.}}]{201757}%
  \BibitemOpen
  \bibfield  {author} {\bibinfo {author} {\bibfnamefont {V.}~\bibnamefont
  {Khachatryan}} \emph {et~al.} (\bibinfo {collaboration} {CMS}),\ }\href
  {\doibase https://doi.org/10.1016/j.physletb.2017.02.010} {\bibfield
  {journal} {\bibinfo  {journal} {Physics Letters B}\ }\textbf {\bibinfo
  {volume} {768}},\ \bibinfo {pages} {57 } (\bibinfo {year}
  {2017})}\BibitemShut {NoStop}%
\bibitem [{\citenamefont {Langacker}(2009)}]{RevModPhys.81.1199}%
  \BibitemOpen
  \bibfield  {author} {\bibinfo {author} {\bibfnamefont {P.}~\bibnamefont
  {Langacker}},\ }\href {\doibase 10.1103/RevModPhys.81.1199} {\bibfield
  {journal} {\bibinfo  {journal} {Rev. Mod. Phys.}\ }\textbf {\bibinfo {volume}
  {81}},\ \bibinfo {pages} {1199} (\bibinfo {year} {2009})}\BibitemShut
  {NoStop}%
\bibitem [{\citenamefont {London}\ and\ \citenamefont
  {Rosner}(1986)}]{PhysRevD.34.1530}%
  \BibitemOpen
  \bibfield  {author} {\bibinfo {author} {\bibfnamefont {D.}~\bibnamefont
  {London}}\ and\ \bibinfo {author} {\bibfnamefont {J.~L.}\ \bibnamefont
  {Rosner}},\ }\href {\doibase 10.1103/PhysRevD.34.1530} {\bibfield  {journal}
  {\bibinfo  {journal} {Phys. Rev. D}\ }\textbf {\bibinfo {volume} {34}},\
  \bibinfo {pages} {1530} (\bibinfo {year} {1986})}\BibitemShut {NoStop}%
\bibitem [{\citenamefont {Butterworth}\ \emph {et~al.}(2016)\citenamefont
  {Butterworth} \emph {et~al.}}]{Butterworth:2015oua}%
  \BibitemOpen
  \bibfield  {author} {\bibinfo {author} {\bibfnamefont {J.}~\bibnamefont
  {Butterworth}} \emph {et~al.},\ }\href {\doibase
  10.1088/0954-3899/43/2/023001} {\bibfield  {journal} {\bibinfo  {journal} {J.
  Phys. G}\ }\textbf {\bibinfo {volume} {43}},\ \bibinfo {pages} {023001}
  (\bibinfo {year} {2016})},\ \Eprint {http://arxiv.org/abs/1510.03865}
  {arXiv:1510.03865 [hep-ph]} \BibitemShut {NoStop}%
\bibitem [{\citenamefont {Ellis}(2017)}]{Ellis:2016jkw}%
  \BibitemOpen
  \bibfield  {author} {\bibinfo {author} {\bibfnamefont {J.}~\bibnamefont
  {Ellis}},\ }\href {\doibase 10.1016/j.cpc.2016.08.019} {\bibfield  {journal}
  {\bibinfo  {journal} {Comput. Phys. Commun.}\ }\textbf {\bibinfo {volume}
  {210}},\ \bibinfo {pages} {103} (\bibinfo {year} {2017})},\ \Eprint
  {http://arxiv.org/abs/1601.05437} {arXiv:1601.05437 [hep-ph]} \BibitemShut
  {NoStop}%
\bibitem [{\citenamefont {Drell}\ and\ \citenamefont
  {Yan}(1970)}]{PhysRevLett.25.316}%
  \BibitemOpen
  \bibfield  {author} {\bibinfo {author} {\bibfnamefont {S.~D.}\ \bibnamefont
  {Drell}}\ and\ \bibinfo {author} {\bibfnamefont {T.-M.}\ \bibnamefont
  {Yan}},\ }\href {\doibase 10.1103/PhysRevLett.25.316} {\bibfield  {journal}
  {\bibinfo  {journal} {Phys. Rev. Lett.}\ }\textbf {\bibinfo {volume} {25}},\
  \bibinfo {pages} {316} (\bibinfo {year} {1970})}\BibitemShut {NoStop}%
\bibitem [{\citenamefont {Ladinsky}\ and\ \citenamefont
  {Yuan}(1994)}]{PhysRevD.50.R4239}%
  \BibitemOpen
  \bibfield  {author} {\bibinfo {author} {\bibfnamefont {G.~A.}\ \bibnamefont
  {Ladinsky}}\ and\ \bibinfo {author} {\bibfnamefont {C.-P.}\ \bibnamefont
  {Yuan}},\ }\href {\doibase 10.1103/PhysRevD.50.R4239} {\bibfield  {journal}
  {\bibinfo  {journal} {Phys. Rev. D}\ }\textbf {\bibinfo {volume} {50}},\
  \bibinfo {pages} {R4239} (\bibinfo {year} {1994})}\BibitemShut {NoStop}%
\bibitem [{\citenamefont {Bal\'azs}\ and\ \citenamefont
  {Yuan}(1997)}]{PhysRevD.56.5558}%
  \BibitemOpen
  \bibfield  {author} {\bibinfo {author} {\bibfnamefont {C.}~\bibnamefont
  {Bal\'azs}}\ and\ \bibinfo {author} {\bibfnamefont {C.-P.}\ \bibnamefont
  {Yuan}},\ }\href {\doibase 10.1103/PhysRevD.56.5558} {\bibfield  {journal}
  {\bibinfo  {journal} {Phys. Rev. D}\ }\textbf {\bibinfo {volume} {56}},\
  \bibinfo {pages} {5558} (\bibinfo {year} {1997})}\BibitemShut {NoStop}%
\bibitem [{\citenamefont {Landry}\ \emph {et~al.}(2003)\citenamefont {Landry},
  \citenamefont {Brock}, \citenamefont {Nadolsky},\ and\ \citenamefont
  {Yuan}}]{PhysRevD.67.073016}%
  \BibitemOpen
  \bibfield  {author} {\bibinfo {author} {\bibfnamefont {F.}~\bibnamefont
  {Landry}}, \bibinfo {author} {\bibfnamefont {R.}~\bibnamefont {Brock}},
  \bibinfo {author} {\bibfnamefont {P.~M.}\ \bibnamefont {Nadolsky}}, \ and\
  \bibinfo {author} {\bibfnamefont {C.-P.}\ \bibnamefont {Yuan}},\ }\href
  {\doibase 10.1103/PhysRevD.67.073016} {\bibfield  {journal} {\bibinfo
  {journal} {Phys. Rev. D}\ }\textbf {\bibinfo {volume} {67}},\ \bibinfo
  {pages} {073016} (\bibinfo {year} {2003})}\BibitemShut {NoStop}%
\bibitem [{\citenamefont {Sirunyan}\ \emph {et~al.}(2017)\citenamefont
  {Sirunyan} \emph {et~al.}}]{Aaboud:2017exx}%
  \BibitemOpen
  \bibfield  {author} {\bibinfo {author} {\bibfnamefont {A.~M.}\ \bibnamefont
  {Sirunyan}} \emph {et~al.} (\bibinfo {collaboration} {ATLAS}),\ }\href
  {\doibase 10.1007/JHEP12} {\bibfield  {journal} {\bibinfo  {journal} {J. High
  Energ. Physics}\ }\textbf {\bibinfo {volume} {2017}},\ \bibinfo {pages} {59}
  (\bibinfo {year} {2017})},\ \Eprint {http://arxiv.org/abs/1710.05167}
  {arXiv:1710.05167 [hep-ex]} \BibitemShut {NoStop}%
\bibitem [{\citenamefont {Boonekamp}\ \emph {et~al.}(2009)\citenamefont
  {Boonekamp}, \citenamefont {Chevallier}, \citenamefont {Royon},\ and\
  \citenamefont {Schoeffel}}]{Boonekamp:2009yd}%
  \BibitemOpen
  \bibfield  {author} {\bibinfo {author} {\bibfnamefont {M.}~\bibnamefont
  {Boonekamp}}, \bibinfo {author} {\bibfnamefont {F.}~\bibnamefont
  {Chevallier}}, \bibinfo {author} {\bibfnamefont {C.}~\bibnamefont {Royon}}, \
  and\ \bibinfo {author} {\bibfnamefont {L.}~\bibnamefont {Schoeffel}},\
  }\href@noop {} {\bibfield  {journal} {\bibinfo  {journal} {Acta Phys.
  Polon.}\ }\textbf {\bibinfo {volume} {B40}},\ \bibinfo {pages} {2239}
  (\bibinfo {year} {2009})},\ \Eprint {http://arxiv.org/abs/0902.1678}
  {arXiv:0902.1678 [hep-ph]} \BibitemShut {NoStop}%
\bibitem [{\citenamefont {Hou}\ \emph {et~al.}(2017)\citenamefont {Hou},
  \citenamefont {Dulat}, \citenamefont {Gao}, \citenamefont {Guzzi},
  \citenamefont {Huston}, \citenamefont {Nadolsky}, \citenamefont {Pumplin},
  \citenamefont {Schmidt}, \citenamefont {Stump},\ and\ \citenamefont
  {Yuan}}]{PhysRevD.95.034003}%
  \BibitemOpen
  \bibfield  {author} {\bibinfo {author} {\bibfnamefont {T.-J.}\ \bibnamefont
  {Hou}}, \bibinfo {author} {\bibfnamefont {S.}~\bibnamefont {Dulat}}, \bibinfo
  {author} {\bibfnamefont {J.}~\bibnamefont {Gao}}, \bibinfo {author}
  {\bibfnamefont {M.}~\bibnamefont {Guzzi}}, \bibinfo {author} {\bibfnamefont
  {J.}~\bibnamefont {Huston}}, \bibinfo {author} {\bibfnamefont
  {P.}~\bibnamefont {Nadolsky}}, \bibinfo {author} {\bibfnamefont
  {J.}~\bibnamefont {Pumplin}}, \bibinfo {author} {\bibfnamefont
  {C.}~\bibnamefont {Schmidt}}, \bibinfo {author} {\bibfnamefont
  {D.}~\bibnamefont {Stump}}, \ and\ \bibinfo {author} {\bibfnamefont {C.-P.}\
  \bibnamefont {Yuan}},\ }\href {\doibase 10.1103/PhysRevD.95.034003}
  {\bibfield  {journal} {\bibinfo  {journal} {Phys. Rev. D}\ }\textbf {\bibinfo
  {volume} {95}},\ \bibinfo {pages} {034003} (\bibinfo {year}
  {2017})}\BibitemShut {NoStop}%
\bibitem [{\citenamefont {Alwall}\ \emph {et~al.}(2014)\citenamefont {Alwall},
  \citenamefont {Frederix}, \citenamefont {Frixione}, \citenamefont {Hirschi},
  \citenamefont {Maltoni}, \citenamefont {Mattelaer}, \citenamefont {Shao},
  \citenamefont {Stelzer}, \citenamefont {Torrielli},\ and\ \citenamefont
  {Zaro}}]{Alwall:2014hca}%
  \BibitemOpen
  \bibfield  {author} {\bibinfo {author} {\bibfnamefont {J.}~\bibnamefont
  {Alwall}}, \bibinfo {author} {\bibfnamefont {R.}~\bibnamefont {Frederix}},
  \bibinfo {author} {\bibfnamefont {S.}~\bibnamefont {Frixione}}, \bibinfo
  {author} {\bibfnamefont {V.}~\bibnamefont {Hirschi}}, \bibinfo {author}
  {\bibfnamefont {F.}~\bibnamefont {Maltoni}}, \bibinfo {author} {\bibfnamefont
  {O.}~\bibnamefont {Mattelaer}}, \bibinfo {author} {\bibfnamefont {H.~S.}\
  \bibnamefont {Shao}}, \bibinfo {author} {\bibfnamefont {T.}~\bibnamefont
  {Stelzer}}, \bibinfo {author} {\bibfnamefont {P.}~\bibnamefont {Torrielli}},
  \ and\ \bibinfo {author} {\bibfnamefont {M.}~\bibnamefont {Zaro}},\ }\href
  {\doibase 10.1007/JHEP07(2014)079} {\bibfield  {journal} {\bibinfo  {journal}
  {JHEP}\ }\textbf {\bibinfo {volume} {07}},\ \bibinfo {pages} {079} (\bibinfo
  {year} {2014})},\ \Eprint {http://arxiv.org/abs/1405.0301} {arXiv:1405.0301
  [hep-ph]} \BibitemShut {NoStop}%
\bibitem [{\citenamefont {{Shalaev, V.}}\ \emph {et~al.}(2018)\citenamefont
  {{Shalaev, V.}}, \citenamefont {{Gorbunov, I.}},\ and\ \citenamefont
  {{Shmatov, S.}}}]{refId0}%
  \BibitemOpen
  \bibfield  {author} {\bibinfo {author} {\bibnamefont {{Shalaev, V.}}},
  \bibinfo {author} {\bibnamefont {{Gorbunov, I.}}}, \ and\ \bibinfo {author}
  {\bibnamefont {{Shmatov, S.}}},\ }\href {\doibase
  10.1051/epjconf/201817704010} {\bibfield  {journal} {\bibinfo  {journal} {EPJ
  Web Conf.}\ }\textbf {\bibinfo {volume} {177}},\ \bibinfo {pages} {04010}
  (\bibinfo {year} {2018})}\BibitemShut {NoStop}%
\bibitem [{\citenamefont {Paukkunen}\ and\ \citenamefont
  {Zurita}(2014)}]{Paukkunen:2014zia}%
  \BibitemOpen
  \bibfield  {author} {\bibinfo {author} {\bibfnamefont {H.}~\bibnamefont
  {Paukkunen}}\ and\ \bibinfo {author} {\bibfnamefont {P.}~\bibnamefont
  {Zurita}},\ }\href@noop {} {\bibfield  {journal} {\bibinfo  {journal} {JHEP}\
  }\textbf {\bibinfo {volume} {1412}},\ \bibinfo {pages} {100} (\bibinfo {year}
  {2014})},\ \Eprint {http://arxiv.org/abs/1402.6623} {arXiv:1402.6623
  [hep-ph]} \BibitemShut {NoStop}%
\end{thebibliography}%
\end{document}